\documentclass[aps,article,twocolumn,longbibliography,citeautoscript,footinbib,superscriptaddress,prb]{revtex4-2}

\pdfoutput=1						

\usepackage[charter]{mathdesign}	
\usepackage{amsmath}					
\usepackage{mathrsfs}					

\usepackage[pdftex]{graphicx}
\usepackage{microtype}
\usepackage{gensymb}
\usepackage{mdframed}
\usepackage[bbgreekl]{mathbbol}

\usepackage[svgnames]{xcolor}
\usepackage[colorlinks=True,linkcolor=DarkRed,citecolor=ForestGreen,urlcolor=MediumBlue, pdfstartview=FitH,bookmarks=False,pdfpagemode=UseNone
]{hyperref}

\usepackage{natbib}
\usepackage[bbgreekl]{mathbbol}

\usepackage{breqn}
\usepackage{caption}
\usepackage{tcolorbox}
\usepackage{listings}

\makeatletter
\let\cat@comma@active\@empty

\begin{document}

\title{Superlinear Hall angle and carrier mobility from non-Boltzmann magnetotransport in the spatially disordered Yukawa-Sachdev-Ye-Kitaev model on a square lattice}

\author{Davide~Valentinis}\email[]{davide.valentinis@kit.edu}
\affiliation{Institut f\"ur Quantenmaterialien und Technologien, Karlsruher Institut
f\"ur Technologie, 76131 Karlsruhe, Germany} 
\affiliation{Institut f\"ur Theorie der Kondensierten Materie, Karlsruher Institut
f\"ur Technologie, 76131 Karlsruhe, Germany}
\author{J\"org Schmalian}
\affiliation{Institut f\"ur Quantenmaterialien und Technologien, Karlsruher Institut
f\"ur Technologie, 76131 Karlsruhe, Germany}
\affiliation{Institut f\"ur Theorie der Kondensierten Materie, Karlsruher Institut
f\"ur Technologie, 76131 Karlsruhe, Germany}
\author{Subir Sachdev}
\affiliation{Department of Physics, Harvard University, Cambridge MA 02138, USA}
\affiliation{Center for Computational Quantum Physics, Flatiron Institute, 162 5th Avenue, New York, NY 10010, USA}
\author{Aavishkar A. Patel}
\affiliation{International Centre for Theoretical Sciences, Tata Institute of Fundamental Research, Bengaluru 560089, India}
\affiliation{Center for Computational Quantum Physics, Flatiron Institute, 162 5th Avenue, New York, NY 10010, USA}

\date{\today}

\begin{abstract}
Exact numerical results for the dc magnetoconductivity tensor of the two-dimensional spatially disordered Yukawa-Sachdev-Ye-Kitaev (2D-YSYK) model on a square lattice, at first order in applied perpendicular magnetic field, are obtained from the self-consistent disorder-averaged solution of the 2D-YSYK saddle-point equations. This system describes fermions endowed with a Fermi surface and coupled to a bosonic scalar field through spatially random Yukawa interactions. The resulting local and energy-dependent fermionic self-energies are employed in the Kubo formalism to calculate the longitudinal and Hall conductivities, the Hall coefficient, the carrier mobility, and the cotangent of the Hall angle, at fixed fermion density. From the interplay between YSYK interactions and square-lattice embedding, and the non-Boltzmann frequency-dependent self energies, we find nontrivial evolution of the magnetotransport coefficients as a function of temperature and YSYK interaction strength, notably a superlinear evolution of the Hall-angle cotangent and the inverse carrier mobility with temperature, concomitant with linear-in-temperature resistivity, in an extended crossover regime above the low-temperature Marginal Fermi Liquid (MFL) ground state. 
Our model and results provide a controlled theoretical framework to interpret magnetotransport experiments, at linear order in applied magnetic field, in strange-metal phases found in strongly correlated solid-state electron systems. 
\end{abstract}

\maketitle

\section{Introduction}

Transport experiments in strongly correlated electron systems display a plethora of anomalous properties, due to the presence of strange-metal and bad-metal phases \cite{Chowdhury-2022,Phillips-2022} \footnote{We use the term ``strange metal'' only for those metals whose resistivity is smaller than the quantum unit ($h/e^2$ in $d=2$ spatial dimensions). When the linear-in-T resistivity of the metal is larger than the quantum unit, the system is a ``bad metal''.}. Hallmarks of strange metallicity, such as linear-in-temperature ($T$-linear) resistivity \cite{Cooper-2009,Giraldo-2018,Keimer-2015,Grissonnanche-2021,Romero-2025_preprint} and single-particle scattering rate \cite{Bruin-2013}, linear frequency/temperature ($\omega/T$) scaling of optical properties \cite{vanderMarel-2003,vanderMarel-2006,Michon-2023,Romero-2025_preprint,Park-2025_preprint}, and logarithmic divergence of heat capacity \cite{Michon-2019}, are observed in materials as diverse as heavy-fermion compounds \cite{Lohneysen-1994,Trovarelli-2000,Schroder-2000,Paglione-2003,Tokiwa-2013}, pnictides \cite{Fang-2009,Palmstrom-2022,Cai-2023,Liu-2024}, organic charge-transfer salts \cite{Doiron-Leyraud-2009}, twisted heterostructures \cite{Lyu-2021,Xia-2025_preprint}, and high-temperature cuprate superconductors \cite{Valla-1999,Bruin-2013,Giraldo-2018,Naqib-2019,Mandal-2019,Michon-2019,Legros-2019,Grissonnanche-2021,Michon-2023}. 

The application of external magnetic fields reveals further striking phenomenology, including anomalous magnetoresistance scalings  \cite{Hayes-2016,Ayres-2021,Giraldo-2018} in the low- and high-field regimes, doping-dependent cyclotron masses \cite{Legros-2022} and superlinear, $T^2$-like dependence of the Hall-angle cotangent as detected through Hall effect \cite{Chien-1991,Harris-1992,Manako-1992,Carrington-1992,Mackenzie-1996,Matthey-2001,Li-2007,Nakajima-2007,Liu-2008,Kokalj-2012,Legros-2019,Huang-2020,Lyu-2021,Xia-2025_preprint}.
In particular, the latter feature implies an apparent dichotomy between the temperature dependencies $\sigma_{xx}(T)\sim 1/T$ and $\sigma_{xy} \sim 1/T^3$ characterizing longitudinal and Hall conductivities \cite{Chien-1991,Harris-1992}, which challenges conventional models of magnetotransport grounded in the Fermi-liquid picture. Indeed, assuming well defined low-energy electronic quasiparticles as charge carriers, Boltzmann semiclassics predicts that a single frequency-independent scattering rate $1/\tau_r$ characterizes longitudinal and Hall conductivities: the anisotropy of such rate, together with Fermi-surface topology, would universally determine the magnetoconductivity tensor, leading to a geometric interpretation of the Hall conductivity in two dimensions (2D) in terms of magnetic flux quanta threading the area that the scattering length sweeps along the Fermi surface \cite{Ong-1991}. 
However, the absence of quasiparticles generated by strange metallicity invalidates the premises of semiclassical treatments \cite{Chowdhury-2022,Phillips-2022}, and calls for innovative theoretical frameworks for non-Fermi liquid phases. 

A crucial aspect in this direction is controlled embedding of non-Fermi liquid states in a lattice potential, which originates nontrivial transport phenomena, especially in the presence of static and oscillating magnetic fields \cite{Nagaosa-2010}. 
In particular, pioneering works using a finite-temperature generalization of Ong's kinetic theory \cite{Bok-2004}, and the Anti de Sitter/Conformal Field Theory (AdS/CFT) correspondence \cite{Blake-2015,Chagnet-2024_preprint}, suggest that geometric constraints on electronic conduction imposed by the lattice are crucial to generate different temperature scalings of transport coefficients in the longitudinal and Hall channels; see also Sec. \ref{Discussion}. 

In this work, we go beyond the Boltzmann approach, and describe magnetotransport with frequency-dependent self energies in the spatially disordered Yukawa-Sachdev-Ye-Kitaev (YSYK) model on the square lattice. 
This is a model of a Fermi surface of electrons coupled to a bosonic scalar field ($\phi$) with a spatially random Yukawa coupling. 
We remain deliberately agnostic about the specific physical nature of the bosonic mode, apart from its scalar character and its assumed dispersion relation: this is because various forms of microscopic bosonic coupling, ranging
from order parameters breaking point-group, time-reversal, or spin-rotation symmetry, to transverse components of emergent gauge fields, to fractionalized particles, to fluctuations engendered by Fermi volume-changing transitions, essentially all map to the same
low-energy theory. In the simplest case, the coupling boson can also represent a (polarization-unresolved) phonon on the underlying lattice. 
Furthermore, due to spatial disorder in our interaction $g'$, finite-momentum bosonic modes, like antiferromagnetic spin fluctuations or charge/spin density-wave fluctuations, map to saddle-point equations analogous to Eqs\@. (\ref{eq:saddle_point_imag_gen}) via reparametrizations of the boson mass and dispersion \footnote{To see the mapping of finite-momentum modes to Eqs\@. (\ref{eq:saddle_point_imag_gen}), assume for simplicity a quadratic boson dispersion, $\left[\omega_{\vec{q}}\right]^2=(m_b^0)^2+K q^2$, so that the momentum-integrated boson propagator is
\begin{equation}\label{eq:D_foot}
\mathscr{D}(i\Omega_n)=\int \frac{d \vec{q}}{(2\pi)^2}\frac{1}{(\Omega_n)^2 +(m_b^0)^2+K q^2-\Pi(i\Omega_n) }. 
\end{equation}
If we now modify the dispersion to $\left[\omega_{\vec{q}}\right]^2=(m_b^0)^2+K (q-Q)^2$ with the characteristic finite momentum $\vec{Q}$, the propagator given in Eq\@. (\ref{eq:D_foot}) can still be written as
\\[
\mathscr{D}(i\Omega_n)=\int \frac{d \vec{q}}{(2\pi)^2}\frac{1}{(\Omega_n)^2 +(\tilde{m}_b^0)^2+K q^2-2 K Q q-\Pi(i\Omega_n) }, 
\\]
with a reparametrized bare boson mass $(\tilde{m}_b^0)^2=(m_b^0)^2+K Q^2$.
}. 
The present work extends an earlier study that examined transport properties in the strange metal and the superconductor in zero magnetic field \cite{Li-2024}.
We provide exact numerical solutions for the longitudinal and transverse (Hall) components of the interacting conductivity tensor, namely, $\sigma_{xx}^{(0)}(T)=\sigma_{yy}^{(0)}(T)$ and $\sigma_{xy}^{(1)}(T)$, of this YSYK model as a function of temperature $T$ and at linear order in applied magnetic field $B$.  Employing the 2$\times$2 conductivity tensor, we compute the Hall coefficient \cite{Morpurgo-2024,Morpurgo-2025} 
\begin{equation}\label{eq:RH_gen}
R_H(T)=\frac{\sigma_{xy}^{(1)}(T)}{B \left[\sigma_{xx}^{(0)}(T)\right]^2},
\end{equation}
and the cotangent of the Hall angle
\begin{equation}\label{eq:cotThetaH_gen}
\cot\left[\Theta_H(T)\right]=\frac{\sigma_{xx}^{(0)}(T)}{\sigma_{xy}^{(1)}(T)}=\frac{1}{\sigma_{xx}^{(0)}(T) B R_H(T)}. 
\end{equation}
We find that $\left|\cot\left[\Theta_H (T)\right]\right| \propto T^\alpha$ acquires superlinear evolution with $T$ ($\alpha>1$) in an intermediate crossover temperature region $0.1 t \lessapprox k_B T \lessapprox t$, concomitantly with $T$-linear resistivity, due to the decrease of the Hall coefficient $R_H(T)$ generated by the lattice embedding of the YSYK system. When we tune the system to quantum criticality at $T=0$ -- see Sec. \ref{Model_methods_YSYK} -- the exponent $1\lessapprox\alpha \lessapprox 1.5$ increases with decreasing fermion-boson interaction $g'$ -- defined below in Eq\@. (\ref{eq:distribution}) -- and decreasing doping, i.e.\@, close to particle-hole symmetry. 
The increase of $\alpha$ with decreasing doping qualitatively matches the experimental phenomenology in archetypical strange metals as found in YBa$_2$Cu$_3$O$_{7-\delta}$ \cite{Harris-1992} and Bi$_2$Sr$_{2-x}$La$_x$CuO$_6$ \cite{Ando-1999}; see also Sec. \ref{Discussion}. 
Further increase in $\alpha$ occurs when we detune the system from quantum criticality -- see Sec. \ref{Distance_QCP} and Fig\@. \ref{fig:rho_RH_cotThetaH_square_T_gp2_fixedn_distQCP} -- and is accompanied by an increased low-$T$ concavity of the longitudinal resistivity curve, even though $T$-linear evolution is recovered at higher temperatures inside the quantum critical fan. 
\begin{figure*}[ht]
  \includegraphics[width=1\textwidth]{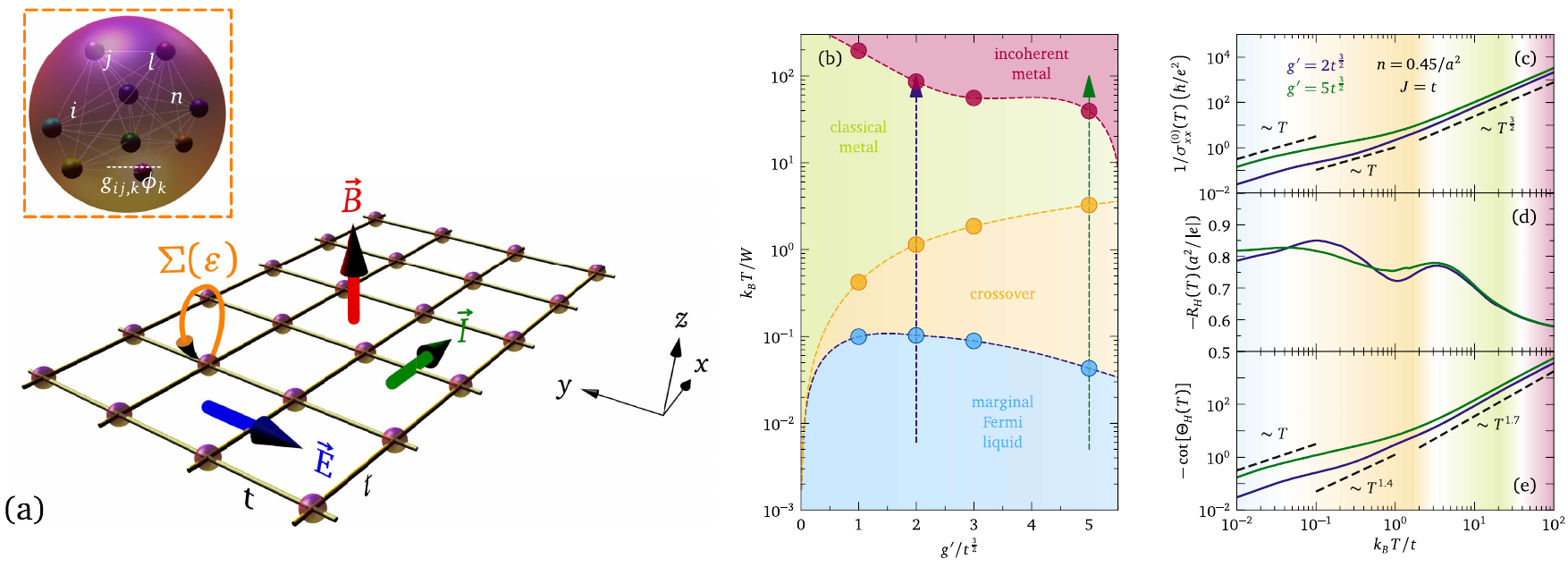}
\caption{\label{fig:Hall_summary}
(a) Schematics of our system's geometry and configuration: a 2D-YSYK square lattice with nearest neighbour hopping amplitude $t$ is subjected to an external electric field along $x$ and out-of-plane magnetic field $\vec{B}$ along $z$, thus inducing a longitudinal current $\vec{I}$ as well as a Hall electric field $\vec{E}$ along $y$; at each lattice site, a local self-energy $\Sigma(\epsilon)$ stems from the disorder-averaged YSYK interactions $g_{ij,k}$ between fermion flavours $i=\left\{i,j,l,n\ldots\right\}$ and a scalar bosonic mode $\Phi_k$. (b) Schematic phase diagram of transport regimes as a function of temperature $k_B T/t$ and interaction strength $g'/t^{\frac{3}{2}}$, for fermion density $n=0.45/a^2$ and boson stiffness $J=t$, normalized to the fermion hopping $t$; coloured circles are numerical estimations of the boundaries between different regimes from the saddle-point solutions, while dashed curves are continuous interpolations; dashed blue and green arrows show the temperature paths scanned in panels (c-e) for $g'/t^{\frac{3}{2}}=\left\{2,5\right\}$. (c-e) Inverse longitudinal conductivity (c), Hall coefficient (d) , and cotangent of the Hall angle (e), as a function of $k_B T/t$, for $g'/t^{\frac{3}{2}}=2$ (blue solid curves) and $g'/t^{\frac{3}{2}}=5$ (green solid curves); colour shadings qualitatively identify the regimes in panel (b). }
\end{figure*}

Technically, we employ the dc ($\omega \rightarrow 0$) limit of the Kubo formula for the homogeneous current-current correlation function, according to Eqs\@. (\ref{eq:sigma_dc_gen}), to calculate $\sigma_{xx}^{(0)}(T)=\sigma_{yy}^{(0)}(T)$ and $\sigma_{xy}^{(1)}(T)$, explicitly given by Eqs\@. (\ref{eq:Kubo_formulae}) \cite{Morpurgo-2024,Morpurgo-2025}, at fixed charge carrier density $e n$ \footnote{By charge density, here we denote the product of the electric charge $e$ and the fermion density $n$, calculated in accordance with Eqs\@. (\ref{eq:n_def}) or (\ref{eq:n_def_imag}). This measure of carrier density is independent from the existence of well-defined electronic quasiparticles, and it also holds for our 2D-YSYK strongly interacting problem.}. The latter constraint implies that the chemical potential $\mu=\mu(n,T)$ adjusts itself self-consistently to keep $n$ fixed at any interaction $g'$ and temperature $T$ \cite{vanderMarel-1990,Valentinis-2016a,Valentinis-2016b,Valentinis-2017,Morpurgo-2024,Morpurgo-2025}. The conductivities depend both on the fermions dispersion $\epsilon_{\vec{k}}$ -- here assumed to be the single-band, nearest-neighbour square-lattice Eq\@. (\ref{eq:square_disp}) -- through their longitudinal $\Phi_{(0)}^{xx}(\epsilon)=\Phi_{(0)}^{yy}(\epsilon)$ and Hall $\Phi_{(1)}^{xy}(\epsilon)$ transport functions, and on fermionic interactions contained in the local (momentum-independent) and retarded self-energy $\Sigma^R(\omega)$; the latter stems from the exact numerical solution of the saddle-point 2D-YSYK equations (\ref{eq:saddle_point_imag_gen}), analytically continued to the real axis \cite{Schmalian-1996}; see Sec. \ref{Model_methods} and App\@. \ref{Real_continuation}. The 2D-YSYK saddle point is the same that was previously shown to lead to $T$-linear resistivity \cite{Patel-2023}, $\omega/T$ scaling of the optical conductivity \cite{Li-2024}, superconductivity with unconventional phase stiffness \cite{Li-2024}, and linear-in-$B$ effective cyclotron resonance frequency \cite{Guo-2024}.

The paper is organized as follows: Sec. \ref{Summary} hosts a brief summary of our main results for the conductivities, the Hall coefficient, and the Hall angle cotangent. Sec. \ref{Model_methods} describes the spatially disordered 2D-YSYK model for fermion-boson interactions, the Kubo formalism for the calculation of the linear-in-$B$ conductivities on the square lattice, and the protocol for the self-adjustment of the chemical potential in fixed-density calculations. Sec. \ref{Imaginary_thermodyn} reports the results for the renormalized boson mass $m_b(T)$ and chemical potential $\mu(n,T)$, stemming from the imaginary-axis solutions of the 2D-YSYK saddle-point equations, and later employed in the calculation of the spectroscopic (real-axis) properties. Details on the results for the conductivities, the Hall coefficient, and the Hall angle cotangent, for different interactions $g'$ and fermion densities $n$, are collected in Sec. \ref{Hall_results}. Sec. \ref{Mobility_results} discusses results for the carrier mobilities, as equivalently calculated from the fermion density $n$ or the Hall coefficient $R_H$. Our results are further discussed in Sec. \ref{Discussion}, with reference to alternative theoretical routes to nontrivial Hall angle, and qualitatively comparing our findings to experimental literature on strange-metal magnetotransport. 
Conclusions and future perspectives of our work are summarized in Sec. \ref{Conclusions_outlook}. Appendixes contain details on solution protocols for the imaginary-axis and real-axis self-consistent loops, derivations of the square-lattice Green's function (for completeness) and of the longitudinal and Hall transport functions, additional plots and information of the real-axis fermion and boson self-energies, and a derivation of analytical results for linear-magnetotransport coefficients of our model at fixed $\mu$ in different temperature regimes. 

\section{Summary of main results}\label{Summary}

Our 2D-YSYK theory produces superlinear Hall-angle cotangent as a result of lattice embedding, which modifies the Hall coefficient $R_H(T)$, in particular decreasing with $T$ in an intermediate temperature regime, with respect to the results for quadratic, free fermion-like dispersion $\epsilon_{\vec{k}}=\hbar^2 k^2/(2m)-\mu$; in the same crossover regime, we still have $\sigma_{xx}^{(0)}(T) \propto T$, as argued below and summarized in Fig\@. \ref{fig:Hall_summary}. In order to better compare our results with the conventional phenomenology stemming from the semiclassical Boltzmann kinetic equation, let us first recall the latter phenomenology in broad strokes. \

For 2D semiclassical Boltzmann theory in a weak out-of-plane magnetic field $B$, the longitudinal and Hall conductivities are given by $\sigma_{xx}^B=\sigma_0/\left[1+(\omega_c \tau_r)^2\right]$ and $\sigma_{xy}^B=\sigma_0 \omega_c \tau_{r}/\left[1+(\omega_c \tau_{r})^2\right]$, where $\sigma_0=ne^2 \tau_{r}/m$ is the Drude conductivity for carrier density $n$, bare electron charge $e$, and carrier mass $m$, and $\omega_c =eB/m$ is the cyclotron frequency, while $B$ always enter through the product $\omega_c \tau_r$ (Kohler's rule \cite{Kohler-1938}). Inverting the conductivity tensor, we find the longitudinal and Hall resistivites, $\rho_{xx}^B=1/\sigma_0$ and $\rho_{xy}^B=B/(n \left|e\right|)$. The Hall coefficient $R_H^B=\rho_{xy}/B=1/(n \left|e\right|)\approx \sigma_{xy}^B/\left[B (\sigma_{xx}^B)^2\right]$ is a measure of electron density, where the latter approximation holds for $\omega_c \tau_r \ll 1$, while the Hall-angle cotangent $\cot\left(\Theta_H^B\right)=\sigma_{xx}^B/\sigma_{xy}^B=1/(\omega_c \tau_r)=m/(\left|e\right| B \tau_r)$ shares the same $\tau_r$ dependence as $\sigma_{xx}^B$. 

Fig\@. \ref{fig:Hall_summary}(a) illustrates the schematics of our system geometry: we construct a square lattice with nearest-neighbour hopping $t$, where on-site spatially disordered YSYK interactions between $\left\{i,j,l,n\right\}=\left\{1,\cdots \mathscr{N}\right\}$ flavours of fermions and $k=\left\{1,\cdots \mathscr{N}\right\}$ flavours of a scalar bosonic mode (see Sec. \ref{Model_methods_YSYK}) produce, in the large-$\mathscr{N}$ limit and after disorder-averaging, a local retarded self-energy $\Sigma^R(\omega)$ for fermions which interplays with the dispersion $\epsilon_{\vec{k}}$ given by Eq\@. (\ref{eq:square_disp}). We apply an external electric field along $x$ and out-of-plane magnetic field $\vec{B}$ along $z$ and investigate the longitudinal (along $x$) and Hall (along $y$) linear current response at first order in magnetic field $B$ \cite{Morpurgo-2024,Morpurgo-2025}. This setup allows us to study the weak-field Hall effect on the square lattice, while the influence of the Hofstadter butterfly fractal spectrum on the fermion propagators appears at order $B^2$ or higher \cite{Hofstadter-1976}. 

Fig\@. \ref{fig:Hall_summary}(b) summarizes the longitudinal conduction regimes encountered in our model as a function of spatially disordered interaction $g'$ and temperature $T$, normalized by hopping $t$, for the exemplary parameters of fermion density $n=0.45$ (doping $\Delta n=0.05$) and boson stiffness $J=t$. When the system is tuned to the quantum critical point (QCP) by varying the bare boson mass $m_b^0$ (see Sec. \ref{Model_methods_YSYK}), at low $T$ we retrieve a ground state exhibiting Marginal Fermi Liquid (MFL) scaling of $\Sigma^R(\omega)$ \cite{Varma-1989} (light-blue shaded region), which is the source of $T$-linear resistivity \cite{Patel-2023} and $\omega/T$ scaling of optical properties \cite{Li-2024} in our model. These low-$T$ features are qualitatively independent of the assumed dispersion $\epsilon_{\vec{k}}$: they universally stem from overdamped dynamics of soft bosons with vanishing renormalized frequency $\lim_{T\rightarrow 0^+} m_b(T)=0$, which inelastically scatter off fermions thus generating MFL phenomenology. 

Fig\@. \ref{fig:Hall_summary}(c) illustrates the $T$-linear resistivity in units of the 2D resistance quantum $\hbar/e^2$ in the light-blue shaded region, for both interactions smaller ($g'=2 t^{\frac{3}{2}}$, blue curve) and larger ($g'=5 t^{\frac{3}{2}}$, green curve) than the half fermion bandwidth $W/2=4t$. Increasing $T$, the self-energy reaches $\mathrm{Im}\left\{\Sigma^R(k_B T_{\mathrm{MFL}})\right\}\lessapprox k_B T$, which signals the competition between the no-longer negligible energy-fluctuation term $\omega$ and $\Sigma^R(\omega)$ in the analytically continued fermion propagator (\ref{eq:Dyson_fermions_gen}). This feature is also reflected by a peak in the Hall coefficient shown in Fig\@. \ref{fig:Hall_summary}(d), the position of which shifts with varying interaction $g'$ (dashed blue curve and light-blue circles in Fig\@. \ref{fig:Hall_summary}(b)). At even higher temperatures $R_H(T)$ decreases with $T$ in the orange-shaded region of Fig\@. \ref{fig:Hall_summary}(c-e). 

\begin{figure}[ht]
  \centering
\includegraphics[width=0.8\linewidth]{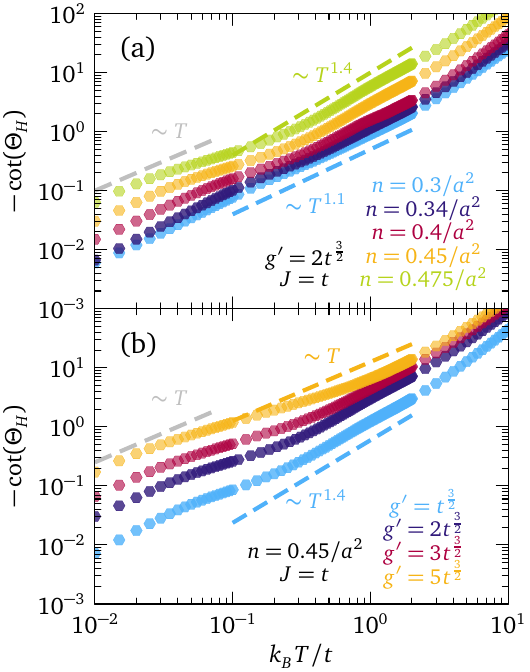}
\caption{\label{fig:cotTheta_fixedn_logscale}
Visual summary in logarithmic scale of the dependence of the Hall-angle cotangent $\cot[\Theta_H(T)]$ on dimensionless temperature $k_B T/t$, for boson stiffness $J=t$. (a) Results at fixed interaction $g'=2t^{3/2}$ and fermion density $n=\left\{0.3,0.35,0.4,0.45,0.475\right\}/a^2$. (b) Results at fixed fermion density $n=0.45/a^2$ $g'=2t^{3/2}$ and interaction $g'=\left\{1,2,3,5\right\}t^{3/2}$. The exponent $\alpha$ of $\left|\cot[\Theta_H(T)]\right|\propto T^\alpha$ is maximized for weak interaction and small doping $\Delta n=0.5/a^2-n$.}
\end{figure}

As further analyzed below, the decrease of $R_H(T)$ at intermediate temperatures is a general bandstructure effect, not directly linked to YSYK interactions: it stems from the sign change and symmetry of the Hall transport function $\Phi_{(1)}^{xy}(\epsilon)$ around $\epsilon=0$ occurring on a square lattice (see Fig\@. \ref{fig:scheme_Hall_T}), {\it i.e.}\@, from the thermally excited counterpropagation of electron and hole excitations under the Hall electric field $\vec{E}=E \hat{u}_y$ generated by the Lorentz force \cite{Bok-2004}. Such phenomenon requires $(g')^2 k_B T \gtrapprox \mu(T)$, consistently with the onset in Fig\@. \ref{fig:Hall_summary}(d).  We call this regime ``crossover" -- see also orange shading in Fig\@. \ref{fig:Hall_summary}(b) -- where the overdamped bosons with sizable self-energy $\left|\Pi^R(\omega)\right|/\left|\omega\right|\gg1$ continue to provide the marginal susceptibility for inelastic scattering off fermions \cite{Chen-2024}, and have renormalized mass $0<m_b(T)<m_b^0$; see also Fig\@. \ref{fig:mb_mu_square_T_varg_fixedn}(a,c). Further increasing $T$, we reach the ``classical metal" regime, where the boson self-energy $\Pi^R(\omega)$ becomes negligible and $m_b(T)\lessapprox m_b^0$ (the condition $\Pi^R(0)=k_B T$ is shown by the orange dashed curve and orange circles in Fig\@. \ref{fig:Hall_summary}(b)): the boson dynamics essentially decouples from the fermions, which interact with essentially free bosons on a square lattice; $\Sigma^R(\omega)$ is thus calculable at ``one-loop" level with standard many-body methods \cite{Mahan-2000,Berthod-2018} and produces $1/\sigma_{xx}(T)\propto T^{3/2}$, as seen in the green-shaded part of Fig\@. \ref{fig:Hall_summary}(c). In the same regime, the Hall coefficient in Fig\@. \ref{fig:Hall_summary}(d) attains a second maximum for temperatures $k_B T \approx W$, before decreasing again. 

Finally, at high temperatures the static fermion self-energy $\Sigma^R(0)$ becomes larger than the fermion bandwidth $W=8t$, as marked by the dashed red curve and red circles in Fig\@. \ref{fig:Hall_summary}(b): here the broad, featureless fermionic spectral functions inspire the label ``incoherent metal", where self-energy effects dominate over the dispersion $\epsilon_{\vec{k}}$ and completely local physics emerges; therefore, this regime does not depend on $\epsilon_{\vec{k}}$ and is universally reached at sufficiently high $g'$ and $T$. Furthermore, we distinguish between an incoherent metal, an extension of the classical-metal regime where $W<\Sigma^R(0)<k_B T$ -- see red-shaded regions in Fig\@. \ref{fig:Hall_summary}(b-e) -- and an actual bad metal, where the self-energy $\Sigma^R(0)>\left\{k_B T,W\right\}$ is the dominant energy scale, and SYK-like local physics with $\Sigma^R(\omega)\propto \sqrt{\omega}$ is at play due to the Parcollet-Georges mechanism \cite{Georges-2000,Georges-2001,Parcollet-1999} (not shown; see App\@. \ref{App:bad} for a sketch of analytical proof); the latter regime thus adiabatically connects to local YSYK models and the Combescot-Allen-Dynes strong-coupling limit of Eliashberg theory \cite{Allen-1975,Combescot-1994,long-paper,short-paper,Heath-2025}. 
Finally, Fig\@. \ref{fig:Hall_summary}(e) shows the Hall-angle cotangent in accordance with Eq\@. (\ref{eq:cotThetaH_gen}). In particular, while the conventional relation $\left|\cot[\Theta_H(T)]\right|\propto T$ is obeyed in MFL regime (light blue-shaded region), as expected, the persistence of $T$-linear resistivity and the concomitant decrease of the Hall coefficient with $T$ in the crossover regime produces a superlinear evolution of $\left|\cot[\Theta_H(T)]\right|$ at weak to moderate interaction (orange-shaded region). 

Further analysis confirms that such superlinearity is enhanced at weak interactions and small doping $\Delta n$, i.e.\@, close to particle-hole symmetry $n=0.5$ according to Eq\@. (\ref{eq:n_def}), as observed from Fig\@. \ref{fig:cotTheta_fixedn_logscale}. Within the scanned parameter space, the maximum exponent is $\alpha\approx 1.45$ of $\left|\cot[\Theta_H(T)]\right|\propto T^\alpha$ in the crossover regime of the present 2D-YSYK model, while keeping the system at criticality at $T=0$. Further increase of $\alpha$ occurs by detuning the system from the QCP; see Sec. \ref{Distance_QCP}. The dependence of $\alpha$ on doping and interactions, with highest values found at weak coupling, is qualitatively consistent with observations in strange metals \cite{Chien-1991,Harris-1992,Ando-1999}, although the lack of quantitative agreement is reflected by an overall smaller superlinearity exponents in the theory, with respect to the ones fitted from experimental data. However, refinements in the employed bandstructure (e.g.\@, introduction of second nearest-neighbour hoppings), finite one-body impurity potential $v$ \cite{Guo-2024,Li-2024}, and temperature-dependent carrier densities \cite{Putzke-2021}, could improve the quantitative comparison of our model with experiments, as further discussed in Sec. \ref{Discussion}. 

In the following, we provide more technical information on the 2D-YSYK interaction model assumed in our calculations, on the structure of spatially disordered bosonic couplings, on the self-consistent adjustment of chemical potential as a function of interaction and temperature, and on the Kubo formalism employed to numerically compute the linear-in-$B$ conductivities at fixed fermion density.  

\section{Models and methods}\label{Model_methods}

\subsection{Fermion self-energy from the 2D-YSYK saddle-point equations}\label{Model_methods_YSYK}
The 2D-YSYK action in the space of imaginary time $\tau$ and positions $\vec{r}$ reads 
\begin{widetext}
\begin{align}
\mathscr{S}& = \int d\tau \sum_{i=1}^{\mathscr{N}} \sum_{\vec{k}}  \sum_{\sigma}\psi^\dagger_{i,\sigma,\vec{k}}(\tau)\left[\partial_\tau+\varepsilon_k\right]\psi_{i,\sigma,\vec{k}}(\tau)
+\frac{1}{2}\int d\tau \sum_{i=1}^{\mathscr{N}} \sum_{\vec{q}} \phi_{i,\vec{q}}(\tau)\left\{-\frac{\partial^2}{\partial\tau^2}+\left[\omega_b(q)\right]^2\right\}\phi_{i,-\vec{q}}(\tau) \nonumber \\ 
&+ \int d^2r \int d\tau  \sum_{\left\{i,j\right\}=1}^{\mathscr{N}}\sum_{\sigma} \frac{v_{ij} (\vec{r}) }{\sqrt{\mathscr{N}}}\psi_{i,\sigma}^{\dagger} (\vec{r},\tau) \psi_{j,\sigma}(\vec{r}, \tau)  +\int d^2 r \int d\tau  \sum_{\left\{i,j,\ell\right\}=1}^{\mathscr{N}}\sum_{\sigma}\frac{g_{ij,\ell}^{'}(\vec{r})}{\mathscr{N}}  \psi^\dagger_{i,\sigma}(\vec{r},\tau)\psi_{j,\sigma}(\vec{r},\tau)\phi_{\ell}(\vec{r},\tau),
\label{eq:latticeaction}
\end{align}
\end{widetext}
where $i,j,\ell$ are flavor indices and $\sigma=\left\{\uparrow,\downarrow\right\}$ is the spin index. We employ the square-lattice dispersion with lattice parameter $a$, nearest-neighbour hoppings $t$, and chemical potential $\mu$,
\begin{equation}\label{eq:square_disp}
\epsilon_{\vec{k}}=-2 t\left[\cos(k_x a)+\cos(k_y a)\right]-\mu,
\end{equation}
together with the associated bosonic dispersion \cite{Esterlis-2021,Guo-2022}
\begin{equation}\label{eq:disp_bosons_squarelike}
\left[\omega_b(\vec{q})\right]^2= -2J\left[\cos(q_x a) +\cos(q_y a) -2\right]+\left(m_b^0\right)^2. 
\end{equation}
$m_b^0$ and $J$ are the bare boson mass and the boson stiffness, respectively.
The random-valued Yukawa couplings $g_{ij,\ell}^{'}(\vec{r})$ read $g_{ij,\ell}^{'}(\vec{r})=g_{1,ij,\ell} (\vec{r}) + i g_{2,ij,\ell} (\vec{r})$, where the real ($g_{1,ij,\ell}(\vec{r})$) and imaginary ($g_{2,ij,\ell}(\vec{r})$) parts follow a Gaussian statistics with zero mean and variances
\begin{align}
\overline{g_{1,ij,\ell}(\vec{r})g_{1,i'j',\ell'}(\vec{r'})} & = \left(1-\frac{\alpha}{2}\right)(g')^2\delta_{\ell \ell'} \nonumber \\ & \times \left(\delta_{ii'}\delta_{jj'}+\delta_{ij'}\delta_{ji'}\right)\delta(\vec{r}-\vec{r'}),\nonumber \\
\overline{g_{2,ij,\ell}(\vec{r})g_{2,i'j',\ell'}(\vec{r'})} & = \frac{\alpha}{2}(g')^2\delta_{\ell\ell'}\left(\delta_{ii'}\delta_{jj'}-\delta_{ij'}\delta_{ji'}\right) \nonumber \\ & \times \delta(\vec{r}-\vec{r'}),\nonumber \\
\overline{g_{1,ij,\ell}(\vec{r})g_{2,i'j',\ell'}(\vec{r'})} & = 0.\label{eq:distribution}
\end{align}
The prefactor $\alpha\in \left[0,1\right]$ acts as a pair-breaking parameter \cite{Hauck-2020}: in the $\alpha=1$ limit, Eqs\@. \eqref{eq:distribution} yield $\overline{g_{ij,\ell}^{'}(\vec{r}) [g_{i'j',\ell'}^{'}(\vec{r'})]^\ast}=(g')^2\delta(\vec{r}-\vec{r'})\delta_{ii'}\delta_{jj'}\delta_{kk'}$ and no superconductivity occurs; the case $\alpha=0$ leads to real-valued coupling constants, which preserve time-reversal symmetry for each random-couplings disorder realization, thus generating Cooper pairing and superconductivity \cite{Esterlis-2019,Hauck-2020,long-paper,short-paper,Li-2024}. Here we analyze the ensuing saddle-point equations (\ref{eq:saddle_point_imag_gen}) in the normal state, independent of $\alpha$.
The spatially random potential, mimicking, e.g.\@, the effect of impurity scattering, satisfies
\begin{align}
    & \overline{v_{ij} (\vec{r})} = 0 \,, \quad \overline{v^\ast_{ij} (\vec{r}) v_{i'j'} (\vec{r'})} = v^2 \, \delta(\vec{r}-\vec{r'}) \delta_{ii'} \delta_{jj'}. 
    \label{eq:vdistribution}
\end{align} 

We perform a disorder average over the random couplings (\ref{eq:distribution}) and potentials (\ref{eq:vdistribution}) in the limit of a large number of flavors $\mathscr{N}\rightarrow +\infty$, where fluctuations with respect to the saddle-point solution are suppressed at least at order $1/\mathscr{N}$ and the mean-field approximation becomes exact. We thus obtain the following SYK-type saddle-point equations on the imaginary axis, written in terms of $\omega_n=(2n+1)\pi k_B T$ and $\Omega_n=2\pi k_B T$ which are fermionic and bosonic Matsubara frequencies respectively:
\begin{subequations}\label{eq:saddle_point_imag_gen}
\begin{align}\label{eq:saddle_point_imag_gen_Sigma}
\Sigma(i \omega_n)&= (g')^2 k_B T\sum_{i \omega_m} \mathscr{G}(i \omega_m) \mathscr{D}(i \omega_n-i\omega_m) \nonumber \\ &+ v^2 \mathscr{G}(i \omega_n),
\end{align}
\begin{equation}\label{eq:saddle_point_imag_gen_Pi}
\Pi(i\Omega_n)= -2 (g')^2 k_B T \sum_{i \omega_m} \mathscr{G}(i \omega_m) \mathscr{G}(i \omega_m+i \Omega_n),
\end{equation}
\begin{equation}\label{eq:Dyson_bosons_gen}
\mathscr{D}(i\Omega_n)=\int \frac{d \vec{q}}{(2\pi)^2} \frac{1}{(\Omega_n)^2+ \left[\omega_b(\vec{q})\right]^2-\Pi(i\Omega_n)}
\end{equation}
\begin{equation}\label{eq:Dyson_fermions_gen}
\mathscr{G}(i \omega_n)=\int \frac{d\vec{k}}{(2 \pi)^2} \frac{1}{i\omega_n-\epsilon_{\vec{k}}-\Sigma(i\omega_n)}.
\end{equation}
\end{subequations}
The spatially uncorrelated nature of the disorder in Eqs\@. (\ref{eq:distribution}) and (\ref{eq:vdistribution}), encoded in the delta functions $\delta(\vec{r}-\vec{r'})$, makes the self-energies (\ref{eq:saddle_point_imag_gen_Sigma}) and (\ref{eq:saddle_point_imag_gen_Pi}) momentum-independent, which in turn restricts bandstructure effects to a dispersion-dependent sum over momenta in the momentum-integrated fermion Green's function $\mathscr{G}(i\omega_n)$ and boson Green's function $\mathscr{D}(i\Omega_n)$. We mention that a translationally invariant model \cite{Esterlis-2021,Guo-2022} devoid of the $\delta(\vec{r}-\vec{r'})$ in the random terms leads to full momentum dependences of the self-energies and propagators. We leave full solutions combining the effect of translationally invariant and spatially disordered interactions to future work, while here we focus on spatial disorder ($g'$) alone and $v=0$. 

To tune the system to criticality, here we vary the bare boson mass $m_b^0$ such that the interacting boson mass $m_b(T)$, renormalized by fermion-boson interactions according to
\begin{equation}\label{eq:m_b_T}
\left[m_b(T)\right]^2=(m_b^0)^2-\Pi(0),
\end{equation}
is null (within numerical accuracy) at zero temperature: $\lim_{T\rightarrow 0} m_b(T)\approx 0$. The tuning (\ref{eq:m_b_T}) is performed at fixed fermion density $n$, calculated according to Eq\@. (\ref{Model_methods_density}). 
Our model cannot self-consistently determine the absolute value of $m_b^0$ (i.e.\@, the absolute distance from the QCP), in terms of an external experimentally accessible parameter, like doping or pressure. However, our theory describes all large-$\mathscr{N}$ thermodynamic and spectroscopic properties of coupled bosons and fermions as a function of relative distance from the QCP, tuned through the bare $m_b^0$. Such tuning self-consistently yields the renormalized boson mass $m_b (T)$ in accordance with Eq\@. (\ref{eq:m_b_T}). \\
Although $m_b^0$ is a theoretical tuning variable, in real experiments it could be accessed and controlled through non-thermal external parameters that change the bosonic ordering tendency. These parameters include: chemical doping/composition, which can change the distance to an ordering instability of the underlying bosonic mode (e.g., spin, charge or lattice order); applied pressure, which can tune lattice constants and electronic overlap, shifting the energy of collective modes and thus the effective boson mass; magnetic field, that can suppress or enhance ordering (e.g., altering spin fluctuations) in magnetic quantum critical systems; carrier density (via gating or doping), which affects the bosonic (spin/charge) susceptibility. \\

Eqs\@. (\ref{eq:saddle_point_imag_gen}) are first self-consistently solved on the imaginary axis. The $T$-dependent renormalized boson mass $m_b(T)$ and chemical potential $\mu(n,T)$ thus obtained are then used as an input, to solve for the spectral properties of fermions and bosons directly on the real axis; see Sec. \ref{Imaginary_thermodyn}. These properties result from the analytic continuation $i\omega\rightarrow \omega+i 0^+$ of Eqs\@. (\ref{eq:saddle_point_imag_gen}), which is performed through the spectral (Lehmann) representation of Green's functions \cite{Mahan-2000,Berthod-2018}, and the representation of convolutions in $\omega$ as products in real time, involving Laplace transforms implemented as Fast Fourier Transforms (FFTs) \cite{Schmalian-1996}; this protocol is analogous to the ones employed in several previous works \cite{long-paper,short-paper,Li-2024}, and is sketched for completeness in App\@. \ref{Real_continuation}. 
In particular, the dynamical retarded (R) fermionic self-energy $\Sigma^R(\omega)$ is then inserted in Eqs\@. (\ref{eq:Kubo_formulae}) to calculate the longitudinal and Hall conductivities. 

\subsection{Conductivity tensor at leading order in magnetic field from the Kubo formula with a local self-energy}\label{Model_methods_Kubo}

To calculate magnetotransport coefficients, we utilize the Kubo formula for the spatially integrated (uniform) conductivity tensor in a translationally invariant system \footnote{The quantity $\sigma_{\alpha \beta}(\omega)$ in Eq\@. (\ref{eq:Kubo_gen}) can be equally understood as the $q=0$ value, $\sigma_{\alpha \beta}(\omega)= \left. \sigma_{\alpha \beta}(\vec{q},\omega)\right|_{q=0}$, of the nonlocal conductivity tensor
\begin{equation}\label{eq:Kubo_foot}
\sigma_{\alpha \beta}(\vec{q},\omega)=\int d\vec{r} \int_{-\infty}^{+\infty} d t \sigma_{\alpha \beta}(\vec{r},t) e^{-i \left(\vec{q} \cdot \vec{r}-\omega t\right)}
\end{equation}
written in Fourier space of transferred momentum $\vec{q}$ and frequency $\omega$, or as the total, spatially integrated conductivity $\sigma_{\alpha \beta}(\omega)=\int d\vec{r} \int_{-\infty}^{+\infty} d t \sigma_{\alpha \beta}(\vec{r},t) e^{i \omega t}$, as also stems from Eq\@. (\ref{eq:Kubo_foot}). }
\begin{equation}\label{eq:Kubo_gen}
\sigma_{\alpha \beta}(\omega)=\frac{i e^2}{\omega}\left[\chi_{\hat{J}\hat{J}}^{R,\alpha \beta}(\omega)-\chi_{\hat{J}\hat{J}}^{R,\alpha \beta}(0)\right],
\end{equation}
where $\chi_{\hat{J}\hat{J}}^{R,\alpha \beta}(\omega)$ is the component in the spatial coordinates $\left\{\alpha,\beta\right\}=\left\{x,y\right\}$ of the retarded current-current correlation function \cite{Morpurgo-2024}. The dc conductivity tensor then results from \cite{Mahan-2000,Berthod-2018}
\begin{equation}\label{eq:sigma_dc_gen}
\sigma_{\alpha \beta}=\lim_{\omega\rightarrow 0}\sigma_{\alpha \beta}(\omega)=-\lim_{\omega \rightarrow 0}\frac{\mathrm{Im}\left\{\sigma_{\alpha \beta}(\omega)\right\}}{\omega}.
\end{equation}
At first order in magnetic field $B$, the structure of the static conductivity tensor is $\sigma_{\alpha \beta}(0)=\delta_{\alpha \beta} \sigma_{\alpha \alpha}+(1-\delta_{\alpha \beta})\sigma_{\alpha \beta}^{(1)}+o(B^2)$, so that Eq\@. (\ref{eq:sigma_dc_gen}) yields the dc longitudinal and Hall conductivities \cite{Hartnoll-2007,Morpurgo-2024,Morpurgo-2025}
\begin{subequations}\label{eq:Kubo_formulae}
\begin{align}\label{eq:sigma_xx_def}
\sigma_{\alpha \alpha}^{(0)}(T)&= 2 e^2\hbar \pi  \int_{-\infty}^{+\infty} d \omega \left[-\frac{\partial f_{FD}(\omega)}{\partial \omega}\right] \nonumber \\ &\times \int_{-\infty}^{+\infty} d\epsilon \Phi_{(0)}^{\alpha \alpha}(\epsilon) \left[A(\epsilon,\omega)\right]^2,   \, \alpha=\left\{x,y\right\},
\end{align}
\begin{align}\label{eq:sigma_xy_def}
\frac{\sigma_{xy}^{(1)}(T)}{B}&= 2 \left|e\right|^3\hbar  \int_{-\infty}^{+\infty} d \omega \left[-\frac{\partial f_{FD}(\omega)}{\partial \omega}\right]  \nonumber \\ &\times \int_{-\infty}^{+\infty} d\epsilon \Phi_{(1)}^{x y}(\epsilon) \left[A(\epsilon,\omega)\right]^3,
\end{align}
\end{subequations}
where $f_{FD}(\omega)=\left[e^{\omega/(k_B T)}+1\right]^{-1}$ is the Fermi-Dirac distribution, and the multiplicative factor of $2$ accounts for twofold spin degeneracy. Let us stress that, at linear order in $B$, Eq\@. (\ref{eq:sigma_xx_def}) corresponds to the longitudinal conductivity in the \emph{absence} of magnetic field, while Eq\@. (\ref{eq:sigma_xy_def}) yields the linear-in-$B$ contribution to the Hall conductivity. 
The integrals over $\epsilon$ in Eqs\@. (\ref{eq:Kubo_formulae})  depend on the fermions dispersion $\epsilon_{\vec{k}}$ through their longitudinal and Hall transport functions \cite{Berthod-2013,Berthod-2018,Morpurgo-2024,Morpurgo-2025}:
\begin{equation}\label{eq:transport_func_long}
\Phi_{(0)}^{\alpha \alpha}(\epsilon)=\frac{1}{\hbar^2}\frac{1}{\mathscr{V}}\sum_{\vec{k}}\left(\frac{\partial \epsilon_{\vec{k}}}{\partial k_\alpha}\right)^2 \delta (\epsilon-\epsilon_{\vec{k}}),
\end{equation}
\begin{align}\label{eq:transport_func_Hall}
\Phi_{(1)}^{x y }(\epsilon)&=\frac{1}{\hbar^3} \frac{\pi^2}{3}\frac{1}{\mathscr{V}}\sum_{\vec{k}}\left[2 \frac{\partial \epsilon_{\vec{k}}}{\partial k_x} \frac{\partial \epsilon_{\vec{k}}}{\partial k_y} \frac{\partial^2 \epsilon_{\vec{k}}}{\partial k_x \partial k_y}  -\left(\frac{\partial \epsilon_{\vec{k}}}{\partial k_x}\right)^2 \right. \nonumber \\  &\left. \times \frac{\partial^2 \epsilon_{\vec{k}}}{\partial k_y^2}-\left(\frac{\partial \epsilon_{\vec{k}}}{\partial k_y}\right)^2 \frac{\partial^2 \epsilon_{\vec{k}}}{\partial k_x^2}\right]\delta(\epsilon-\epsilon_{\vec{k}}),
\end{align}
where $\mathscr{V}$ is the system volume (which is an area in 2D). 
For the employed square-lattice fermionic dispersion (\ref{eq:square_disp}), the transport functions (\ref{eq:transport_func_long}) \cite{Li-2024} and (\ref{eq:transport_func_Hall}) admit fully analytical expressions, derived in App\@. \ref{Square_transport_long} and \ref{Square_transport_trans}, and corresponding to Eqs\@. (\ref{eq:Phi_xx_square_final}) and (\ref{eq:Phi_xy_square_final}) respectively. Moreover, to analyze the spectral and transport properties in the low-$T$ MFL regime it is convenient to linearize the transport functions as
\begin{subequations}\label{eq:transport_Phi_linearized}
\begin{equation}
\overline{\Phi_{(0)}^{xx}(\mu)}=\Phi_{(0)}^{xx}(\mu)- \left.    \frac{d \Phi_{(0)}^{xx}(\epsilon)}{d \epsilon}\right|_{\epsilon=\mu}\mathrm{Re}\left\{\Sigma^R(0)\right\},
\end{equation}
\begin{equation}
\overline{\Phi_{(1)}^{xy}(\mu)}=\Phi_{(1)}^{xy}(\mu) -\left. \frac{d \Phi_{(1)}^{xy}(\epsilon)}{d \epsilon}\right|_{\epsilon=\mu} \mathrm{Re}\left\{\Sigma^R(0)\right\}.
\end{equation}
\end{subequations}
The integrals over $\omega$ in Eqs\@. (\ref{eq:Kubo_formulae}) are functions of the (local) fermionic interactions. 
Local correlations are encoded in $\Sigma^R(\omega)$, yielding the spectral function
\begin{align}\label{eq:spectr_local}
&A(\epsilon,\omega)=-\frac{\mathrm{Im}\left\{G^R(\epsilon,\omega)\right\}} {\pi} \nonumber \\ &=\frac{-\mathrm{Im}\left\{\Sigma^R(\omega)\right\}/\pi}{\left[\omega-\epsilon+\mu-\mathrm{Re}\left\{\Sigma^R(\omega)\right\}\right]^2+\left[\mathrm{Im}\left\{\Sigma^R(\omega)\right\}\right]^2},
\end{align}
where $G^R(\epsilon,\omega)=\left[\omega+i0^+-\epsilon+\mu-\Sigma^R(\omega)\right]^{-1}$ is the retarded fermionic Green's function.

\subsection{Self-adjusting chemical potential at fixed fermion density}\label{Model_methods_density}

Keeping track of the fermion density $n$ is essential to let the chemical potential $\mu(n,T)$ self-adjust with interaction and temperature, and also to calculate the carrier mobility $\mu_n(T)$ defined in Eq\@. (\ref{eq:mu_n}) \cite{Morpurgo-2024,Morpurgo-2025}. On the real axis, the fermion density \emph{per spin} is
\begin{align}\label{eq:n_def}
n&=  \int_{-\infty}^{+\infty} d \omega f_{FD}(\omega) \int_{-\infty}^{+\infty} d\epsilon N_0(\epsilon) A(\epsilon,\omega) \nonumber \\ &=\int_{-\infty}^{+\infty} d \omega  f_{FD}(\omega) N(\omega),
\end{align}
where $N_0(\epsilon)=\mathscr{V}^{-1}\sum_{\vec{k}}\delta(\epsilon-\epsilon_{\vec{k}})$ is the noninteracting density of states, which is given by Eq\@. (\ref{eq:DOS_square}) for our square-lattice dispersion.
At the second step in Eq\@. (\ref{eq:n_def}) we have defined the interacting density of states \cite{Morpurgo-2024},
\begin{equation}\label{eq:interacting_DOS}
N(\omega)=\int_{-\infty}^{+\infty} d\epsilon N_0(\epsilon) A(\epsilon,\omega),
\end{equation}
which reduces to its noninteracting expression $N_0(\epsilon)$ for a corresponding spectral function $A(\epsilon,\omega)=\delta(\omega-\epsilon)$. The spin-resolved density (\ref{eq:n_def}) thus acquires values $0\leq n(T)\leq 1/a^2$ as $-\infty<\mu(n,T)<+\infty$, where we recall that $a$ is the lattice constant. 

The same density can also be calculated on the imaginary axis, through
\begin{equation}\label{eq:n_def_imag}
n=  \lim_{\tau\rightarrow 0^-}\mathscr{G}(\tau)=k_B T\sum_{i\omega_n} \mathscr{G}(i\omega_n)e^{i\omega_n0^+}, 
\end{equation}
where $\mathscr{G}(\tau)$ is the fermion Green's function (here assumed to be local) in imaginary time $\tau\in\left[0,1/(k_BT)\right]$, and $\mathscr{G}(i\omega_n)$ is the Fourier-transformed Matsubara Green's function written in terms of fermionic Matsubara frequencies $\omega_n$. The equivalence between the formulations (\ref{eq:n_def}) and (\ref{eq:n_def_imag}) is shown through the spectral representation of the interacting Green's function $\mathscr{G}^R(\omega)$, as shown by the following steps ($\beta=1/(k_B T))$: 
\begin{align}\label{eq:n_Mats_displess2}
n&=\frac{1}{\beta} \sum_{i \omega_n} \mathscr{G}(i \omega_n) e^{i \omega_n 0^+} \nonumber \\ &=\frac{1}{\beta} \sum_{i \omega_n} \left(-\frac{1}{\pi}\right)  \int_{-\infty}^{+\infty} d \epsilon \frac{\mathrm{Im}\left\{\mathscr{G}^R(\epsilon)\right\}}{i \omega_n-\epsilon} e^{i \omega_n 0^+} \nonumber \\ &= \int_{-\infty}^{+\infty} d \epsilon   \left[-\frac{1}{\pi} \mathrm{Im}\left\{G^R(\epsilon)\right\} \right]  \underbrace{ \frac{1}{\beta}\sum_{i \omega_n} \frac{e^{i \omega_n 0^+}}{i \omega_n-\epsilon} }_{f_{FD}(\epsilon)e^{i \epsilon 0^+} } \nonumber \\ &=\int_{-\infty}^{+\infty} d \epsilon N(\epsilon) f_{FD}(\epsilon),
\end{align}
where we recognized the interacting density of states (\ref{eq:interacting_DOS}). We checked the equivalence of the two representations (\ref{eq:n_def}) and (\ref{eq:n_def_imag}) for the fermion density, by comparing our results on the real and on the imaginary axis for the 2D-YSYK problem of Sec. \ref{Model_methods_YSYK}.
From an operational standpoint, we keep density $n$ and temperature $T$ as independent variables, and we fix $n$ to be constant, independent of $T$ and of interaction $g'$. Then, to calculate the chemical potential $\mu(n,T)$ self-consistently, on the imaginary axis we employ Eqs\@. (\ref{eq:n_def_imag}) and (\ref{eq:Dyson_fermions_gen}), which explicitly yield
\begin{align}\label{eq:n_mu_explicit}
n&=k_B T \sum_{i\omega_n}\left[\mathscr{G}(i\omega_n)-\frac{1}{i\omega_n-\mu}\right]+f_{FD}(\mu) \nonumber \\ & = k_B T \sum_{i\omega_n}\left\{\int \frac{d\vec{k}}{(2\pi)^2}
\frac{1}{i\omega_n-\xi_{\vec{k}}+\mu-\Sigma(i\omega_n)}  \right. \nonumber \\ &  \, \, \, \, \, \, \, \, \, \, \, \, \, \, \, \, \, \, \, \, \, \, \, \, \, \, \,\, \, \, \, \left.-\frac{1}{i\omega_n+\mu}\right\}+f_{FD}(\mu) \nonumber \\ & =k_B T \sum_{i\omega_n} \left\{\frac{2}{\pi}\frac{K_E\left[\frac{4 t}{i\omega_n+\mu-\Sigma(i\omega_n)}\right]}{i\omega_n+\mu-\Sigma(i\omega_n)}  -\frac{1}{i\omega_n+\mu}\right\} \nonumber \\ &  \, \, \, \, \, \, \, \, \, \, \, \, \, \, \, \, \, \, \, \, \, \, \, \, \, \, \,\, \, \, \, +f_{FD}(\mu).
\end{align}
At the second step of Eq\@. (\ref{eq:n_mu_explicit}) we have used $\xi_{\vec{k}}=\epsilon_{\vec{k}}-\mu$, while at the third step we have performed the momentum sum over $\vec{k}$ for the square-lattice dispersion (\ref{eq:square_disp}), written in terms of the complete elliptic integral of the first kind $K_E(z)$, in accordance with Eq\@. (\ref{eq:G0_int_cont_big}). Eq\@. (\ref{eq:n_mu_explicit}) also explicitly shows the simultaneous addition and subtraction of a free-fermion term $f_{FD}(\mu)$ to the Matsubara sum over $i\omega_n$, which improves numerical convergence \cite{Ferrari-2019,Smit-2021}. 
Since the self-energy $\Sigma(i\omega_n)$ also implicitly depends on the chemical potential through Eq\@. (\ref{eq:saddle_point_imag_gen_Sigma}), Eq\@. (\ref{eq:n_mu_explicit}) must be solved self-consistently to keep $n$ fixed. In practice, we opt for first solving the saddle-point problem (\ref{eq:saddle_point_imag_gen}) numerically at fixed chemical potential $\mu$, temperature $T$, and interaction $g'$, starting from an initial guess for $\mu$. We embed these saddle-point numerics into another self-consistent loop on $\mu(n,T)$, with the latter adjusted so that $n$ is kept constant at each $g'$ and $T$. Convergence of both embedded loops finally yields saddle-point self-energies and propagators, as well as the interacting $\mu(n,T)$, at fixed density. 
Further details on the numerical algorithm are collected in App\@. \ref{App:Numerics}. We can define a ``doping" level of the system as the difference between the set density and its value at particle-hole symmetry, i.e.\@, $\Delta n=n-0.5/a^2$. For the nearest-neighbour square-lattice dispersion $\epsilon_{\vec{k}}$ here employed, the results at $\pm \Delta n$ only differ in the sign, but not the absolute value, of the Hall conductivity \footnote{Notice that the zero-temperature Hall coefficient, as given by Eq\@. (\ref{eq:RH_gen}), is $R_H^{(0)}(0)=1/(\left|e\right| n)$ only in the semiclassical regime at low density, where the chemical potential approaches the band edge \cite{Morpurgo-2024}. Away from this band-edge limit, $\left|R_H(0)\right|<\left|R_H^{(0)}\right|$.  This is consistent with the expectation that band-structure effects
reduce the Hall constant in the semiclassical approximation \cite{Ong-1991}. }. 
\begin{figure}[ht]
  \includegraphics[width=1.0\linewidth]{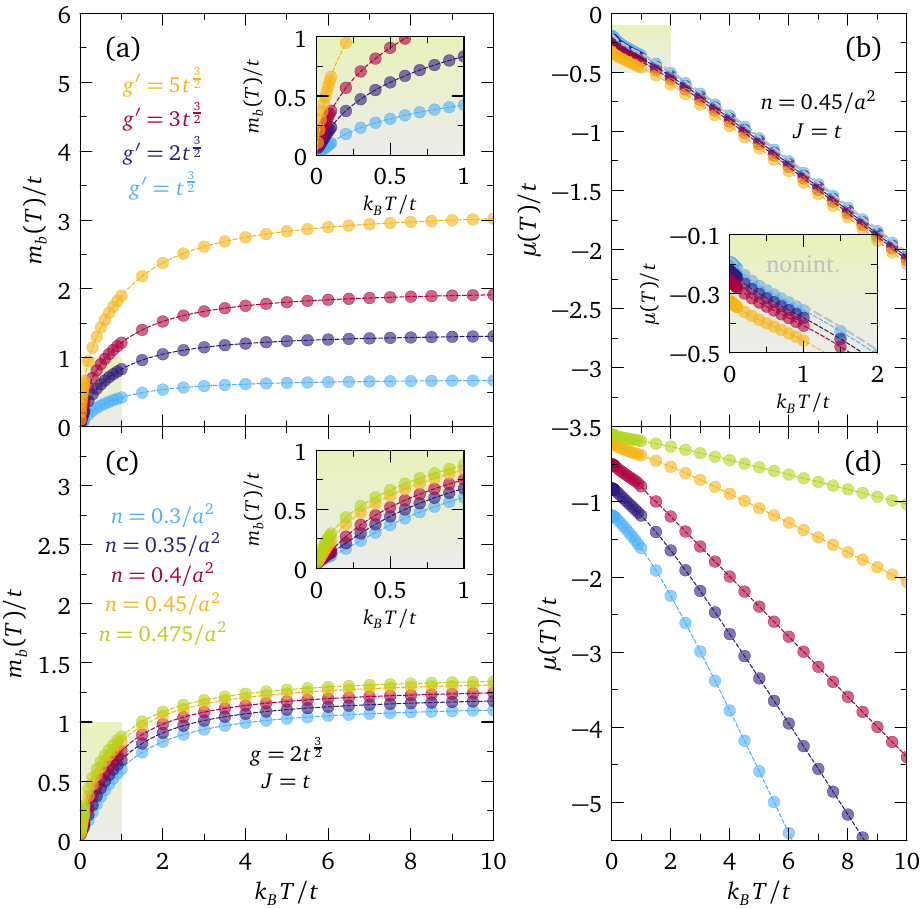}
\caption{\label{fig:mb_mu_square_T_varg_fixedn} Dependence of the renormalized boson mass $m_b(T)$ and of the chemical potential $\mu(T)$ on dimensionless temperature $k_B T/t$ for boson stiffness $J=t$. (a) $m_b(T)$ and (b) $\mu(T)$ at fixed density $n=0.45/a^2$ and different interactions $g'$.  (c) $m_b(T)$ and (d) $\mu(T)$ at fixed interaction $g'=2 t^{3/2}$ and different densities $n$. The insets in panels (a), (b), and (c) zoom on the low-$T$ regime. The gray dashed curves in panel (b) show the noninteracting ($g'=0$) limit of $\mu(T)$.
}
\end{figure}

\section{Thermodynamic properties at fixed fermion density: renormalized boson mass and chemical potential}\label{Imaginary_thermodyn}

We first self-consistently solve the 2D-YSYK saddle-point equations (\ref{eq:saddle_point_imag_gen}) on the imaginary axis, using the numerical protocols described in App\@. \ref{App:Numerics}; see also Refs\@. \cite{Esterlis-2021,Guo-2022,Patel-2023,Li-2024}. These solutions grant us access to all thermodynamic properties of the 2D-YSYK system, and specifically yield the temperature-dependent renormalized boson mass $m_b(T)$, in accordance with Eq\@. (\ref{eq:m_b_T}), and the interaction- and temperature-dependent chemical potential $\mu(n,T)$, as (implicitly) determined by Eq\@. (\ref{eq:n_def_imag}). 
Fig\@. \ref{fig:mb_mu_square_T_varg_fixedn}(a) shows the renormalized boson mass $m_b(T)$ as a function of dimensionless temperature $k_B T/t$, at fixed density $n=0.45/a^2$, and for boson stiffness $J=t$, for different spatially disordered interactions $g'=\left\{1,2,3,5\right\}t^{\frac{3}{2}}$. The corresponding values of the bare boson masses at $T\rightarrow 0^+$, used to tune the system to criticality, are $(m_b^0/t)^2\approx \left\{0.5,1.625, 4.1, 10.305\right\}$, respectively. The inset in Fig\@. \ref{fig:mb_mu_square_T_varg_fixedn}(a) zooms on the low-temperature region, where it is indeed seen that the renormalized mass decreases with decreasing $T$ as $m_b(T)\propto T\ln(T)$ \cite{Esterlis-2021, Patel-2023,Li-2024}.
At high temperature, $m_b(T)$ reaches an asymptotic plateau, which is less than the corresponding value of $m_b^0$ for the same interaction; this discrepancy stems from the fact that $\mu(n,0^+)$ at $T\rightarrow 0^+$, self-adjusted at constant density $n$ to get close to criticality, significantly differs from its high-temperature values, as seen in Fig\@. \ref{fig:mb_mu_square_T_varg_fixedn}(b). Since $n=0.45/a^2$, the system has a density lower than the particle-hole symmetry condition $n=0.5/a^2$, so carriers are electron-like and the chemical potential decreases with $T$. For comparison, the dashed gray curve shows the noninteracting ($g'=0$) limit of $\mu(T)$, obtained through numerical inversion of Eq\@. (\ref{eq:n_def}) with the noninteracting density of states $N_0(\epsilon)$, i.e., $n=\int_{-\infty}^{+\infty}d \omega f_{FD}(\omega)N_0(\omega)$. The temperature dependence of the noninteracting $\mu(T)$ is approximately linear at high $T$, while in the limit $k_B T \ll 1$ a Sommerfeld expansion reveals a quadratic evolution: 
\begin{equation}\label{eq:mu_nonint_Sommerfeld}
\mu(T)\approx \mu(0) +\frac{\pi^2}{6} \frac{(k_B T)^2}{\mu(0) \ln[16 t/\left|\mu(0)\right|]}.
\end{equation}
The same qualitative trends as a function of temperature are followed by the interacting chemical potential ($g'>0$), with the noninteracting limit correctly approached as $g'\rightarrow 0$.

The inset of Fig\@. \ref{fig:mb_mu_square_T_varg_fixedn}(b), zooming in on the low-$T$ region, shows that the main effect of increasing interaction $g'$ is a rigid shift of the $\mu(T)$ curves downwards, as stronger interactions broaden the high-energy tails of the interacting density of states $N(\epsilon)$, thus pushing $\mu(T)$ down to keep $n$ constant. 

\begin{figure*}[ht]
  \includegraphics[width=1\textwidth]{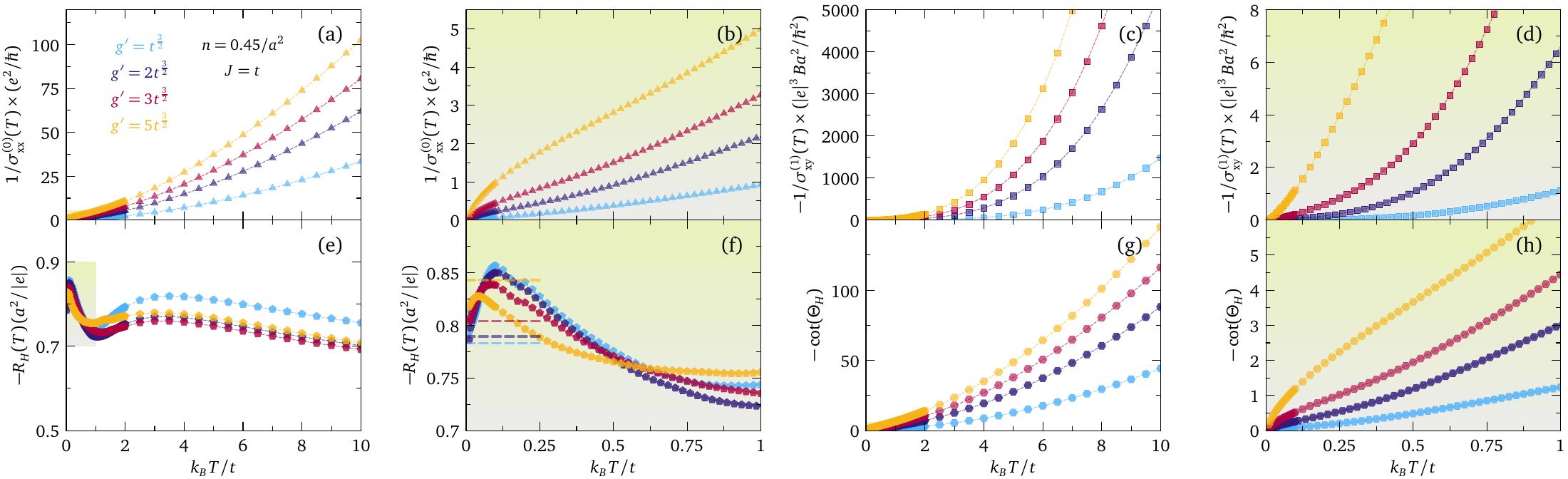}
\caption{\label{fig:rho_RH_cotThetaH_square_T_vargp_fixedn}
Linear dc magnetotransport coefficients as a function of temperature $k_B T/t$ normalized by the fermion hopping $t$, for fermion density $n=0.45/a^2$ and boson stiffness $J=t$, for different interactions $g'$. The system is tuned to the QCP at $T\rightarrow 0^+$ for each $g'$ at fixed density. }
\end{figure*}
Fig\@. \ref{fig:mb_mu_square_T_varg_fixedn}(c) displays the renormalized boson mass $m_b(T)$ as a function of $k_B T/t$, at fixed interaction $g'=2 t^{\frac{3}{2}}$, boson stiffness $J=t$, and for different densities $n=\left\{0.3,0.35,0.4,0.45,0.475\right\}/a^2$. The corresponding values of the bare boson masses at $T\rightarrow 0^+$ are here $(m_b^0/t)^2\approx \left\{1.375,1.565,1.74, 1.918, 2.0\right\}$, respectively. The inset in Fig\@. \ref{fig:mb_mu_square_T_varg_fixedn}(c) illustrates that the low-$T$ slope of $m_b(T)$ increases with increasing $n$. However, the high-temperature behaviour of $m_b(T)$ is less sensitive to density changes than interaction changes, since the latter cause stronger variations of the value $m_b^0$ that tunes the system towards the QCP. 
Fig\@. \ref{fig:mb_mu_square_T_varg_fixedn}(d) shows the chemical potential $\mu(T)$ for the same parameters as in Fig\@. \ref{fig:mb_mu_square_T_varg_fixedn}(c): we see that the temperature dependence is less marked when moving density towards particle-hole symmetry. This feature will be seen in Sec. \ref{Hall_results} to impact the results for the Hall coefficient $R_H(T)$. 

\section{Conductivities, Hall coefficient, and Hall angle as a function of disordered interactions and fermion density}\label{Hall_results}

The imaginary-axis results for the renormalized boson mass $m_b(T)$ and chemical potential $\mu(n,T)$, described in Sec. \ref{Imaginary_thermodyn}, constitute inputs for the real-axis solutions of the 2D-YSYK saddle-point equations (\ref{eq:saddle_point_imag_gen}), obtained after analytic continuation as described in App\@. \ref{Real_continuation}. The numerically exact solutions for the retarded fermion and boson self-energies, $\Sigma^R(\omega)$ and $\Pi^R(\omega)$, are described and explicitly shown in App\@. \ref{Num_real_plots}. The ensuing low-$T$ phenomenology is completely analogous to the MFL regime mentioned in Refs\@. \onlinecite{Patel-2023,Guo-2024,Li-2024}. 
In particular, the fermion self-energy $\Sigma^R(\omega)$ is then inserted into Eqs\@. (\ref{eq:Kubo_formulae}), to obtain the longitudinal and Hall conductivities at linear order in $B$. 

\begin{figure*}[ht]
  \includegraphics[width=1\linewidth]{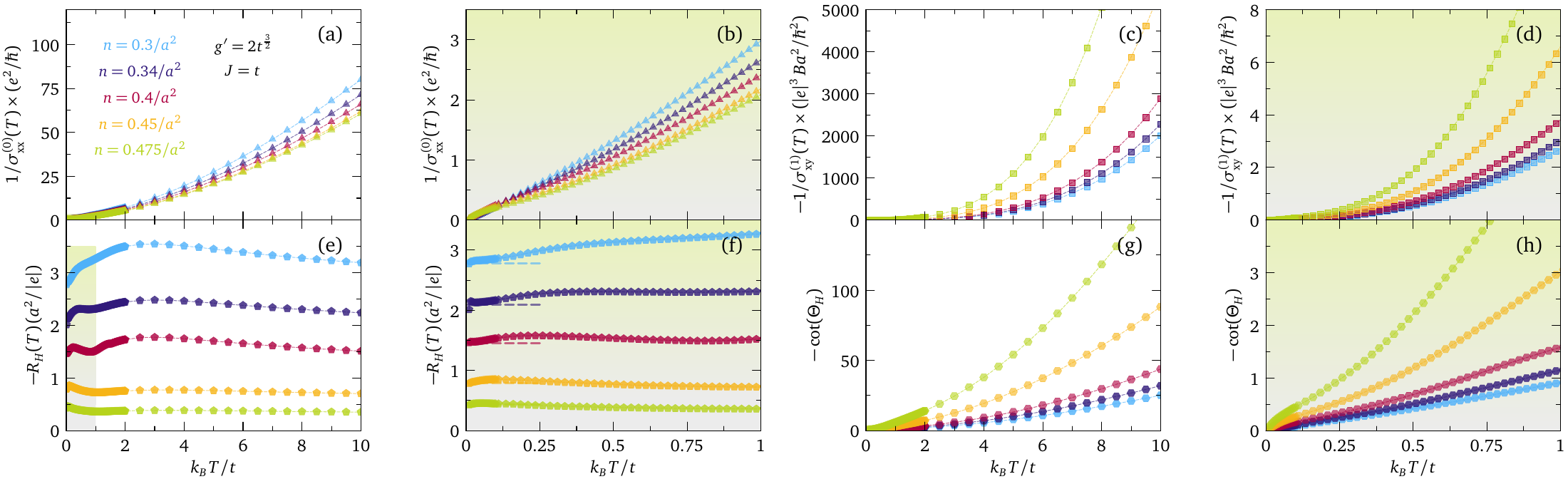}
\caption{\label{fig:rho_RH_cotThetaH_square_T_gp2_fixedn}
Linear dc magnetotransport coefficients as a function of temperature $k_B T/t$ normalized by the fermion hopping $t$, for spatially disordered interaction $g'=2 t^{3/2}$ and boson stiffness $J=t$, for different fermion density $n$. The system is tuned to the QCP at $T\rightarrow 0^+$ for each density. }
\end{figure*}

Let us further substantiate the results in Fig\@. \ref{fig:Hall_summary} by a quantitative analysis of the conductivities, first at fixed fermion density and then at fixed spatially disordered interaction. Fig\@. \ref{fig:rho_RH_cotThetaH_square_T_vargp_fixedn}(a-h) shows all $T$-dependent linear magnetotransport coefficients at fixed density $n=0.45/a^2$, boson stiffness $J=t$, and for different interaction $g'=\left\{1,2,3,5\right\}t^{3/2}$. At each interaction value, the system is tuned to the QCP at $T\rightarrow 0^+$, as detailed in Sec. \ref{Imaginary_thermodyn}. The low-$T$ dependence of the longitudinal resistivity $1/\sigma_{xx}^{(0)}(T)$, normalized by the 2D conductivity quantum $e^2/\hbar$, corresponds to the MFL phenomenology \cite{Varma-1989,Patel-2023,Li-2024}: the resistivity is $T$-linear within logarithmic corrections that are amplified with increasing $g'$, as seen in Fig\@. \ref{fig:rho_RH_cotThetaH_square_T_vargp_fixedn}(b); this evolution is due to the increasing low-$T$ slope of the renormalized boson mass $m_b(T)$ with increasing $g'$, as previously commented on with reference to Fig\@. \ref{fig:mb_mu_square_T_varg_fixedn}(a). As temperature is further increased, the $1/\sigma_{xx}^{(0)}(T)$ curves develop a low-$T$ concavity, which signals the crossover from a MFL to Fermi-liquid physics as the system is progressively detuned from criticality \cite{Patel-2023,Li-2024}; see Fig\@. \ref{fig:mb_mu_square_T_varg_fixedn}(a). 
The corresponding linear-in-$B$ Hall resistivity $1/\sigma_{xy}^{(1)}(T)$, normalized to $\left|e\right|^3 B/\hbar$, is displayed in Fig\@. \ref{fig:rho_RH_cotThetaH_square_T_vargp_fixedn}(c) within a large temperature range, and in Fig\@. \ref{fig:rho_RH_cotThetaH_square_T_vargp_fixedn}(d) that zooms on the low-temperature region. There we have $1/\sigma_{xy}^{(1)}(T)\propto T^2$, with an increasing slope for increasing $g'$. 
Indeed, in MFL regime analytical estimations of the conductivities at fixed chemical potential $\mu$ read $1/\sigma_{xx}^{(0)}(T)=e^2 \hbar \overline{\Phi_{(0)}^{xx}(\mu)}/(2\left|\mathrm{Im}\left\{\Sigma^R(0)\right\}\right|)$ and $1/\sigma_{xy}^{(1)}(T)/B=\left|e\right|^3 \hbar e^2 \hbar \overline{\Phi_{(1)}^{xy}(\mu)} 3/(8\pi^2\left|\mathrm{Im}\left\{\Sigma^R(0)\right\}\right|^2)$ -- see App\@. \ref{App:analysis_magneto} -- which are respectively linear and quadratic in temperature if $\left|\mathrm{Im}\left\{\Sigma^R(0)\right\}\right|\propto k_B T$ as it occurs in MFL regime. These estimations, through Eq\@. (\ref{eq:RH_gen}), yield the low-$T$ Hall coefficient
\begin{equation}\label{eq:RH_1stB_MFL_T0}
R_H(T)\approx \frac{1}{\left|e\right| \hbar} \frac{3}{2\pi^2} \frac{\overline{\Phi_{(1)}^{xy}(\mu)}}{\left[\overline{\Phi_{(0)}^{xx}(\mu)} \right]^2}, \, T\rightarrow 0^+,
\end{equation}
where $\overline{\Phi_{(0)}^{xx}(\mu)}$ and $\overline{\Phi_{(1)}^{xy}(\mu)}$ are energy-linearizations of the transport functions at $\mu$, as in Eqs\@. (\ref{eq:transport_Phi_linearized}). The linear terms in Eqs\@.  (\ref{eq:transport_Phi_linearized}) are nonnull because the retarded fermionic self-energy $\Sigma^R(\omega)$ is not symmetric around $\omega=0$ for $\mu \neq 0$, with its static real part $\mathrm{Re}\left\{\Sigma^R(0)\right\}$ acting as a renormalization of chemical potential; see Fig\@. \ref{fig:Sigma_YSYK_square_T_fixedn}. Eq\@. (\ref{eq:RH_1stB_MFL_T0}) gives the dashed horizontal lines in Fig\@. \ref{fig:rho_RH_cotThetaH_square_T_vargp_fixedn}(f). Notice that the MFL Hall coefficient depends on 2D-YSYK interactions through $\mathrm{Re}\left\{\Sigma^R(0)\right\}$, and is thus renormalized (specifically, decreased) with respect to its noninteracting value on the square lattice. 

At higher temperatures, approaching the crossover regime, the numerical Hall coefficients in \ref{fig:rho_RH_cotThetaH_square_T_vargp_fixedn}(f) first traverse a $g'$-dependent maximum, which occurs at $g'k_B T \approx \mu(T)$, before decreasing due to the sign-changing Hall-transport function $\Phi_{(1)}^{xy}(\epsilon)$, as analyzed in App\@. \ref{App:analysis_magneto}. At still higher temperatures $k_B T\gtrapprox t$, $R_H(T)$ increases again due to boson dynamics progressively decoupling from fermions, and then we observe another decrease of $R_H(T)$ throughout the crossover between the classical-metal and incoherent-metal regimes, as shown in Fig\@. \ref{fig:rho_RH_cotThetaH_square_T_vargp_fixedn}(e). Finally, Fig\@. \ref{fig:rho_RH_cotThetaH_square_T_vargp_fixedn}(g) displays the results for the cotangent of the Hall angle, computed from the data in Fig\@. \ref{fig:rho_RH_cotThetaH_square_T_vargp_fixedn}(a-d) through Eq\@. (\ref{eq:cotThetaH_gen}). While the high-$T$ concavity of the curves in classical-metal and incoherent-metal regimes is expected, since the longitudinal resistivity itself in Fig\@. \ref{fig:rho_RH_cotThetaH_square_T_vargp_fixedn}(a) becomes superlinear (Fermi liquid-like) in the same regime, at low temperatures $k_B T <t$ we observe nontrivial behaviour of the Hall angle cotangent, as shown by Fig\@. \ref{fig:rho_RH_cotThetaH_square_T_vargp_fixedn}(h) which zooms the same data of Fig\@. \ref{fig:rho_RH_cotThetaH_square_T_vargp_fixedn}(g) on the low-$T$ region. 
In MFL regime at the lowest temperatures, $\cot\left[\Theta_H(T)\right]\propto T$ as expected for all $g'$ values. In fact, utilizing the same estimations as for Eq\@. (\ref{eq:RH_1stB_MFL_T0}), we can derive the MFL analytical approximation
\begin{equation}\label{eq:cotthetaH_1stB_MFL_T0}
\cot\left[\theta_H(T)\right]=\frac{1}{\hbar B}\frac{3 \hbar}{4 \pi^2}\frac{\overline{\Phi_{(0)}^{xx}(\mu)}}{\overline{\Phi_{(1)}^{xy}(\mu)}} \left|\Sigma^R(0)\right|, \, T\rightarrow 0^+.
\end{equation}
Due to the low-$T$ MFL-like static self-energy $\mathrm{Im}\left\{\Sigma^R(0)\right\}\propto k_B T$ retrieved in our model, Eq\@. (\ref{eq:cotthetaH_1stB_MFL_T0}) indeed yields $\cot\left[\theta_H(T)\right]\propto T$. However, in the intermediate-$T$ crossover regime, the simultaneous presence of $T$-linear longitudinal resistivity from Fig\@. \ref{fig:rho_RH_cotThetaH_square_T_vargp_fixedn}(b) and the decrease of the Hall coefficient in Fig\@. \ref{fig:rho_RH_cotThetaH_square_T_vargp_fixedn}(f) determines an upward concavity of the $\left|\cot\left[\Theta_H (T)\right]\right|\propto T^\alpha$ curves, with $\alpha>1$. The exponent $\alpha$ is maximized at weak coupling $g'$, as also seen in Fig\@. \ref{fig:cotTheta_fixedn_logscale}(a).  

All the above analysis is confirmed when we keep the spatially disordered interaction fixed and we vary the fermion density. Fig\@. \ref{fig:rho_RH_cotThetaH_square_T_gp2_fixedn}(a-h) shows the linear magnetotransport coefficients at interaction $g'=2 t^{3/2}$, boson stiffness $J=t$, and for different fermion densities $n=\left\{0.3,0.35,0.4,0.45,0.475\right\}/a^2$. Again, at each fixed density the system is tuned to the QCP at $T\rightarrow 0^+$. Since the interpretation is similar to the previously discussed fixed-density calculations, here we just comment on the main differences with respect to Fig\@. \ref{fig:rho_RH_cotThetaH_square_T_vargp_fixedn}. First, the variation in density has less impact on the longitudinal resistivity, and more impact on the Hall coefficient, with respect to variations in interactions, as appreciated by comparing Figs.\@. \ref{fig:rho_RH_cotThetaH_square_T_gp2_fixedn}(b,f) with Figs.\@. \ref{fig:rho_RH_cotThetaH_square_T_vargp_fixedn}(b,f). Specifically, the maximum in $R_H(T)$ shifts to higher temperatures with decreasing $n$ (i.e.\@ increasing doping $\Delta n$), so that the low-$T$ MFL regime extends to increasingly high temperatures. However, the $T\rightarrow 0^+$ value of the Hall coefficient still obeys the estimation (\ref{eq:RH_1stB_MFL_T0}), as shown by the dashed horizontal lines in Fig\@. \ref{fig:rho_RH_cotThetaH_square_T_gp2_fixedn}(f). 
Thus, the Hall-angle cotangent is still linear in $T$ up to $k_B T/t=1$ for the lowest densities (highest doping), while it is superlinear with exponent $\alpha \approx 1.4$ close to particle-hole symmetry (lowest doping), as illustrated in Fig\@. \ref{fig:rho_RH_cotThetaH_square_T_gp2_fixedn}(h). 
When plotted in logarithmic scale, the results in Figs.\@. \ref{fig:rho_RH_cotThetaH_square_T_vargp_fixedn}(h) and \ref{fig:rho_RH_cotThetaH_square_T_gp2_fixedn}(h) yield Fig\@. \ref{fig:cotTheta_fixedn_logscale}, which reveals the precise exponents $1<\alpha\lessapprox 1.4$ of the Hall-angle cotangent at intermediate temperatures. 
\begin{figure*}[ht]
  \centering
\includegraphics[width=0.65\linewidth]{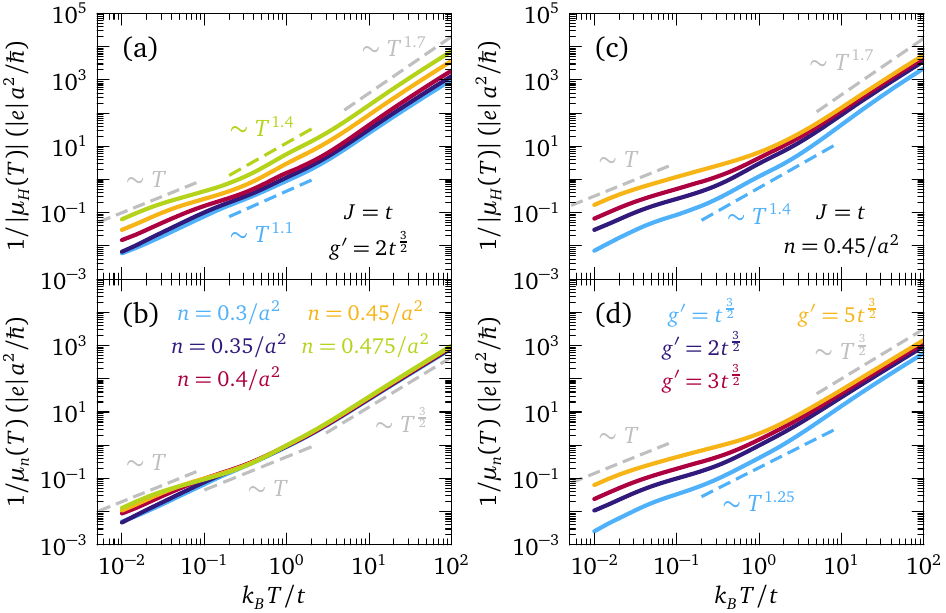}
\caption{\label{fig:mu_compare}
Inverse carrier mobilities as a function of normalized temperature $k_B T/t$, for boson stiffness $J=t$, estimated from (a,c) the Hall effect $\mu_H(T)$ according to Eq\@. (\ref{eq:mu_H}) and (b,d) from the fermion density $\mu_n(T)$ in accordance with Eq\@. (\ref{eq:mu_n}). Results are shown at (a,b) fixed interaction $g'=2 t^{\frac{3}{2}}$ and different fermion densities, and at (c,d) fixed density $n=0.45/a^2$ and different interactions.  }
\end{figure*}

Moreover, the computations in the crossover regime of Figs.\@. \ref{fig:rho_RH_cotThetaH_square_T_vargp_fixedn} and \ref{fig:rho_RH_cotThetaH_square_T_gp2_fixedn} are qualitatively robust with respect to varying the distance from the QCP (tuning the bare boson mass $m_B^0$); see also Sec. \ref{Distance_QCP}. All analyses so far in turn imply the qualitative robustness of our results for the superlinear Hall angle against detuning from the QCP and varying density/interaction. 

\section{Carrier mobility}\label{Mobility_results}

To assess the origin of the apparent experimental difference of scattering rates for the longitudinal and Hall channels \cite{Chien-1991,Anderson-1991,Anderson-1991theory,Ayres-2021,Grissonnanche-2021}, we also compute the carrier mobility. Two definitions of carrier mobility are frequently utilized \cite{Morpurgo-2025}: the first is connected to fermion density and longitudinal conductivity,
\begin{subequations}\label{eq:mu_defs}
\begin{equation}\label{eq:mu_n}
\mu_n(T)=\frac{\sigma_{xx}^{(0)}(T)}{\left|e\right| n},
\end{equation}
while the second involves the Hall coefficient, 
\begin{align}\label{eq:mu_H}
\mu_H(T)&=\sigma_{xx}^{(0)}(T) R_H(T)= \frac{\sigma_{xy}^{(1)}(T)}{B \sigma_{xx}^{(0)}(T)} \nonumber \\ & \equiv \frac{1}{B \cot\left[\Theta_H(T)\right]}.
\end{align}
\end{subequations}
The definitions (\ref{eq:mu_defs}) are equivalent in the semiclassical regime at low density $n$, where $R_H(T)\equiv 1/(\left|e\right| n)$ \cite{Morpurgo-2025}. Away from this regime, even for noninteracting fermions on the square lattice we have $R_H(T)\neq  1/(\left|e\right| n)$ -- see App\@. \ref{App:analysis_magneto} -- and fermionic interactions further renormalize the Hall coefficient: in particular, increasing $g'$ the zero-temperature Hall coefficient increases in magnitude; see Fig\@. \ref{fig:mb_mu_square_T_varg_fixedn}(f). Therefore, it is not guaranteed that Eqs\@. (\ref{eq:mu_H}) and (\ref{eq:mu_n}) have the same qualitative evolution with temperature. 
In particular, the experimental determination of the mobility often relies on the definition (\ref{eq:mu_H}), where the Hall coefficient $R_H(T)=1/[n^\ast(T) e]$ is taken as an estimation of carrier density. By definition, this would imply that the effective carrier density $n^\ast(T)$ is temperature-dependent. 

Fig\@. \ref{fig:mu_compare}(a,c) show the inverse ``Hall" mobility $1/\mu_H(T)$ according to Eq\@. (\ref{eq:mu_H}), which is equivalent to Hall angle cotangent at fixed magnetic field; cfr. Figs.\@. \ref{fig:rho_RH_cotThetaH_square_T_vargp_fixedn}(g,h) and \ref{fig:rho_RH_cotThetaH_square_T_gp2_fixedn}(g,h). Therefore, all comments made with respect to $\cot[\Theta_H(T)]$ identically apply to $1/\mu_H(T)$, including the superlinearity in crossover regime, both at fixed interaction $g'=2 t^{3/2}$ and different fermion densities $n=\left\{0.3,0.35,0.4,0.45,0.475\right\}/a^2$ -- see Fig\@. \ref{fig:mu_compare}(a) -- and at fixed density $n=0.45/a^2$ and varying interaction $g'=\left\{1,2,3,5\right\}t^{3/2}$ -- see Fig\@. \ref{fig:mu_compare}(c). Remarkably, significative differences in the qualitative evolution with temperature are obtained in Fig\@. \ref{fig:mu_compare}(b,d), which display the inverse mobility $1/\mu_n(T)$ stemming from Eq\@. (\ref{eq:mu_n}). As for the Hall angle cotangent, we find that the superlinearity of the mobilities in the intermediate-$T$ crossover regime is enhanced in the weak-interaction limit, as shown in Fig\@. \ref{fig:mu_compare}(d), but the apparent exponent is lower than the corresponding one for $1/\mu_H(T)$. Conversely, $1/\mu_n(T)$ is much less sensitive to density variations than to interaction tuning, as illustrated in Fig\@. \ref{fig:mu_compare}(b), as compared to $1/\mu_H(T)$; such robustness stems from the fact that $R_H(T)$ varies superlinearly with density $n$ close to particle-hole symmetry, so that $1/\mu_H(T)$ is more sensitive to density variations than its counterpart $1/\mu_n(T)$. 
In crossover regime, $1/\mu_n(T)$ still displays the approximate $T$-linearity that it inherits from $1/\sigma_{xx}^{(0)}(T)$, in qualitative contrast with $1/\mu_H(T)$

Overall, from Fig\@. \ref{fig:mu_compare} we conclude that the estimations (\ref{eq:mu_defs}) of carrier mobilities yield qualitatively different results in crossover regime, where a superlinear relation with temperature is retrieved for $1/\mu_H(T)$ (the more experimentally oriented definition) similarly to $\left|\cot[\Theta_H(T)]\right|$, but not for $1/\mu_n(T)$ (where the theoretical fixing of carrier density plays a crucial role). 
These considerations reflect the absence in our model of a true separation between distinct scattering rates in the longitudinal and Hall conductivities: both conductivities are analogously affected by YSYK interactions, and the superlinearity of the Hall-angle cotangent results from embedding this interacting system on a lattice; see also Sec. \ref{eq:phys_origin}. Our analysis leads to the conclusion that, even in the absence of different scattering mechanisms for the longitudinal and Hall responses, experimental protocols using Eq\@. (\ref{eq:mu_H}) would still observe a superlinear evolution of the mobility with temperature in crossover regime, due to lattice effects. Hence, such effects might need to be reevaluated in the light of their impact on Hall measurements in strongly interacting systems.

\section{Discussion}\label{Discussion}

\subsection{Physical origin of Hall-angle cotangent superlinearity: synergy of interactions and square-lattice embedding}\label{eq:phys_origin}

The main result of our 2D-YSYK lattice computation is the concomitant $T$-superlinearity of the Hall-angle cotangent and $T$-linearity of the longitudinal resistivity in crossover regime. In essence, this synergy stems from two distinct effects: on one hand, disordered interactions $g'$ generate $T$-linear longitudinal resistivity $1/\sigma_{xx}(T)$; on the other hand, square-lattice embedding, i.e.\@, a sign-changing Hall transport function $\Phi_{(1)}^{xy}(\epsilon)$, produces a Hall coefficient $R_H(T)$ that decreases with $T$.

Specifically, the interactions $g'$ strongly break translational invariance and they provide the source of inelastic boson-fermion scattering that produces $T$-linear longitudinal resistivity $1/\sigma_{xx}(T)\propto T$, both in the low-$T$ MFL and intermediate-$T$ crossover regimes \cite{Patel-2023,Li-2024,Guo-2024}. This feature qualitatively holds true at all values of $g'$, apart from enhanced logarithmic corrections in $1/\sigma_{xx}(T)$ at large $g'$; see Figs.\@. \ref{fig:Hall_summary}(c) and \ref{fig:rho_RH_cotThetaH_square_T_vargp_fixedn}(a,b). In this sense, the breaking of translational invariance positively contributes to strange-metal physics in the longitudinal channel. 

However, as shown in Figs.\@. \ref{fig:Hall_summary}(d) and \ref{fig:rho_RH_cotThetaH_square_T_vargp_fixedn}(f), increasing $g'$ also makes the Hall coefficient $R_H(T)$  decrease less with temperature $T$ in the crossover regime, thus effectively \emph{weakening} the superlinearity of $\cot[\Theta_H(T)]=1/[R_H(T) B \sigma_{xx}(T)]$. This weakening occurs because the decrease of $R_H(T)$ with $T$ in crossover regime is not due to YSYK interactions, but it is essentially a lattice effect -- in our case, due to the square-lattice fermionic dispersion (\ref{eq:square_disp}): the embedding on the square lattice directly implies the sign change of the Hall transport function $\Phi^{xy}_{(1)}(\epsilon)$ at $\epsilon=0$ -- see Fig\@. \ref{fig:Phixx_xy_example}(b) -- which makes $R_H(T)$ decrease with $T$ in crossover regime. A qualitatively analogous effect occurs on a square lattice using a constant (impurity-like) scattering rate -- see, e.g.\@, Fig\@. 7 in Ref. \onlinecite{Morpurgo-2024} -- instead of our YSYK interactions. Conversely, this effect on $R_H(T)$ would not occur in our 2D-YSYK system if we assumed a quadratic dispersion $\epsilon_{\vec{k}}=\hbar^2 k^2/(2m)$ (with ultraviolet cutoff $\Lambda$ for momentum integrals), since in that case the Hall transport function does not change sign.  

In fact, increasing interaction $g'$ broadens the spectral functions and washes out lattice-related features even at low $T$; see, e.g.\@, Fig\@. \ref{fig:DOS_YSYK_square_T} in our manuscript that compares the noninteracting spectral functions -- dashed gray lines -- with the results at finite $g'$ and different temperatures. Hence, larger values of $g'$ are detrimental towards the decrease of $R_H(T)$ with $T$. 
This is the reason why superlinearity in the Hall-angle cotangent is stronger at \emph{weak coupling}, i.e.\@, small $g'$, in our model, as seen from Figs.\@. \ref{fig:Hall_summary} and \ref{fig:rho_RH_cotThetaH_square_T_vargp_fixedn}.

The above observations indicate the absence of two intrinsically different scattering rates in our model: the same spatially disordered YSYK interactions affect the longitudinal and Hall conductivities on equal footing. Thus, we demonstrate that even in the absence of an additional scattering channel for the Hall response, the Hall-angle cotangent can become superlinear due to lattice embedding. However, this mechanism might cross over into a true separation between scattering channels for the longitudinal and Hall response when Mott physics (not included in our model) dominates transport; see also Sec. \ref{Compare_theory}.
\begin{figure*}[ht]
  \includegraphics[width=1\textwidth]{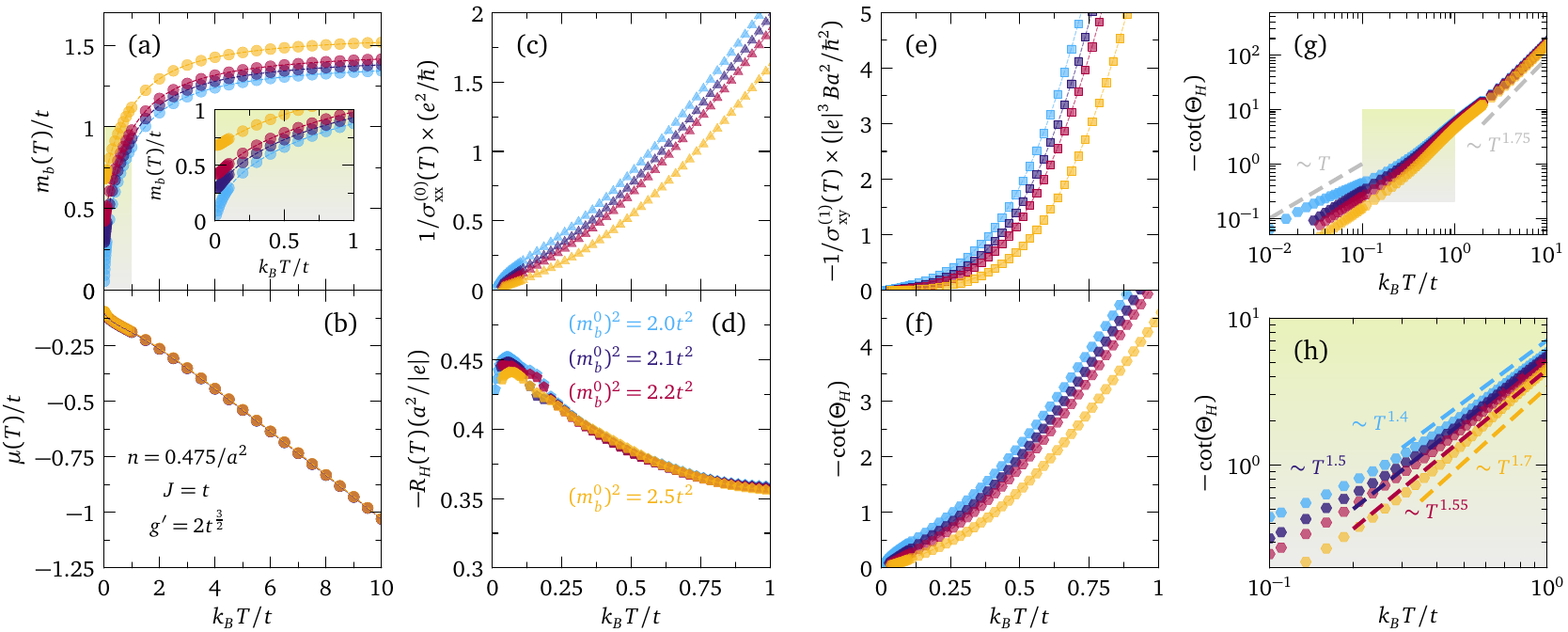}
\caption{\label{fig:rho_RH_cotThetaH_square_T_gp2_fixedn_distQCP}
Thermodynamic variables and magnetotransport coefficients as a function of temperature $k_B T/t$ normalized by the fermion hopping $t$, for spatially disordered interaction $g'=2 t^{\frac{3}{2}}$ , boson stiffness $J=t$, density $n=0.475/a^2$, and for various distances from the QCP, tuned by the bare boson mass $m_b^0$. (a) Renormalized boson mass $m_b(T)/t$; the inset shows a low-temperature zoom. (b) Chemical potential $\mu(T)/t$. (c-f) Magnetotransport coefficients. (g) Hall-angle cotangent $\cot\left[\Theta_H(T)\right]$ in logarithmic scale; the shaded area in crossover regime is zoomed on in panel (h).}
\end{figure*}

Furthermore, we osberve that superlinearity of the Hall-angle cotangent persist in the high-temperature classical-metal regime, as seen for instance in Fig\@. \ref{fig:Hall_summary}(e). However, at these high temperatures $k_B T \gtrapprox t$, the bosons are not capable anymore of providing the marginal susceptibility for the fermions, and the longitudinal resistivity essentially stems from free bosons (with no self-energy) quasi-elastically scattering off fermions; see App\@. \ref{eq:classical_metal}. Due to the temperature dependence of $\mu(n,T)$ in this regime we have $1/\sigma_{xx}(T)\sim T^{3/2}$; see, e.g.\@, Fig\@. \ref{fig:Hall_summary}(c). Therefore, although in the classical-metal regime we still find superlinearity of the Hall-angle cotangent, this is not accompanied by $T$-linear strange-metal resistivity, contrarily to what happens in the intermediate-temperature crossover regime.

\subsection{Effect of variable distance from the QCP}\label{Distance_QCP}

The qualitative trends for $\cot\left[\Theta_H(T)\right]$ shown in Figs.\@. \ref{fig:cotTheta_fixedn_logscale}, \ref{fig:rho_RH_cotThetaH_square_T_vargp_fixedn}, and \ref{fig:rho_RH_cotThetaH_square_T_gp2_fixedn} demonstrate that the lattice effects underlying the superlinearity in crossover regime are amplified at weak coupling (lower $g'$) and close to particle-hole symmetry (low doping $\Delta n$), when the system is tuned to the QCP at $T=0$. However, this superlinearity persists, and is even enhanced, when we increase the distance from the QCP by increasing the bare boson mass $m_b^0$, keeping all other parameters fixed. 
This robustness is illustrated in Fig\@. \ref{fig:rho_RH_cotThetaH_square_T_gp2_fixedn_distQCP}, where the numerically calculated linear magnetotransport coefficients are shown for interaction $g'=2 t^{3/2}$, boson stiffness $J=t$, density $n=0.475/a^2$, and for different bare boson masses. The resulting renormalized boson masses $m_b(T)$ are reported in Fig\@. \ref{fig:rho_RH_cotThetaH_square_T_gp2_fixedn_distQCP}(a), with the low-$T$ zoom in the inset highlighting that by increasing $m_b^0$, the bosons harden at $T=0$, which translates as an increased distance from the QCP. The chemical potential $\mu(T)$ is less sensitive to such distance, as illustrated in Fig\@. \ref{fig:rho_RH_cotThetaH_square_T_gp2_fixedn_distQCP}(b). Fig\@. \ref{fig:rho_RH_cotThetaH_square_T_gp2_fixedn_distQCP}(g) reports the associated changes in the Hall angle cotangent, while detuning from quantum criticality, in logarithmic scale for better comparison: while the low-$T$ regime progressively develops an upward concavity like in conventional Fermi liquids, due to the corresponding changes in the longitudinal resistivity in Fig\@. \ref{fig:rho_RH_cotThetaH_square_T_gp2_fixedn_distQCP}(c), the superlinearity in crossover regime is preserved, and even enhanced, with increasing distance from the QCP. This ehnancement still occurs simultaneously with $T$-linear resistivity inside the quantum critical fan, and it is highlighted by Fig\@. \ref{fig:rho_RH_cotThetaH_square_T_gp2_fixedn_distQCP}(h), which zooms on the crossover regime and shows how the apparent power law has an increasing exponent for larger $m_b^0$. Therefore, the superlinarity effect is more pronounced when detuning from quantum criticality. 

In reality, in the scenario where a single QCP is found at a specific doping $\Delta n=\overline{\Delta n}$, varying $\Delta n$ simultaneously modifies the distance from the QCP, the fermion density $n$, and the interaction $g'$ (e.g.\@, due to screening). These three parameters are intertwined in realistic experiments. Our model allows us to controllably disentangle the effect of individual changes in such parameters, thus showing that the exponent $\alpha>1$ of $\left|\cot\left[\Theta_H (T)\right]\right|\propto T^\alpha$ in the intermediate-$T$ regime is a robust feature of the theory. However, the precise value of $\alpha$ depends on $n$, $g'$, and bare boson mass $m_b^0$. Such variability, and specifically the increase of $\alpha$ with decreasing doping, is experimentally confirmed as commented upon in Sec. \ref{Compare_experiments}.

\subsection{Comparison to other theoretical approaches}\label{Compare_theory}

The results of our model stem from strong spatially disordered electronic interactions engendered by the proximity to a QCP, and from the embedding of the quantum critical system into a crystalline lattice (a square lattice with nearest-neighbour hoppings, in our example). Specifically, interactions -- exactly treated in the 2D-YSYK framework -- yield $T$-linear resistivity in MFL regime that extends at higher temperatures into the crossover regime, while lattice embedding allows for the decrease of the Hall coefficient with temperature and for the separation of longitudinal and Hall transport phenomenologies. 

\begin{figure}[ht]
  \centering
\includegraphics[width=1\linewidth]{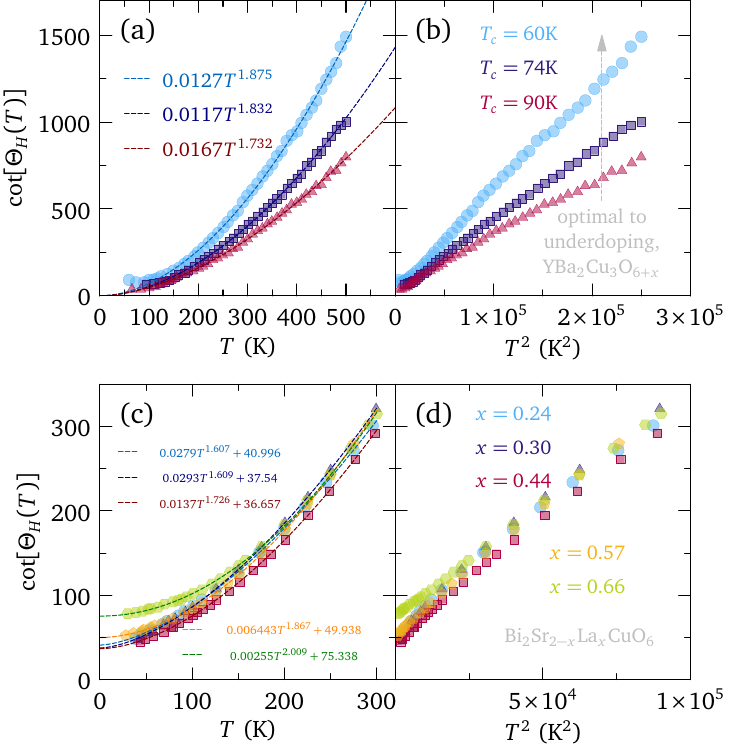}
\caption{\label{fig:Hall_data}
(a) Experimental data (solid symbols) from Ref. \onlinecite{Harris-1992} for $\cot\left[\Theta_H(T)\right]$ as a function of $T$ in YBa$_2$Cu$_3$O$_{6+x}$, together with power-law fits (dashed curves). (b) Same data as in panel (a), plotted as a function of $T^2$. (c) Experimental data (solid symbols) from Ref.~\onlinecite{Ando-1999} for $\cot\left[\Theta_H(T)\right]$ as a function of $T$ in Bi$_2$Sr$_{2-x}$La$_x$CuO$_{6}$, together with power-law fits (dashed curves). (b) Same data as in panel (a), plotted as a function of $T^2$.}
\end{figure}

This mechanism for superlinear Hall angle can be compared with earlier groundbreaking works on bosonization in ``tomographic" 2D Luttinger liquids, leading to separation between spin (spinons) and charge (holons) excitations \cite{Anderson-1991,Anderson-1991theory}. In particular, in these models the longitudinal conductivity $\sigma_{xx}(T)\propto \tau_{\rho}$, with the scattering time $\tau_{\rho}\propto 1/T$ due to the decay of physical electrons into spinons and holons, while  the Hall channel is governed by $\sigma_{xy}(T)\propto \tau_{\rho} \tau_{ss}$ that includes the spinon-spinon scattering time $\tau_{ss}\propto 1/T^2$; the ratio (\ref{eq:cotThetaH_gen}) then only depends on $1/\tau_{ss}\propto T^2$, thus giving a Hall-angle exponent $\alpha=2$. This perspective influenced the graphical presentations of experimental data for $\cot\left[\Theta_H(T)\right]$, plotted as a function of $T^2$ \cite{Anderson-1991,Chien-1991}; see also Fig\@. \ref{fig:Hall_data}. Although these models depart from qualitatively different premises, we mention that spinon-determined low-temperature spectral properties
are similar to a MFL, which also corresponds to the low-$T$ phenomenology in our 2D-YSYK theory. The latter also finds an increase in the Hall-angle exponent $\alpha>1$ towards particle-hole symmetry, which suggests a possible crossover towards Mott-insulating and spinon-related pictures, not described in our model, in the low-doping regime \cite{Bonetti-2025_preprint}. In these pictures, our strange-metal results, which imply a common YSYK scattering mechanism for both longitudinal and Hall responses, would evolve into a true separation between distinct scattering channels for the longitudinal and Hall conductivities in proximity to the Mott transition.

Increasingly accurate numerical techniques, such as numerically exact Quantum Monte Carlo (QMC) solvers \cite{Assaad-1995,Wang-2020c}, or Dynamical Mean Field Theory (DMFT) methods \cite{Pruschke-1995,Markov-2019,Vucicevic-2021,Khait-2023}, allow the computation of the Hall response of the 2D Hubbard model in transverse magnetic fields, even beyond the linear magnetotransport regime considered in this paper. Going beyond the qualitative agreement in the general nonmonotonic shape of $R_H(T)$, it would be interesting to perform an in-depth comparison with our 2D-YSYK results in the relevant regime of doping and on-site interaction $U$ where strange-metal physics occurs. 

On the other side of the theoretical spectrum, phenomenological scaling theories can reconcile a substantial portion of the thermodynamic and spectroscopic properties of strange metals starting from three quantities: the dynamical critical exponent, the hyperscaling violation exponent, and the charge density anomalous exponent \cite{Hartnoll-2015}. As the Hall angle cotangent, the longitudinal resistivity, and the Lorentz ratio provide initial inputs for this analysis, one could investigate how the numerical scalings of aforementioned quantities in our 2D-YSYK theory affect the predictions for other scaling exponents. 

Our work is also in close conceptual proximity to other approaches invoking the crucial role of  lattice effects for the Hall conductivity. 
Boltzmann calculations using a finite-temperature generalization of Ong's geometric construction, and taking into account the qualitative difference between electron- and hole-like states separated by the van Hove singularity on a square lattice, pointed out that at high $T$ electron- and hole-like orbits both contribute to the Hall conductivity, thus determining a decrease of the Hall coefficient with $T$; this is a lattice effect, similar in spirit to our approach except for the assumed existence of well-defined quasiparticle states \cite{Bok-2004}. 
More recently, Boltzmann-based computations in the Shockley–Chambers tube-integral formalism (SCTIF) were successfully fitted to experimentally measured magnetotransport coefficients in cuprate superconductors, including optimally doped  La$_{1.6-x}$Nd$_{0,4}$Sr$_x$CuO$_4$ (Nd-LSCO) \cite{Grissonnanche-2021}, heavily (Pb/La)-doped Bi$_2$Sr$_2$CuO$_{6+\delta}$ (Bi2201), and Tl$_2$Ba$_2$CuO$_{6+\delta}$ (Tl2201) \cite{Ayres-2021}. The success of these kinetic formulations in modeling the in-plane resistivity at zero applied field, the interlayer angle-dependent magnetoresistance (ADMR), and the in-plane T-dependent Hall coefficient $R_H(T)$, is grounded in the inclusion of two different scattering rates: $1/\tau_{\mathrm{is}}$, assumed isotropic in 2D momentum space, and $1/\tau_{\mathrm{an}}(\phi)$, taken to be anisotropic with respect to the angle $\phi$ between the chosen in-plane direction $\vec{k}$ and the $\vec{k}_x$ axis. Qualitatively, our 2D-YSYK formalisms could map $1/\tau_{\mathrm{is}}$ and $1/\tau_{\mathrm{an}}(\phi)$ onto the effects of spatially disordered ($g'$) and translationally invariant ($g$) interactions \cite{Esterlis-2021,Guo-2024}, with associated temperature dependencies of scattering rates determined by the magnitude of the respective inelastic fermion-boson interaction $g'/g$. This is a very interesting generalization of our computations, which we reserve for future investigations; see also Sec. \ref{Conclusions_outlook}. However, notice that in Nd-LSCO the SCTIF fits imply that an isotropic $1/\tau_{\mathrm{is}}(T) \propto T$ dominates the in-plane zero-field resistivity: this situation is qualitatively similar to the effect of spatial disorder ($g'$) in our 2D-YSYK formalism. 

On the other hand, holographic models of quantum critical metals in the presence of a spatially modulated chemical potential predict an anomalous cyclotron resonance already in the hydrodynamic regime of weak lattice perturbations, suggesting that stronger lattice effects would affect the Hall response \cite{Hartnoll-2007,Blake-2015,Chagnet-2024_preprint}. As YSYK models admit holographic duals \cite{Inkof-2022,Stangier-2025}, a systematic mapping of our magnetotransport theory through the AdS/CFT correspondence could shed new light into both sides of the holographic duality. 

Furthermore, it would be interesting to map our results onto hydrodynamic theories, which found success in explaining qualitative features of strange-metal magnetotransport in local equilibrium \cite{Hartnoll-2007,Patel-2017a,Gouteraux-2023}. 

Recently, Hall-angle cotangent calculations for  the vector-field 2D-YSYK model at $T=0$ were performed \cite{Wang-2025a,Wang-2025b}; these go beyond the linear-in-$B$ regime employing a sum over Landau levels, similarly to the scalar cases of Ref. \cite{Guo-2024, Aldape-2022}. These computations confirm the $T$-linearity of longitudinal resistivity in the presence of $B$, and find $\cot\left[\Theta_H(T)\right]\propto T$ through an energy/temperature scaling, in agreement with our results in MFL regime and at quantum criticality. Our present 2D-YSYK computations also explore the crossover regime at higher temperatures, where superlinearity of $\cot\left[\Theta_H(T)\right]$ is found. 

\subsection{Comparison to experimental data}\label{Compare_experiments}

As analyzed in Sec. \ref{Hall_results}, our theory implies a variable temperature exponent of the Hall-angle cotangent, which increases for decreasing interaction $g'$ and decreasing doping $\Delta n$. 

Such variability implies less universality of the exponent $\alpha$ of $\left|\cot\left[\Theta_H (T)\right]\right| \propto T^\alpha$ in strange metals, with respect to the robust $T$-linear evolution of the longitudinal resistivity, 
as opposed to the predictions of theories where $\alpha$ is fixed, e.g.\@, $\alpha=2$ in bosonization theories for 2D Luttinger liquids \cite{Anderson-1991}; see Sec. \ref{Compare_theory}. 
In fact, numerous experimental datasets on strange metals confirm the change of the exponent $\alpha$ with doping level, as exemplified by Fig\@. \ref{fig:Hall_data} that reproduces data on YBa$_2$Cu$_3$O$_{6+x}$ \cite{Harris-1992} and Bi$_2$ Sr$_{2-x}$La$_x$CuO$_{6}$ \cite{Ando-1999}. Power-law fits of the data -- see dashed curves in Fig\@. \ref{fig:Hall_data}(a,c) -- highlight the increase of the exponent $\alpha$ with decreasing hole doping, in qualitative agreement with our predictions. Similar deviations from quadratic evolution have been reported at low $T$ for overdoped Tl$_2$Ba$_2$CuO$_{6+\delta}$ \cite{Mackenzie-1996}. A suitable platform to systematically investigate the variation of the exponent $\alpha$ with doping could be twisted bilayers \cite{Lyu-2021,Xia-2025_preprint}, which allow for controlled tuning of carrier density through applied gate voltages. 

\section{Conclusions}\label{Conclusions_outlook}

In conclusion, we have shown the emergence of superlinear Hall-angle cotangent and carrier mobility in the 2D-YSYK model on a square lattice, through exact self-consistent solutions of the saddle-point equations combined with the Kubo formula for $B$-linear magnetotransport. In this model, the superlinear dependence  $\left|\cot\left[\Theta_H (T)\right]\right|\propto T^{\alpha}$ in the intermediate-$T$ crossover regime, with $1 \lessapprox \alpha \lessapprox 1.4$ for the scanned parameter space when the system is tuned to the QCP, stems from the sign-changing Hall transport function that results from the lattice embedding of the 2D-YSYK system. 
In this picture, a $T^2$-like relation would result from the ``artifact" of apparent linearity when the Hall-angle cotangent is plotted as a function of $T^2$. The superlinear temperature window extends more for weak interactions $g'$ below the fermion half-bandwidth $W/2=4t$, and is more pronounced at low doping $\Delta n$, that is, close to particle-hole symmetry. Our results are robust against varying distance from the QCP, and the exponent $\alpha$ at intermediate temperatures further increases with distance from the QCP. 

More generally, the found superlinearity is suggestive that lattice effects can provide a natural mechanism to decouple the longitudinal and Hall transport dynamics, since the Hall conductivity is more sensitive to the lattice embedding through its sign-changing transport function, with respect to the longitudinal conductivity, thus determining nontrivial temperature evolutions of the Hall angle cotangent. Still, quantitative agreement with experiments on strange metals, e.g., in the normal state of cuprate high-temperature superconductors, remains elusive and requires further investigations to assess whether generalizations of the present model could lead to stronger superlinearities and a ``true" $T^2$ relation for the Hall-angle cotangent in the temperature regime relevant for available experiments, and especially at low doping. 

Our results lead to several perspectives. At very strong coupling $g'$ in the bad-metal regime, we expect the Hall coefficient to reach an asymptotic constant, lower in magnitude than the MFL estimation (\ref{eq:RH_1stB_MFL_T0}) \cite{Aldape-2022}; see also App\@. \ref{App:bad}. Spatially disordered two-body potentials $v$ \cite{Patel-2023,Li-2024}, introduced in accordance with Eq\@. (\ref{eq:vdistribution}), perturbatively affect the low-$T$ MFL regime, leading to a constant contribution to $1/\sigma_{xx}(T)$ and therefore leading to constant Hall angle; these effects will not affect the crossover regime unless $v>g'$. However, the contribution of $v$ could increase the $T=0$ value of the Hall coefficient, while not affecting the $T$-linearity of the longitudinal resistivity (apart from a constant offset), which could further enhance the superlinearity of $\cot\left[\Theta_H(T)\right]$.

The present calculations can adiabatically be extended to finite frequency $\omega$, in order to consider the ac Hall effect and compare to spectroscopic experiments \cite{Kaplan-1996,Parks-1997,Grayson-2002,Schmadel-2007,Jenkins-2010}. 

Furthermore, it would be interesting to complete the magnetotransport phenomenology with magnetoresistance computations. Longitudinal magnetoresistivity appears only at order $B^2$ and therefore needs the implementation of the fractal Hofstadter butterfly spectrum \cite{Hofstadter-1976,Hatsuda-2016} on a square lattice. However, at small fillings the Fermi surface should be almost isotropic and the dispersion could be assumed to be quadratic; we leave an explicit computation of this case to a separate future work, along the same lines of Ref. \onlinecite{Guo-2024}.

In the context of magnetoresistance analyses it would be interesting to generalize our model to include translationally invariant interactions $g$ \cite{Esterlis-2021,Guo-2024}, in addition to spatially disordered couplings $g'$: the interplay between the two interaction channels could introduce additional anisotropies in the longitudinal and Hall scattering rates, and contribute to the separation of timescales for the longitudinal and Hall response observed in strange metals \cite{Anderson-1991,Ayres-2021,Grissonnanche-2021}. 

Another tantalizing research direction is modeling the finite-temperature Hall response in a quantum critical superconductor. Such considerations are inspired by experimental reports of sign change in the Hall resistivity and conductivity of cuprates \cite{Ginsberg-1994,Ginsberg-1995} as a function of applied magnetic field $B$ for temperatures $T\lessapprox T_c$, where $T_c$ is the superconducting critical temperature. A candidate explanation for such sign change is the traction of vortices by the superflow in the mixed (Abrikosov) state of a type-II superconductor \cite{Ullah-1992,Dorsey-1992,Kopnin-1993}. A detailed comparison of our 2D-YSYK theory with sign-changing Hall-conductivity measurements requires the generalization of our theory to simultaneously include a superconducting state \cite{Li-2024} and an applied magnetic field \cite{Guo-2024}. Very interesting related questions concern the corrections to the Hall response upon entering the superconducting state, as well as the structure and dynamics of a vortex-lattice state, in a quantum critical superconductor.

Finally, it is expected that the effects of lattice embedding of 2D-YSYK models will be qualitatively analogous for dispersions $\epsilon_{\vec{k}}$ leading to a sign-changing, $\epsilon$-symmetric Hall transport function (\ref{eq:transport_func_Hall}). Explicit computations for other lattice geometries and multiband configurations could controllably reveal further interplays between non-Fermi liquidness and lattice effects, such as in twisted bilayers and moir\'{e} heterostructures, as recently experimentally observed in twisted bilayer graphene \cite{Lyu-2021} and WSe$_2$ \cite{Xia-2025_preprint}. 

\section{acknowledgments}
We thank Ilya Esterlis, Haoyu Guo, and Chenyuan Li for insightful discussions and collaborations on related work. 
D.V\@. acknowledges enlightening discussions with Christophe Berthod, Sergio Caprara, Andrey Chubukov, Antoine Georges, Giacomo Ghiringhelli, Blaise Gout\'{e}raux, Tamaghna Hazra, Iksu Jang, Matthieu Le Tacon, Christoph Renner, Koenraad Schalm, Daniel Schultz, Dirk van der Marel, and Jan Zaanen. S. S. was supported  by the U.S. National Science Foundation grant No. DMR-2245246 and by the Simons Collaboration on Ultra-Quantum Matter that is a grant from the Simons Foundation (651440, S.S\@.). The Flatiron Institute is a division of the Simons Foundation. This work was also supported by the German Research Foundation (DFG) through CRC TRR 288 ``Elasto-Q-Mat,'' project A07 (D.V\@. and J.S\@.) and the Simons Foundation Collaboration on New Frontiers in Superconductivity (Grant No\@. SFI-MPS-NFS-00006741-03, J.S.).

\section{Data availability}
The data that support the findings of this article are openly available \cite{Radar4KIT-data-2026}.
 
\vspace{0.5cm}

\appendix
\onecolumngrid

\section{Numerical methods}\label{App:Numerics}

Our numerical algorithm to exactly solve the large-$\mathscr{N}$ saddle-point Eliashberg equations (\ref{eq:saddle_point_imag_gen}) first involves finding the solution for the propagators $\mathscr{G}(i\omega_n)$ and $\mathscr{D}(i\Omega_n)$, and for the associated self-energies $\Sigma(i\omega_n)$ and $\Pi(i\Omega_n)$, on the imaginary axis and at fixed fermion density $n$ according to Eq\@. (\ref{eq:n_def_imag}). This protocol is described in Sec. \ref{Saddle_selfcons_num}. The iteration method relies on Fast Fourier Transforms (FFTs), as sketched in Sec. \ref{App:Fourier}. The knowledge of the imaginary-axis solutions enables the calculations of all thermodynamic properties of our 2D Yukawa-SYK system \cite{Smit-2021,long-paper,short-paper}. In particular, we keep track of the temperature dependences of the renormalized boson mass $m_b(T)$, according to Eq\@. (\ref{eq:m_b_T}), and of the chemical potential $\mu(T)$; these quantities are subsequently feeded into the real-axis self-consistent iterations, described in Sec. \ref{Real_continuation}, to obtain the spectral properties on the real axis, \textit{i.e.}, $\mathscr{G}^R(\omega)$, $\mathscr{D}^R(\omega)$, $\Sigma^R(\omega)$, and $\Pi^R(\omega)$. 

\subsection{Matsubara Fourier transforms on the imaginary axis}\label{App:Fourier}

 The imaginary-axis saddle-point equations (\ref{eq:saddle_point_imag_gen}) depend on Matsubara frequencies $i\omega_n$ and $i\Omega_n$, for fermions and bosons, respectively. Schematically, the saddle-point problem involves convolutions of the kind 
 \begin{equation}\label{eq:Pi_imag}
\mathbb{\Pi}(i\Omega_n)=- k_B T \sum_{i\omega_m} G_1(i\omega_m+i\Omega_n) G_2(i\omega_m)
\end{equation}
and
\begin{equation}\label{eq:Sigma_imag}
\mathbb{\Sigma}(i\omega_n)= k_B T \sum_{i\Omega_m}D (i\Omega_m) G(i\omega_n+i\Omega_m).
\end{equation}
An efficient computation of the quantities (\ref{eq:Pi_imag}) and (\ref{eq:Sigma_imag}) results from the transformation to imaginary time $\tau \in \left[0, \beta\right]$ with $\beta=(k_B T)^{-1}$, which yields $\mathbb{\Pi}(\tau)=-G_1(\tau)G_2(-\tau)$ and $\mathbb{\Sigma}(\tau)=D (\tau) G(\tau)$. To perform the numerical transformation, we discretize the imaginary-time interval in equal steps according to $\tau_l=l/(2 N_f k_B T)$ with $l \in \left[0, 2N_f-1\right]$, so that the discretized Matsubara frequencies are $\omega_n=(2 n+1) \pi k_B T$ and $\Omega_n=2n \pi k_B T$, with $n \in \left[-N_f,N_f-1\right]$. $N_f$ is the high-frequency cutoff, chosen to lie well into the ultraviolet regime of the theory, where self-energies are negligible and the propagators have already decayed from their low-frequency evolution. Then, the propagators and self-energies become finite discrete lists of values. The resulting discrete lists are algebraically manipulated (using circshifts), to be adapted to the built-in FFT implementations provided by the optimized \verb|Fourier[]| and \verb|InverseFourier[]| functions of Mathematica. Such implementation is similar to many other self-consistent loops used to solve Shwinger-Dyson saddle-point equations of SYK-like models \cite{Azeyanagi-2018,Patel-2018,Ferrari-2019,Patel-2019,Wang-2020b,Sorokhaibam-2020,Sachdev-1993,Maldacena-2016a,Song-2017,Smit-2021,Grunwald-thesis-2022, long-paper,short-paper,Guo-2024,Li-2024,Wang-2025a,Wang-2025b,Chowdhury-2022}. 

\subsection{Self-consistent loops for the saddle-point equations on the imaginary axis}\label{Saddle_selfcons_num}

We perform calculations at fixed density $n$. Notice that the QCP position, i.e.\@, the condition $\lim_{T\rightarrow 0}m_b(T)=0$, depends simultaneously on interaction $g'$ and density $n$. This condition can be reached through different global minimization procedures. We elect for first solving the saddle-point problem, Eqs\@. (\ref{eq:saddle_point_imag_gen}), at fixed chemical potential $\mu$, and nesting this problem into another self-consistent loop over $\mu=\mu(n,T)$. Assuming $n<0.5/a^2$, a suitable first guess on $\mu$ is its noninteracting value at the given $n$ and $T$, since interactions $g'$ are seen to decrease (increase in magnitude) the value of $\mu<0$; see, e.g.\@, Fig\@. \ref{fig:mb_mu_square_T_varg_fixedn}. Conversely, for $n>0.5/a^2$, we have instead $\mu>0$ that increases with interactions, specularly to the case below particle-hole symmetry. 

At the first iteration $j=0$, our algorithm to solve Eqs\@.  (\ref{eq:saddle_point_imag_gen}) starts with a guess on $\Sigma_0(i \omega_n)$ and $\Pi_0(i\Omega_n)$: to reach convergence, we find it sufficient to input $\Sigma_0(i \omega_n)=i \Gamma \, \forall \omega_n$ and $\Pi_0(i \omega_n)=0 \, \forall \omega_n$, with $\Gamma=0.1$, in the scanned parameter space. To improve convergence at the lowest temperatures, it is also useful to employ an \emph{annealing} approach, where previously converged solutions at higher temperatures are fed as inputs for lower-temperature computations. \\
At the end of the current iteration $j>0$, the fermion self-energy list $\Sigma_{j}(i \omega_n)$ is updated with a weighted sum of the solution $\bar{\Sigma}_j(i \omega_n)$ of Eq\@. (\ref{eq:saddle_point_imag_gen_Sigma}) and the solution $\Sigma_{j-1}(i \omega_n)$ at the previous iteration $j-1$, according to
\begin{equation}\label{eq:Sigma_mix}
\Sigma_{j}(i \omega_n)=\alpha_s \bar{\Sigma}_j(i \omega_n)+ (1-\alpha_s) \Sigma_{j-1}(i \omega_n).
\end{equation}
The mixing factor $\alpha_s \in \left(0, 1\right)$ \cite{Grunwald-thesis-2022,long-paper,short-paper,Guo-2024,Li-2024}, and for the present problem we find it sufficient to keep $\alpha_s=0.1$ at any considered temperature $T$ and coupling $g'$.
The error between the current iteration $j>0$ and the previous one $j-1$ is monitored by $\epsilon_\Sigma=\sum_{i \omega_n} \left| \Sigma_{j}(i \omega_n)-\Sigma_{j-1}(i \omega_n)\right|$. We also keep track of the error on the boson self-energy, $\epsilon_\Pi=\sum_{i \omega_n} \left| \Pi_{j}(i \omega_n)-\Pi_{j-1}(i \omega_n)\right|$, which monotonically decreases together with $\epsilon_\Sigma$ throughout the self-consistent iterations. 
Convergence is reached when $\epsilon_\Sigma$ falls below a user-imposed threshold. 

When convergence is reached on $\Sigma_{j}(i \omega_n)$ and $\Pi_{j}(i \omega_n)$, the fermion density is calculated according to Eq\@. (\ref{eq:n_def_imag}). The relation $\mu(n)$ is monotonic, so if the calculated density is above the desired value $n$, we decrease $\mu-\Delta \mu$ of one step $\Delta \mu<\mu$ at the next iteration of the chemical potential loop. Conversely, if the calculated density is above $n$, we increase $\mu+\Delta \mu$ and halve $\Delta\mu$ at the next iteration. Convergence on $\mu$ is completed when $\Delta \mu$ becomes lower than a user-imposed threshold. 

\subsection{Stability of the renormalized boson mass in the zero-temperature limit near the QCP}\label{Stability_mb}

Searching for the QCP position, as identified by $\lim_{T\rightarrow 0^+} m_b(T)\approx 0$ according to Eq\@. (\ref{eq:m_b_T}), one encounters a numerical difficulty at very low temperatures. As mentioned in Ref. \onlinecite{Esterlis-2021}, the low-temperature thermal boson mass $m_b(T)$ follows the finite temperature correction to the free fermion compressibility: the latter is very small, which makes the self-consistent condition for $m_b(T)$ numerically challenging; for the nearest-neighbor square lattice considered here, this numerical instability can produce first-order transitions for the bosons where $m_b(T)$ is purely imaginary. This effect is mitigated by increasing the number $2N_f$ of used Matsubara frequencies; for the calculations shown in the present paper, we use $N_f=10^4$. To stabilize the search for the QCP at very low $T$, one could introduce a fixed-length constraint on the bosons, $\sum_{\vec{q}}\sum_{i=1}^{\mathscr{N}}\phi_{i,\vec{q}}(\tau) \phi_{i,-\vec{q}}(\tau)=\mathscr{N}/\gamma$, with $\gamma$ tuning parameter \cite{Esterlis-2021,Li-2024}.
However, for the present problem and to the given resolution of $N_f=10^4$, we were able to tune $m_b(T)$ at fixed bare boson mass $m_b^{0}$ down to $T\approx 0.01 t/k_B$, finding a monotonically decreasing $m_b(T)$ for decreasing $T$ close to the QCP as shown in Fig\@. \ref{fig:mb_mu_square_T_varg_fixedn}. Numerically, we achieved $m_b(T)/t<0.1$ for all scanned densities and interactions at $T\approx 0.01 t/k_B$, which enables the MFL regime at low temperatures.   

\subsection{Analytic continuation of frequency convolutions on the imaginary axis}\label{Real_continuation}

For the real-axis solution of the saddle-point equations (\ref{eq:saddle_point_imag_gen}) we closely follow Ref. \onlinecite{Schmalian-1996}. Using the spectral (Lehmann) representation \cite{Berthod-2018}, one can rewrite Eq\@. (\ref{eq:Pi_imag}) on the real axis as
\begin{equation}\label{eq:Pi_partial_spectr}
\mathbb{\Pi}^R(\omega)=\int \frac{d\epsilon}{\pi} f_{FD}(\epsilon) \left[G^R_1(\epsilon+\omega)\mathrm{Im}\left\{G^R_2(\epsilon)\right\}+\mathrm{Im}\left\{G^R_1(\epsilon)\right\}\underbrace{G^{R,\ast}_2(\epsilon-\omega)}_{G_2^A(\epsilon-\omega)} \right],
\end{equation}
where $f_{FD}(\epsilon)=\left(e^{\epsilon/(k_B T)}+1\right)^{-1}$. 
We dub Eq\@. (\ref{eq:Pi_partial_spectr}) as the ``partial spectral representation''. Using again the spectral representation, as
\begin{equation}\label{eq:spectral_G}
G^R(\omega)=-\frac{1}{\pi}\int_{-\infty}^{+\infty} d \epsilon \frac{\mathrm{Im}\left\{ G^R(\epsilon)\right\}}{\omega+i 0^+-\epsilon},
\end{equation}
in Eq\@. (\ref{eq:Pi_partial_spectr}), we further obtain
\begin{align}\label{eq:Pi_full_spectr_deriv}
\mathbb{\Pi}^R(\omega)&=\int \frac{d\epsilon}{\pi} f_{FD}(\epsilon) \left[\left(-\frac{1}{\pi}\right) \int_{-\infty}^{+\infty} d \epsilon' \frac{\mathrm{Im}\left\{G_1^R(\epsilon')\right\}}{\omega+i 0^+-\epsilon'+\epsilon} \mathrm{Im}\left\{G_2^R(\epsilon)\right\}\right. \nonumber \\ & \left. +\left(-\frac{1}{\pi}\right) \int_{-\infty}^{+\infty} d \epsilon' \frac{\mathrm{Im}\left\{G_1^2(\epsilon')\right\}}{-\omega-i 0^+-\epsilon'+\epsilon} \mathrm{Im}\left\{G_1^R(\epsilon)\right\}\right] \nonumber \\ &=-\frac{1}{\pi^2} \int d\epsilon f_{FD}(\epsilon) \int d\epsilon' \left[\frac{\mathrm{Im}\left\{G_1^R(\epsilon')\right\}\mathrm{Im}\left\{G_2^R(\epsilon)\right\} }{\omega+\epsilon-\epsilon'+i0^+}-\frac{\mathrm{Im}\left\{G_2^R(\epsilon')\right\}\mathrm{Im}\left\{G_1^R(\epsilon)\right\} }{\omega+\epsilon'-\epsilon+i0^+}\right] \nonumber \\&=  \int_0^{+\infty} dt e^{i (\omega+i 0^+)t} \mathbb{\Pi}(t),
\end{align}
where we have defined the analytic continuation of the convolution (\ref{eq:Pi_imag}) in real time $t$ 
\begin{equation}\label{eq:Pi_convolv_t}
\mathbb{\Pi}(t) =i (2\pi)^2 \left[a_{G_2}^\ast(t) A_{G_1}(t)-a_{G_1}(t) A_{G_2}^\ast(t)\right],
\end{equation}
as well as the following functions in the time domain:
\begin{equation}\label{eq:A_x}
A_X(t)=-\int_{-\infty}^{+\infty}\frac{d\epsilon}{2\pi}\frac{\mathrm{Im}\left\{X^R(\epsilon)\right\}}{\pi} e^{-i \epsilon t},
\end{equation}
\begin{equation}\label{eq:a_x}
a_X(t)=-\int_{-\infty}^{+\infty}\frac{d\epsilon}{2\pi} f_{FD}(\epsilon) \frac{\mathrm{Im}\left\{X^R(\epsilon)\right\}}{\pi} e^{-i \epsilon t}.
\end{equation}

We also need to analytically continue convolutions in bosonic Matsubara frequencies as in Eq\@. (\ref{eq:Sigma_imag}), analogously to for the fermion self-energy (\ref{eq:saddle_point_imag_gen_Sigma}).  Employing the spectral (Lehmann) representation \cite{Berthod-2018}, one can rewrite Eq\@. (\ref{eq:Sigma_imag}) on the real axis as
\begin{equation}\label{eq:Sigma_partial_spectr}
\mathbb{\Sigma}^R(\omega)=-\int \frac{d\epsilon}{\pi} \left[f_{FD}(\epsilon)\mathrm{Im}\left\{G^R(\epsilon)\right\}D^R(\omega-\epsilon)-f_{BE}(\epsilon)\mathrm{Im}\left\{D^R(\epsilon)\right\}G^{R}_2(\omega+\epsilon)\right],
\end{equation}
where $f_{BE}(\epsilon)=\left(e^{\epsilon/(k_B T)}-1\right)^{-1}$. 
Eq\@. (\ref{eq:Sigma_partial_spectr}) is the ``partial spectral representation'' of Eq\@. (\ref{eq:Sigma_imag}). Using again the spectral representations (\ref{eq:spectral_G}) and 
\begin{equation}\label{eq:spectral_D}
D^R(\omega)=-\frac{1}{\pi}\int_{-\infty}^{+\infty} d \epsilon \frac{\mathrm{Im}\left\{ D^R(\epsilon)\right\}}{\omega+i 0^+-\epsilon},
\end{equation}
in Eq\@. (\ref{eq:Sigma_partial_spectr}), we can further analyze
\begin{align}\label{eq:Sigma_full_spectr_deriv}
\mathbb{\Sigma}^R(\omega)&=\int \frac{d\epsilon}{\pi^2} f_{FD}(\epsilon) \left[\int_{-\infty}^{+\infty} d \epsilon' \frac{\mathrm{Im}\left\{G^R(\epsilon)\right\}\mathrm{Im}\left\{D^R(\epsilon')\right\}}{\omega+i 0^+-\epsilon'-\epsilon}\right]-\int \frac{d\epsilon}{\pi^2} f_{BE}(\epsilon) \left[\int_{-\infty}^{+\infty} d \epsilon' \frac{\mathrm{Im}\left\{G^R(\epsilon')\right\}\mathrm{Im}\left\{D^R(\epsilon)\right\}}{\omega+i 0^++\epsilon-\epsilon'}\right] \nonumber \\ &=\frac{1}{\pi^2}  \int d\epsilon \int d\epsilon' (-i) \int_0^{+\infty} dt e^{i (\omega-\epsilon-\epsilon'+i0^+)t} f_{FD}(\epsilon) \mathrm{Im}\left\{G^R(\epsilon')\right\} \mathrm{Im}\left\{D^R(\epsilon)\right\} \nonumber \\ &- \frac{1}{\pi^2}  \int d\epsilon \int d\epsilon' (-i) \int_0^{+\infty} dt e^{i (\omega+\epsilon-\epsilon'+i0^+)t} f_{BE}(\epsilon) \mathrm{Im}\left\{G^R(\epsilon')\right\} \mathrm{Im}\left\{D^R(\epsilon)\right\} \nonumber \\ &= \int_0^{+\infty} dt e^{i(\omega+i0^+)t}\mathbb{\Sigma}(t).
\end{align}
Here we have defined the analytic continuation of the convolution (\ref{eq:Sigma_imag}) in real time $t$,
\begin{equation}\label{eq:Sigma_convolv_t}
\mathbb{\Sigma}(t) =-i (2\pi)^2\left[a_{G}(t) A_{D}(t)-b_{D}^\ast(t) A_{G}(t)\right],
\end{equation}
where we employed the functions (\ref{eq:A_x}), (\ref{eq:a_x}), and
\begin{equation}\label{eq:b_x}
b_X(t)=-\int_{-\infty}^{+\infty}\frac{d\epsilon}{2\pi} f_{BE}(\epsilon) \frac{\mathrm{Im}\left\{X^R(\epsilon)\right\}}{\pi} e^{-i \epsilon t}.
\end{equation}

\section{Derivation of the square-lattice Green's function}\label{Square_Green}

For completeness, here we show a derivation of the square-lattice Green's function and density of states. The momentum-integrated Green's function reads
\begin{equation}\label{eq:G_int_def}
\mathscr{G}(z)=\frac{1}{\mathscr{V}}\sum_{\vec{k}} \frac{1}{z-\epsilon_{\vec{k}}-\Sigma(z)},
\end{equation}
for generic complex argument $z \in \mathbb{C}$, fermionic self-energy $\Sigma(z)$, and single-particle dispersion $\epsilon_{\vec{k}}$. Here $\mathscr{V}$ is the system volume (which is a two-dimensional area, for a square lattice). 
The noninteracting Green's function corresponds to Eq\@. (\ref{eq:G_int_def}) without self-energy, 
\begin{equation}\label{eq:G0_int_def}
\mathscr{G}_0(z)=\frac{1}{\mathscr{V}}\sum_{\vec{k}} \frac{1}{z-\epsilon_{\vec{k}}},
\end{equation}
and the analytic continuation $\mathscr{G}_0^R(\omega)=\lim_{z \rightarrow \omega+i 0^+} \mathscr{G}_0(z)$ yields the retarded noninteracting Green's function, from which we can deduce the noninteracting density of states
\begin{equation}\label{eq:DOS_nonint_def}
N_0(\omega)=\frac{1}{\mathscr{V}}\sum_{\vec{k}} \delta(\omega-\epsilon_{\vec{k}})=-\frac{1}{\pi}\mathrm{Im}\left\{\mathscr{G}_0^R(\omega) \right\}.
\end{equation}
As such, the density of states in Eq\@. (\ref{eq:DOS_nonint_def}) has the dimensions of $\left[1/(L^d \mathscr{E})\right]$, where $L$ is a length, $d \in \mathbb{N}^+$ is the system dimensionality, and $\mathscr{E}$ is an energy. 
In the following, we derive the above quantities for the square-lattice dispersion (\ref{eq:square_disp}).

Passing to a continuum description, where we sum over wave vectors $\vec{k}$ in the first Brillouin zone, Eq\@. (\ref{eq:G0_int_def}) translates as
\begin{equation}\label{eq:G0_int_cont}
\mathscr{G}_0(z)=\frac{a^2}{(2\pi)^2} \int_{-\frac{\pi}{a}}^{\frac{\pi}{a}} d k_x \int_{-\frac{\pi}{a}}^{\frac{\pi}{a}} d k_y \frac{1}{z+2 t\left[\cos(k_x a)+\cos(k_y a)\right]+\mu}.
\end{equation}
Performing the change of variables $\alpha=(k_x+k_y) a/2$ and $\beta=(k_x-k_y) a/2$, and using $\cos(k_x a)+\cos(k_y a)=2 \cos\left(\frac{k_x a +k_y a}{2}\right)\cos\left(\frac{k_x a -k_y a}{2}\right)=2 \cos\left(\alpha\right)\cos\left(\beta\right)$, Eq\@. (\ref{eq:G0_int_cont}) translates into 
\begin{equation}\label{eq:G0_int_cont3}
\mathscr{G}_0(z)= \frac{2}{(2\pi)^2} \int_{-\pi}^{\pi} d \alpha \int_{0}^{\pi} d \beta \frac{1}{z+4 t \cos\alpha \cos\beta +\mu},
\end{equation}
where the factor of $2$ at the numerator stems from the Jacobian determinant of the variable transformation.
Using the identity
\begin{equation}\label{eq:id1}
\int_0^{\pi} \frac{d x}{a+b \cos x}=\frac{\pi}{\sqrt{a^2-b^2}}, \, a^2>b^2 \vee b>0 \vee (b<0 \wedge b+a <0)
\end{equation}
in Eq\@. (\ref{eq:G0_int_cont3}), we obtain \footnote{For $\sqrt{(z+\mu)^2}=\pm (z+\mu)$ in the complex plane, with $z \in \mathbb{C}$, we pick the ``principal branch'' $\sqrt{(z+\mu)^2}=(z+\mu)$.}
\begin{equation}\label{eq:G0_int_cont4}
\mathscr{G}_0(z)= \frac{1}{2\pi} \int_{-\pi}^{\pi} d \alpha \frac{1}{\sqrt{(z+\mu)^2-(4 t \cos \alpha)^2}}= \frac{1}{2\pi} \frac{2}{z +\mu} \int_{0}^{\pi} d \alpha \frac{1}{\sqrt{1-\left(\frac{4 t}{z+\mu}\right)^2 (\cos \alpha)^2}}. 
\end{equation}
Now we employ the identities
\begin{equation}\label{eq:KE_manip}
\int_0^{\pi} d x \frac{1}{\sqrt{1-\kappa^2 (\cos x)^2}} =2 \int_0^{\frac{\pi}{2}} d x \frac{1}{\sqrt{1-\kappa^2 (\cos x)^2}} \underbrace{=}_{y=\frac{\pi}{2}-x}  2 \int_{0}^{\frac{\pi}{2}} (d y) \frac{1}{\sqrt{1-\kappa^2 (\sin y)^2}} \equiv 2 K_E(\kappa),
\end{equation}
where at the last line we have recognized the complete elliptic integral of the first kind, 
\begin{equation}\label{eq:KE}
K_E(\kappa)=\int_0^{\frac{\pi}{2}} d \theta \frac{1}{\sqrt{1-\kappa^2 (\sin\theta)^2}}=\int_0^1 \frac{1}{\sqrt{(1-t^2)(1-\kappa^2 t^2)}},
\end{equation}
with argument $\kappa \in \left[0,1\right]$ usually. Equating $\kappa=4t/\left|z+\mu\right|$, and using Eqs\@. (\ref{eq:KE_manip}) and (\ref{eq:G0_int_cont4}), we obtain 
\begin{equation}\label{eq:G0_int_cont_big}
\mathscr{G}_0(z)=\frac{2}{\pi (z+\mu)}K_E\left(\frac{4 t}{ z+\mu}\right) \, : \, \left|z+\mu\right|>4 t. 
\end{equation}
Eq\@. (\ref{eq:G0_int_cont_big}) is equivalent to Eq\@. (14) of Ref. \onlinecite{Li-2024}. 

For $\left|z+\mu\right|< 4 t$, we can employ a modified version of the identity (\ref{eq:id1}), namely 
\begin{equation}\label{eq:id1bis}
\int_0^{\pi} \frac{d x}{a+b \cos x}=-i \frac{\pi}{\sqrt{b^2-a^2}}, \, b^2>a^2 \vee b>0 \vee (b<0 \wedge b+a <0).
\end{equation}
Using $a=z+\mu$, $b= 4t \cos \alpha$, and $x \equiv \beta$, from Eqs\@. (\ref{eq:G0_int_cont3}) and (\ref{eq:id1bis}) we obtain (remember that $\cos\alpha \geq0$ for $\alpha \in \left[-\pi,\pi\right]$)
\begin{equation}\label{eq:G0_int_cont5}
\mathscr{G}_0(z)= \frac{-i}{2\pi} \int_{-\pi}^{\pi} d \alpha \frac{1}{\sqrt{(4 t \cos \alpha)^2-(z+\mu)^2}}=\frac{1}{\mathscr{N}} \frac{-i}{\pi} \frac{1}{4 t} \int_{0}^{\pi} d \alpha \frac{1}{\sqrt{(\cos \alpha)^2-\left(\frac{z+\mu}{4t}\right)^2}}.
\end{equation}
Using the identities
\begin{multline}\label{eq:KE_manip2}
\int_0^{\pi} d x \frac{1}{\sqrt{ (\cos x)^2-\kappa^2}} =2 \int_0^{\pi} d x \frac{1}{\sqrt{1-\kappa^2- (\sin x)^2}} =\frac{2}{\sqrt{1-\kappa^2}}\int_0^\pi d x \frac{1}{\sqrt{1-\frac{1}{1-\kappa^2} (\sin x)^2}}\\ =\frac{2}{\sqrt{1-\kappa^2}} \int_0^{\frac{\pi}{2}} d x \frac{1}{\sqrt{1-\frac{1}{1-\kappa^2}(\sin x)^2}} K_E\left(\frac{1}{\sqrt{1-\kappa^2}}\right),
\end{multline}
with $k \equiv (z+\mu)/(4 t)$, Eq\@. (\ref{eq:G0_int_cont5}) transforms into 
\begin{equation}\label{eq:G0_int_cont6}
\mathscr{G}_0(z)= \frac{-2 i}{\pi} \frac{1}{4 t \sqrt{1-\left(\frac{z+\mu}{4t}\right)^2}} K_E\left(\frac{1}{\sqrt{1-\left(\frac{z+\mu}{4t}\right)^2}} \right).
\end{equation}
Now, using the property of the elliptic integrals for $\mathrm{Re}\left\{\kappa\right\}>0$ \cite{Fettis-1970},
\begin{equation}\label{eq:proprinv_KE}
K_E\left(\frac{1}{\kappa}\right)=\kappa\left[K_E(\kappa) \mp i K_E(\sqrt{1-\kappa^2})\right] \, :\, \mathrm{Im}\left\{\kappa^2\right\}\gtrless 0,
\end{equation}
and choosing the branch according to a factor $\mathrm{sign}\left\{z+\mu\right\}$, since $\mathrm{Im}\left\{1-\left[(z+\mu)/(4t)\right]^2\right\}\gtrless0$ for $z+\mu \lessgtr 0$ with analytic continuation $z \rightarrow \omega+i 0^+$, we obtain from Eqs\@. (\ref{eq:proprinv_KE}) and (\ref{eq:G0_int_cont6}) that 
\begin{multline}\label{eq:G0_int_cont6_bis}
\mathscr{G}_0(z)= \frac{-2 i}{\pi} \frac{1}{4 t \sqrt{1-\left(\frac{z+\mu}{4t}\right)^2}} \sqrt{1-\left(\frac{z+\mu}{4t}\right)^2}\left[ K_E\left(\sqrt{1-\left(\frac{z+\mu}{4t}\right)^2} \right)+ i \mathrm{sign}\left\{z+\mu\right\} K_E \left(\frac{z +\mu}{4 t}\right)\right]\\ =\frac{1}{\mathscr{N}} \frac{2}{4 \pi t} \left[-i K_E\left(\sqrt{1-\left(\frac{z+\mu}{4t}\right)^2} \right)+\mathrm{sign}\left\{z+\mu\right\} K_E\left(\frac{z +\mu}{4 t}\right)\right];
\end{multline}
see also Refs\@. \onlinecite{Mathlib} and \onlinecite{Kogan-2020_preprint}. 
We therefore have
\begin{equation}\label{eq:G0_int_cont_small}
\mathscr{G}_0(z)= \frac{1}{2 \pi t }\left[-i K_E\left(\sqrt{1-\left(\frac{z+\mu}{4t}\right)^2} \right)+ \mathrm{sign}\left\{z+\mu\right\} K_E\left(\frac{z +\mu}{4 t}\right)\right] \, : \, \left|z+\mu\right|<4 t. 
\end{equation}
Eq\@. (\ref{eq:G0_int_cont_small}) is found to be formally equivalent to Eq\@. (\ref{eq:G0_int_cont_big}) through a numerical comparison in \verb|Mathematica|; however, Eq\@. (\ref{eq:G0_int_cont_small}) explicitly separates the imaginary part of the expression (in the first term). 
\begin{figure}[h]
  \includegraphics[width=0.8\linewidth]{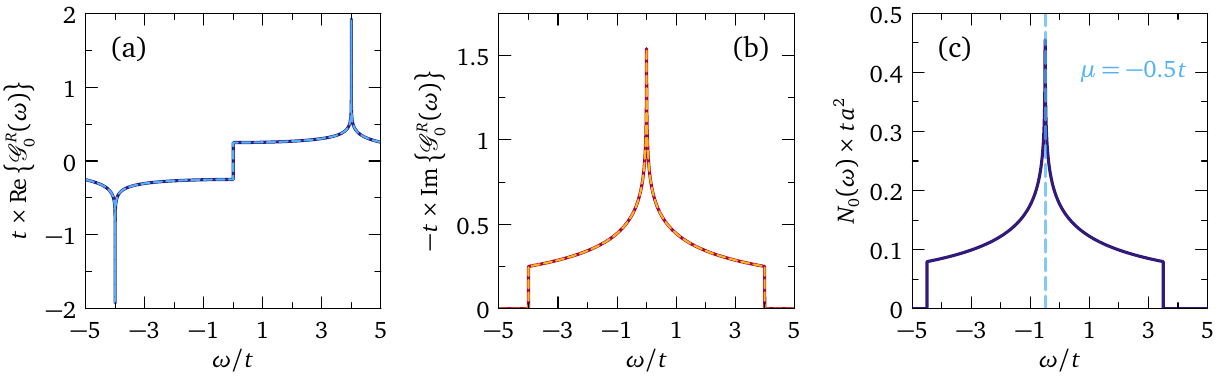}
\caption{\label{fig:Square_G}
(a) Real and (b) imaginary parts of the noninteracting retarded Green's function $\mathscr{G}_0^R(\omega)$, as a function of real frequency $\omega/t$, normalized by hopping $t$ and for $\mu=0$; solid lines are calculated using Eq\@. (\ref{eq:G0_int_cont_big}) for $\left|\omega+\mu\right|>4 t$ and Eq\@. (\ref{eq:G0_int_cont_small}) for $\left|\omega+\mu\right|<4 t$, while dashed lines stem from Eq\@. (\ref{eq:G0_int_cont_big}) for all frequencies. The retarded Green's function is $\mathscr{G}_0^R(\omega)=\lim_{z\rightarrow \omega+i \delta}\mathscr{G}_0(z)$, where here we set $\delta=10^{-9}t$. (c) Noninteracting density of states of the square lattice, corresponding to Eq\@. (\ref{eq:DOS_square}), for $\mu=-0.5t$ (dashed light-blue line) as a function of normalized frequency $\omega/t$.}
\end{figure}

Fig\@. \ref{fig:Square_G} shows the real and imaginary parts of the analytically continued Green's function $\mathscr{G}_0^R(\omega)$, using Eq\@. (\ref{eq:G0_int_cont_big}) for $\left|\omega+\mu\right|>4 t$ and Eq\@. (\ref{eq:G0_int_cont_small}) for $\left|\omega+\mu\right|<4 t$ (solid lines), and using Eq\@. (\ref{eq:G0_int_cont_big}) for all values of $\omega$. We can see the perfect agreement between the two formulations.

\section{Derivation of the square-lattice noninteracting density of states}\label{Square_DOS}

\subsection{Derivation from the noninteracting lattice Green's function}

The density of states (\ref{eq:DOS_nonint_def}) is nonnull only when the imaginary part of the Green's function is finite. Therefore, we only need to employ the analytically continued version of Eq\@. (\ref{eq:G0_int_cont_small}), where the first term provides the needed imaginary part; explicitly, 
\begin{equation}\label{eq:DOS_square}
N_0(\epsilon)=-\frac{1}{\pi \mathscr{A}}\mathrm{Im}\left\{\mathscr{G}_0^R(\epsilon)\right\}=\frac{1}{2 \pi^2 t \mathscr{A}} K_E\left(\sqrt{1-\left(\frac{\epsilon+\mu}{4t}\right)^2} \right) \Theta(4 t-\left|\epsilon+\mu\right|),
\end{equation}
where $\mathscr{A}=a^2$ is the unit cell area, and $\Theta(x)$ is the Heavyside theta function. The density of states is displayed in Fig\@. \ref{fig:Square_G}(c): it is positive semidefinite, and it shows the van Hove singularity at $\omega=\mu$. For the numerical implementation of Eq\@. (\ref{eq:DOS_square}) in \verb|Mathematica|, we have to be aware of the slightly different definition of elliptic integrals employed by \verb|Mathematica|, which is detailed in App\@. \ref{DOS_square_Mathematica}.

\subsection{Derivation from the dispersion relation using Fourier transforms}

An alternative, very useful method to calculate the square-lattice density of states relies on its definition (\ref{eq:DOS_nonint_def}), on the dispersion relation (\ref{eq:square_disp}), and on the use of Fourier transforms between energy $\omega$ and real-valued time $\tau$. For these Fourier transforms we adopt the following definitions, valid for a real-valued function $f(\tau)$ that transforms into $F(\omega)=\mathscr{F}\left\{f(\tau)\right\}$:
\begin{subequations}\label{eq:Fourier_def}
\begin{equation}\label{eq:Fourier}
F(\omega)=\mathscr{F}\left\{f(\tau)\right\}=\int_{-\infty}^{+\infty} d \tau f(\tau)e^{-i \omega \tau},
\end{equation}
\begin{equation}\label{eq:Fourier_inv}
f(\tau)=\mathscr{F}^{1}\left\{F(\omega)\right\}=\frac{1}{2\pi}\int_{-\infty}^{+\infty} d \omega f(\omega)e^{i \omega \tau}.
\end{equation}
\end{subequations}
Employing the inverse Fourier transform (\ref{eq:Fourier_inv}), we have for the density of states (\ref{eq:DOS_nonint_def}) that
\begin{multline}\label{eq:DOS_tau}
N_0(\tau)=\mathscr{F}^{-1}\left\{N_0(\epsilon)\right\}=\frac{1}{2\pi}\int d \epsilon N_0(\epsilon)e^{i \epsilon \tau}=\int \frac{d \epsilon}{2\pi} e^{i \epsilon \tau} \frac{1}{\mathscr{V}}\sum_{\vec{k}} \delta(\epsilon-\epsilon_{\vec{k}})\equiv \frac{1}{2\pi} \frac{1}{\mathscr{V}}\sum_{\vec{k}} e^{i \epsilon_{\vec{k}}\tau} \\ =\frac{1}{(2\pi)^3 \mathscr{A}} \int_{-\pi}^{\pi} d(k_x a) \int_{-\pi}^{\pi} d(k_y a) e^{-i 2 t \left[\cos(k_xa)+\cos(k_y a)\right] \tau}=\frac{1}{(2\pi)^3 \mathscr{A}} \underbrace{\int_{-\pi}^{\pi} d(k_x a) e^{-i 2 t \cos(k_x a)\tau}}_{2 \pi J_0(2 t \tau)} \underbrace{\int_{-\pi}^{\pi} d(k_y a) e^{-i 2 t \cos(k_y a)\tau}}_{2 \pi J_0(2 t \tau)}\\=\frac{1}{2\pi \mathscr{A}} \left[J_0(2 t \tau)\right]^2,
\end{multline}
where at the last steps we have employed one of the integral representations of the Bessel function of the first kind and of zeroth order:
\begin{equation}\label{eq:Bessel_J0}
J_0(z)=\frac{1}{2\pi} \int_{-\pi}^{\pi} d \theta e^{-i z \cos\theta}.
\end{equation}
Performing the Fourier transformation (\ref{eq:Fourier}) of the final result in Eq\@. (\ref{eq:DOS_tau}), one recovers Eq\@. (\ref{eq:DOS_square}), as it should be. The Fourier transformation method is especially efficient to calculate the transport functions of the square lattice, as we do in Secs. \ref{Square_transport_long} and \ref{Square_transport_trans}. 

\section{Derivation of the square-lattice longitudinal transport function using Fourier transforms}\label{Square_transport_long}

An analytical expression for the longitudinal transport function (\ref{eq:transport_func_long}) can be efficiently obtained by the method of the Fourier transforms (\ref{eq:Fourier_inv}) and (\ref{eq:Fourier}). We have
\begin{align}\label{eq:transport_long_tau}
\Phi_{(0)}^{xx}(\tau)=\mathscr{F}^{-1}\left\{\Phi_{(0)}^{xx}(\epsilon)\right\}&=\frac{1}{2\pi}\int d \epsilon \Phi_{(0)}^{xx}(\epsilon)e^{i \epsilon \tau}=\int \frac{d \epsilon}{2\pi} e^{i \epsilon \tau} \frac{1}{\mathscr{V}}\sum_{\vec{k}} (v_{\vec{k}})^2 \delta(\epsilon-\epsilon_{\vec{k}})\equiv \frac{1}{2\pi} \frac{1}{\mathscr{V}}\sum_{\vec{k}} (v_{\vec{k}})^2 e^{i \epsilon_{\vec{k}}\tau} \nonumber \\ &=\frac{1}{(2\pi)^3 a^2} \int_{-\pi}^{\pi} d(k_x a) \int_{-\pi}^{\pi} d(k_y a) \left(\frac{2 t a}{\hbar}\right)^2 \left[\sin(k_x a)\right]^2 e^{-i 2 t \left[\cos(k_xa)+\cos(k_y a)\right] \tau} \nonumber \\ &=\frac{1}{(2\pi)^3 a^2} \left(\frac{2 t a}{\hbar}\right)^2 \underbrace{\int_{-\pi}^{\pi} d(k_x a) \left[\sin(k_x a)\right]^2 e^{-i 2 t \cos(k_x a)\tau}}_{2 \pi \frac{J_1(2 t \left|\tau\right|)}{2 t \left|\tau\right|}} \underbrace{\int_{-\pi}^{\pi} d(k_y a) e^{-i 2 t \cos(k_y a)\tau}}_{2 \pi J_0(2 t \tau)} \nonumber \\ &=\frac{2}{2\pi} \frac{t}{\hbar^2}\frac{J_1(2 t \tau)}{\tau} J_0(2 t \tau),
\end{align}
where at the last steps we have employed one of the integral representations of the Bessel function of the first kind and of first order:
\begin{equation}\label{eq:Bessel_J1}
J_1(z)=-\frac{1}{2\pi i } \int_{-\pi}^{\pi} d \theta e^{i \theta} e^{-i z \cos\theta}=-\frac{1}{2\pi i } \int_{-\pi}^{\pi} d \theta \left[\cos\theta +i \sin\theta\right] e^{-i z \cos\theta} \equiv -\frac{1}{2 \pi i} \int_{-\infty}^{+\infty} d \theta \cos\theta e^{-i z \cos \theta}. 
\end{equation}
A general definition of the Bessel functions of first kind is recalled in App\@. \ref{app:Bessel}. We also have $J_1(\left|x\right|)/\left|x\right|\equiv J_1(x)/x \, \forall x\in \mathbb{R}$. We could now just Fourier-transform Eq\@. (\ref{eq:transport_long_tau}) directly using Eq\@. (\ref{eq:Fourier}), but let us notice a useful connection with the density of states $N_0(\epsilon)$, calculated in Sec. \ref{Square_DOS}.
Using the derivative property of Bessel functions 
\begin{equation}\label{eq:derivative_Bessel_1}
J_0'(z)=\frac{d J_0(z)}{dz}=-J_1(z),
\end{equation}
as verified by deriving Eq\@. (\ref{eq:Bessel_J0}) and comparing with Eq\@. (\ref{eq:Bessel_J1}), we notice a connection between the last result in Eq\@. (\ref{eq:transport_long_tau}) and Eq\@. (\ref{eq:DOS_tau}) for the density of states:
\begin{equation}\label{eq:Phi_xx_square_DOS_tau}
\Phi_{(0)}^{xx}(\tau)=-\frac{2}{2\pi} \frac{t}{\hbar^2}\frac{J_0'(2 t \tau)}{\tau} J_0(2 t \tau)\equiv -\frac{a^2}{2 \hbar^2} \frac{d N_0(\tau)}{d\tau} \frac{1}{\tau}.
\end{equation}
Now, using the derivative property of the Fourier transforms (\ref{eq:Fourier}), Eq\@. (\ref{eq:Phi_xx_square_DOS_tau}) translates in frequency space as 
\begin{equation}\label{eq:Phi_xx_square_DOS}
\frac{d \Phi_{(0)}^{xx}(\epsilon)}{d \epsilon}=-i \left(-\frac{t a^2}{\hbar^2}\right) i \frac{\epsilon}{2} N_0(\epsilon)=-\frac{a^2}{\hbar^2} \frac{\epsilon}{2} N_0(\epsilon). 
\end{equation}
Eq\@. (\ref{eq:Phi_xx_square_DOS}) proves useful in general, to calculate the optical conductivity in Eliashberg theory on a square lattice \cite{Li-2024}. In our specific case, knowing the density of states (\ref{eq:DOS_square}), we can just integrate the latter over energy $\epsilon$ in accordance with Eq\@. (\ref{eq:Phi_xx_square_DOS}), to find \cite{Li-2024}
\begin{equation}\label{eq:Phi_xx_square_final}
\Phi_{(0)}^{xx}(\epsilon)=\frac{4 t}{\pi^2\hbar^2} \left\{E_E\left(\sqrt{1-\left(\frac{\epsilon}{4 t}\right)^2}\right)-\left(\frac{\epsilon}{4 t}\right)^2 K_E\left(\sqrt{1-\left(\frac{\epsilon}{4 t}\right)^2}\right)\right\}.
\end{equation}
 
\begin{figure}[h]
  \includegraphics[width=0.6\linewidth]{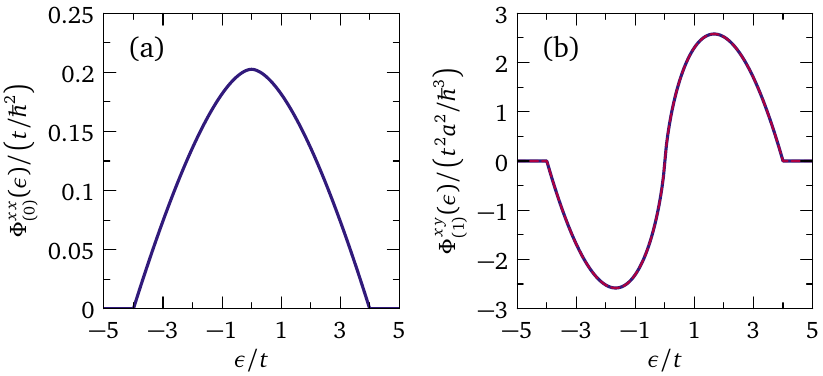}
\caption{\label{fig:Phixx_xy_example}
(a) Longitudinal transport function from Eq\@. (\ref{eq:Phi_xx_square_final}) and (b) transverse (Hall) transport function from Eq\@. (\ref{eq:Phi_xy_square_final}) (blue solid curve) on a square lattice, as a function of $\epsilon/t$. The dashed red curve in panel (b) shows the numerical quadrature given by Eq\@. (\ref{eq:Phi_xy_square_quadrature}).}
\end{figure}

Eq\@. (\ref{eq:Phi_xx_square_final}) coincides with Eq\@. (S20) in Ref. \onlinecite{Li-2024}, modulo the prefactor $2 t a^2/(\hbar^2)$ and the different convention on elliptic integrals adopted by \verb|Mathematica| and aforementioned reference, as detailed in App\@. \ref{DOS_square_Mathematica}. By symmetry, we also have $\Phi_{(0)}^{yy}(\epsilon)=\Phi_{(0)}^{xx}(\epsilon)$. The function written in Eq\@. (\ref{eq:Phi_xx_square_final}) is traced in Fig\@. \ref{fig:Phixx_xy_example}(a) as a function of $\epsilon/t$. 
An additional consistency check stems from the direct comparison of Eq\@. (\ref{eq:Phi_xx_square_final}) with the equivalent representation in terms of an angular integral: 
\begin{equation}\label{eq:Phi_xx_square_angle}
\Phi_{(0)}^{xx}(\epsilon)=\frac{2 t}{\pi^2 \hbar^2} \int_a^b d \theta \sqrt{1-\left(\frac{\epsilon}{2t}-\cos\theta\right)^2}, \, a=\mathrm{arccos}\left(\mathrm{min}\left\{1,\frac{\epsilon}{2t}+1\right\}\right), \, b=\mathrm{arccos}\left(\mathrm{max}\left\{-1,\frac{\epsilon}{2t}-1\right\}\right).
\end{equation}

\section{Derivation of the square-lattice transverse (Hall) transport function at first order using Fourier transforms}\label{Square_transport_trans}

We can also obtain an analytical expression for the Hall transport function (\ref{eq:transport_func_Hall}) employing the Fourier transforms (\ref{eq:Fourier_inv}) and (\ref{eq:Fourier}). Here, for the dispersion (\ref{eq:square_disp}) we have
\begin{equation}\nonumber
\frac{\partial \epsilon_{\vec{k}}}{\partial k_x} \frac{\partial \epsilon_{\vec{k}}}{\partial k_y} \frac{\partial^2 \epsilon_{\vec{k}}}{\partial k_x \partial k_y}=0,
\end{equation}
so that 
\begin{multline}\label{eq:transport_trans_tau}
\Phi_{(1)}^{xy}(\tau)=\mathscr{F}^{-1}\left\{\Phi_{(1)}^{xy}(\epsilon)\right\}=\frac{1}{2\pi}\int d \epsilon \Phi_{(1)}^{xy}(\epsilon)e^{i \epsilon \tau}\\ \equiv\frac{1}{\hbar^3}\frac{\pi^2}{3}\frac{1}{2\pi \mathscr{A}}\int d \epsilon e^{i \epsilon \tau}\sum_{\vec{k}} \left[-\left(\frac{\partial \epsilon_{\vec{k}}}{\partial k_x}\right)^2 \frac{\partial^2 \epsilon_{\vec{k}}}{\partial k_y^2}-\left(\frac{\partial \epsilon_{\vec{k}}}{\partial k_y}\right)^2 \frac{\partial^2 \epsilon_{\vec{k}}}{\partial k_x^2}\right]\delta(\epsilon-\epsilon_{\vec{k}}) \\ = -\frac{(2t)^3}{(2\pi \hbar)^3} \frac{\pi^2 a^4}{3 \mathscr{A}} \int_{-\pi}^{\pi} d(k_x a) \int_{-\pi}^{\pi} d(k_y a) \left\{ \left[\sin(k_x a)\right]^2 \left[-\cos(k_y a)\right] +\left[\sin(k_y a)\right]^2 \left[-\cos(k_x a)\right]\right\}  e^{-i 2 t \left[\cos(k_xa)+\cos(k_y a)\right] \tau} \\ = \frac{1}{(2\pi)^3} \frac{\pi^2 a^4}{3 \mathscr{A}} \left(\frac{2 t}{\hbar}\right)^3 \left\{ \int_{-\pi}^{\pi} d(k_x a) \left[\sin(k_x a)\right]^2 e^{-i 2 t \cos(k_x a)\tau} \int_{-\pi}^{\pi} d(k_y a) \cos(k_y a) e^{-i 2 t \cos(k_y a)\tau}\right. \\ \left. +\int_{-\pi}^{\pi} d(k_y a) \left[\sin(k_y a)\right]^2 e^{-i 2 t \cos(k_y a)\tau} \int_{-\pi}^{\pi} d(k_x a) \cos(k_x a) e^{-i 2 t \cos(k_x a)\tau} \right\} \\= \frac{1}{(2\pi)^3} \frac{\pi^2 a^4}{3 \mathscr{A}} \left(\frac{2 t}{\hbar}\right)^3 \left\{ \frac{2\pi J_1(2 t \tau)}{2 t \tau} \left[-2 i \pi J_1(2 t \tau)\right]+ \frac{2\pi J_1(2 t \tau)}{2 t \tau} \left[-2 i \pi J_1(2 t \tau)\right]\right\} \\= -\frac{i}{2\pi} \frac{\pi^2 a^4}{3 \mathscr{A}} \left(\frac{2 t}{\hbar}\right)^3 2 \frac{J_1(2 t \tau)}{2 t \tau} J_1(2 t \tau)=-i \frac{4 \pi a^4}{3 \mathscr{A}} \left(\frac{t}{\hbar}\right)^3 \frac{\left[J_1(2 t \tau)\right]^2}{t \tau}.
\end{multline}
We could now employ the derivative property (\ref{eq:derivative_Bessel_1}) of Bessel functions, but this time the connection between Eq\@. (\ref{eq:transport_trans_tau}) and the density of states (\ref{eq:DOS_tau}) seems not very useful. Instead, we directly Fourier-transform Eq\@. (\ref{eq:transport_trans_tau}) using Eq\@. (\ref{eq:Fourier}), with the result (verified by \verb|Mathematica|)
\begin{equation}\label{eq:J_1_square_transform}
\mathscr{F}\left\{\frac{\left[J_1(2 t \tau)\right]^2}{t \tau}\right\}=-\frac{i \left[\Theta (-\epsilon)-\Theta (\epsilon)\right] \left[\Theta \left(\frac{1}{16}-\frac{t^2}{\epsilon^2}\right)-1\right] G_{2,2}^{2,0}\left(\frac{\epsilon^2}{16 t^2}|
\begin{array}{c}
 1,2 \\
 \frac{1}{2},\frac{1}{2} \\
\end{array}
\right)}{t}.
\end{equation}
Here $G_{p, q}^{m, n}\left(z\right.| \left.
\begin{array}{c}
 a_1, \cdots a_p \\
 b_1, \cdots b_q \\
\end{array}
\right)$ is the Meijer G function, and $\Theta(x)$ is the Heavyside step function. The Meijer G function is
\begin{equation}
G_{p, q}^{m, n}\left(z|
\begin{array}{c}
 a_1, \cdots a_p \\
 b_1, \cdots b_q \\
\end{array}\right)=
\frac{1}{2 \pi  i}\int d s z^{-s} \frac{ \Gamma  \left(1-a_1-s\right) \cdots \Gamma  \left(1-a_n-s\right) \Gamma  \left(b_1+ s\right) \cdots \Gamma  \left(b_m+s\right)}{\left(a_{n+1}+ s\right) \cdots \Gamma  \left(a_p+ s\right) \Gamma  \left(1-b_{m+1}- s\right) \cdots \Gamma  \left(1-b_q-s\right)}. 
\end{equation}
Eq\@. (\ref{eq:J_1_square_transform}) seemingly cannot be decomposed into simpler elliptic integrals. Eq\@. (\ref{eq:transport_trans_tau}) then becomes in frequency space
\begin{equation}\label{eq:Phi_xy_square_final}
\Phi_{(1)}^{xy}(\epsilon)=\frac{4 \pi a^2}{3} \left(\frac{t}{\hbar}\right)^3 \frac{ \left[\Theta (\epsilon)-\Theta (-\epsilon)\right] \left[\Theta \left(\frac{1}{16}-\frac{t^2}{\epsilon^2}\right)-1\right] G_{2,2}^{2,0}\left(\frac{\epsilon^2}{16 t^2}|
\begin{array}{c}
 1,2 \\
 \frac{1}{2},\frac{1}{2} \\
\end{array}
\right)}{t}.
\end{equation}
The validity of Eq\@. (\ref{eq:Phi_xy_square_final}) can be verified by comparing it with a single-quadrature alternative representation of $\Phi_{(1)}^{xy}(\epsilon)$ \cite{Morpurgo-2024}:
\begin{multline}\label{eq:Phi_xy_square_quadrature}
\Phi_{(1)}^{xy}(\epsilon)=\frac{4 \pi a^2}{3 } \frac{t^2}{\hbar^3}\int_a^b d\theta \frac{\left[1-\left(\frac{\epsilon}{4 t}-\cos\theta\right)^2\right]\cos\theta+\left(\frac{\epsilon}{4 t}-\cos\theta\right)\left(\sin \theta\right)^2}{\sqrt{1-\left(\frac{\epsilon}{4 t}-\cos\theta \right)^2}},  \\ a=\mathrm{arccos}\left(\mathrm{min}\left\{1,\frac{\epsilon}{4 t}+1\right\}\right), \, b=\mathrm{arccos}\left(\mathrm{max}\left\{-1,\frac{\epsilon}{4 t}-1\right\}\right).
\end{multline}
The comparison between Eqs\@. (\ref{eq:Phi_xy_square_final}) and (\ref{eq:Phi_xy_square_quadrature}) is shown in Fig\@. \ref{fig:Phixx_xy_example}(b) as a function of $\epsilon/t$.

\section{Mathematica implementation of the Green's function and density of states}\label{DOS_square_Mathematica}

Notice that \verb|Mathematica| implements a slightly different built-in definition of the complete elliptic integral of the first kind, with respect to Eq\@. (\ref{eq:KE}): it is
\begin{equation}\label{eq:KE_Mth}
\tilde{K}_E(\kappa)=\int_0^{\frac{\pi}{2}} d \theta \frac{1}{\sqrt{1-\kappa (\sin\theta)^2}},
\end{equation}
so that the relation between the two conventions is 
\begin{equation}\label{eq:K_K_E_tilde}
 K_E(\kappa)=\tilde{K}_E(\kappa^2).
\end{equation}
Therefore, the numerical implementation of Eqs\@. (\ref{eq:G0_int_cont_big}), (\ref{eq:G0_int_cont_small}), and (\ref{eq:DOS_square}) in Mathematica results
\begin{equation}\label{eq:G0_int_cont_big_Math}
\mathscr{G}_0(z)=\frac{1}{\mathscr{N}} \frac{2}{\pi (z+\mu)}\tilde{K}_E\left[\left(\frac{4 t}{ z+\mu}\right)^2\right] \, : \, \left|z+\mu\right|>4 t, 
\end{equation}
\begin{equation}\label{eq:G0_int_cont_small_Math}
\mathscr{G}_0(z)=\frac{1}{\mathscr{N}} \frac{1}{2 \pi t }\left\{-i \tilde{K}_E\left[1-\left(\frac{z+\mu}{4t}\right)^2 \right]+\tilde{K}_E\left[\left(\frac{4 t}{ z+\mu}\right)^2\right]\right\} \, : \, \left|z+\mu\right|<4 t,
\end{equation}
\begin{equation}\label{eq:DOS_square_Math}
N_0(\omega)=\frac{1}{\mathscr{N}} \frac{1}{2 \pi^2 t} \tilde{K}_E\left[1-\left(\frac{z+\mu}{4t}\right)^2\right] \Theta(4 t-\left|\omega+\mu\right|).
\end{equation}

\section{Definition of Bessel functions}\label{app:Bessel}

Bessel functions of the first kind, $J_n(z)$, are defined as the solution of the differential equation
\begin{equation}\label{eq:Bessel_eqs}
y \left(z^2-n^2\right)+z y'+z^2 y''=0. 
\end{equation}
An alternative integral representation of the same functions reads 
\begin{equation}\label{eq:Bessel_integral}
J_n(z)=\frac{1}{2 \pi i^n} \int_0^{2\pi} d\phi e^{i n \phi} e^{i z \cos \phi} \underbrace{=}_{\phi=\theta+\pi} \frac{1}{2 \pi i^n} \int_{-\pi}^{\pi} d\phi e^{i n (\theta+\pi)} e^{i z \cos (\theta+\pi)}= -\frac{1}{2 \pi i^n} \int_{-\pi}^{\pi} d\phi e^{i n \theta} e^{-i z \cos \theta}.
\end{equation}
Eqs\@. (\ref{eq:Bessel_J0}) and (\ref{eq:Bessel_J1}) represent the cases $n=0$ and $n=1$ of Eq\@. (\ref{eq:Bessel_integral}).

\begin{figure*}[ht]
  \includegraphics[width=0.65\textwidth]{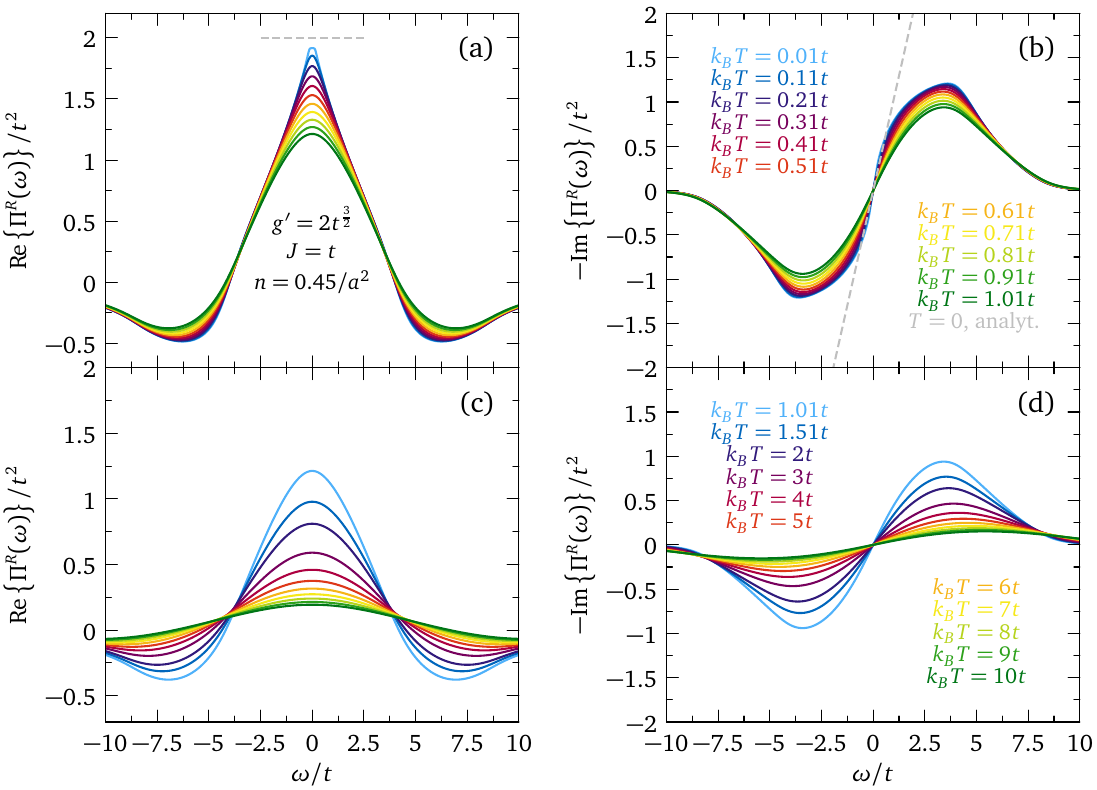}
\caption{\label{fig:Pi_YSYK_square_T} Real (a,c) and imaginary (b,d) parts of the retarded boson self-energy $\Pi^R(\omega)/t^2$ as a function of frequency $\omega/t$, for interaction $g'=2 t^{3/2}$, boson stiffness $J=t$, and density $n=0.45/a^2$, at different temperatures. Dashed gray lines show a the semi-analytical $T=0$ estimation derived in Sec. \ref{App:semianalyt}.
}
\end{figure*}

\section{Spectral properties on the real axis}\label{Num_real_plots}

\subsection{Numerical results at finite temperature}

Here we explicitly show numerical results for the retarded fermion and boson self-energy and Green's functions, obtained by exactly solving the saddle-point equations (\ref{eq:saddle_point_imag_gen}), analytically continued to the real axis using the protocol of Sec. \ref{Real_continuation}. We take the exemplary parameters of interaction $g'=2 t^{3/2}$, boson stiffness $J=t$, and density $n=0.45/a^2$. 

Fig\@. \ref{fig:Pi_YSYK_square_T} displays the real (a,c) and imaginary (b,d) parts of the retarded boson self-energy $\Pi^R(\omega)/t^2$ as a function of frequency $\omega/t$, for different temperatures. At low $T$ and $\omega$, we have $\Pi^R(0)\approx 2$, as shown by the semi-analytical estimation at $T=0$ from Sec. \ref{App:semianalyt} -- dashed gray line in Fig\@. \ref{fig:Pi_YSYK_square_T}(a) -- so that $\sqrt{\Pi^R(0)}\approx \sqrt{2}=1.414 \approx \lim_{T\rightarrow +\infty}m_b(T)$: this reflects the condition $m_b^0\approx \sqrt{\Pi^R(0)}$ for the position of the QCP. Correspondingly, the imaginary part in Fig\@. \ref{fig:Pi_YSYK_square_T}(b) is linear in $\omega$ at small frequencies, as highlighted by the $T=0$ semi-analytical result given by the dashed gray line, and derived in Sec. \ref{App:semianalyt}. At higher temperatures, we see from Fig\@. \ref{fig:Pi_YSYK_square_T}(c) that $\Pi^R(0)<k_B T$ for $k_B T \gtrapprox 1$, which is consistent with the criterion for the crossover/classical metal boundary in Fig\@. \ref{fig:Hall_summary}(b). The imaginary part is also progressively suppressed at high $T$, as seen in Fig\@. \ref{fig:Pi_YSYK_square_T}(d). Therefore, at high temperatures the bosons asymptotically become free, with negligible self-energy.

\begin{figure*}[ht]
  \includegraphics[width=0.65\textwidth]{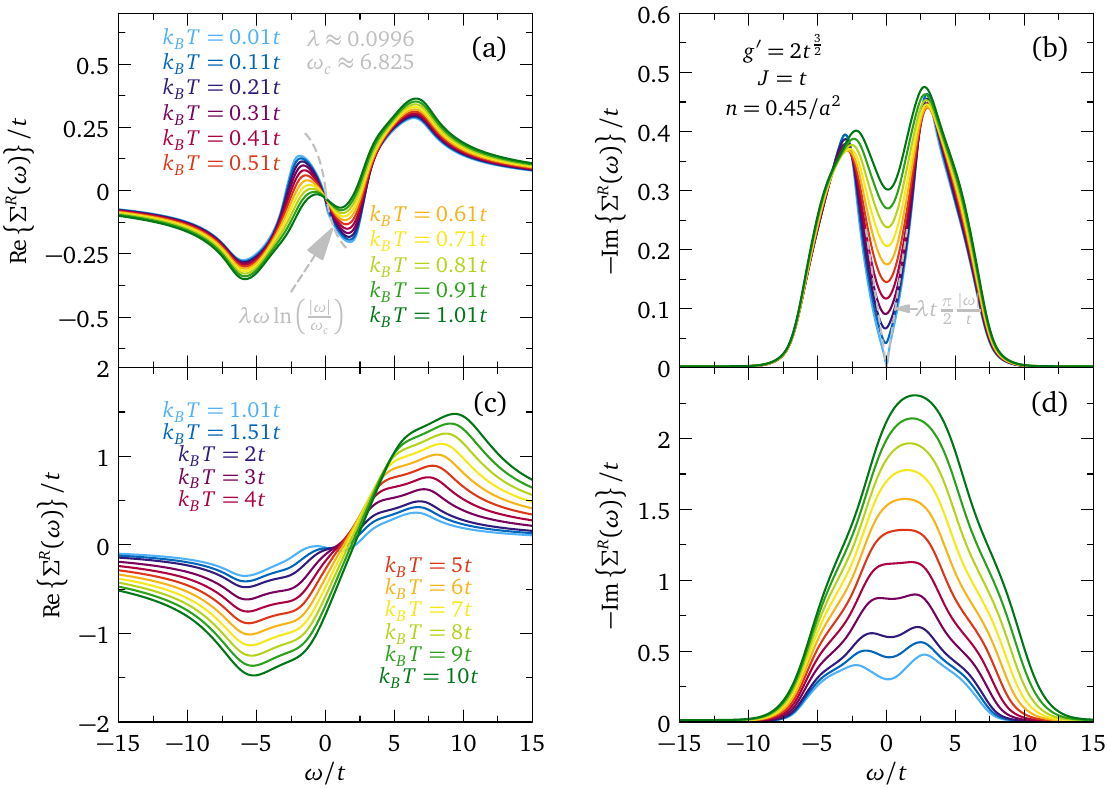}
\caption{\label{fig:Sigma_YSYK_square_T_fixedn} Real (a,c) and imaginary (b,d) parts of the retarded fermion self-energy $\Sigma^R(\omega)/t$ as a function of frequency $\omega/t$, for interaction $g'=2 t^{3/2}$, boson stiffness $J=t$, and density $n=0.45/a^2$, at different temperatures. Dashed gray curves show a the semi-analytical $T=0$ estimation derived in Sec. \ref{App:semianalyt}.
}
\end{figure*}

Fig\@. \ref{fig:Sigma_YSYK_square_T_fixedn} shows the real (a,c) and imaginary (b,d) parts of the retarded fermion self-energy $\Sigma^R(\omega)/t$ as a function of dimensionless frequency $\omega/t$, for different temperatures. At the lowest temperatures, the low-frequency evolution of the self-energy is consistent with the MFL phenomenology \cite{Varma-1989}: in fact, the imaginary part is consistent with the expression $-\mathrm{Im}\left\{\Sigma^R(\omega)\right\}=\lambda \pi/2 \left|\omega\right|$ -- see dashed gray line in Fig\@. \ref{fig:Sigma_YSYK_square_T_fixedn}(b), derived in Sec. \ref{App:semianalyt} -- while the real part is more sensitive to particle-hole asymmetry at $\mu\neq 0$, so that the analytical expression $\mathrm{Re}\left\{\Sigma^R(\omega)\right\}=\lambda \omega \ln \left( \left|\omega\right|/\omega_c\right)$ is only in qualitative agreement with the numerics, but does not capture well the asymmetry of $\mathrm{Re}\left\{\Sigma^R(\omega)\right\}$ with respect to $\omega=0$. At large frequencies $\omega/t \gtrapprox 4$ we see the expected decay of $\mathrm{Re}\left\{\Sigma^R(\omega)\right\}$ and $-\mathrm{Im}\left\{\Sigma^R(\omega)\right\}$ related to the half fermion bandwidth $W/2=4t$.
Further increasing temperature, the zero crossing of the real part of the self-energy in Fig\@. \ref{fig:Sigma_YSYK_square_T_fixedn}(c) moves to higher frequencies, due to the temperature dependence of the chemical potential -- cfr. Fig\@. \ref{fig:mb_mu_square_T_varg_fixedn}(b,d) -- and we also observe the progressive smearing of the logarithmic MFL feature that was present at low $T$. Correspondingly, the imaginary part of the self-energy develops a peak at $\omega \approx -\mu$ that increases in magnitude for increasing $T$. 

\begin{figure*}[ht]
  \includegraphics[width=0.65\textwidth]{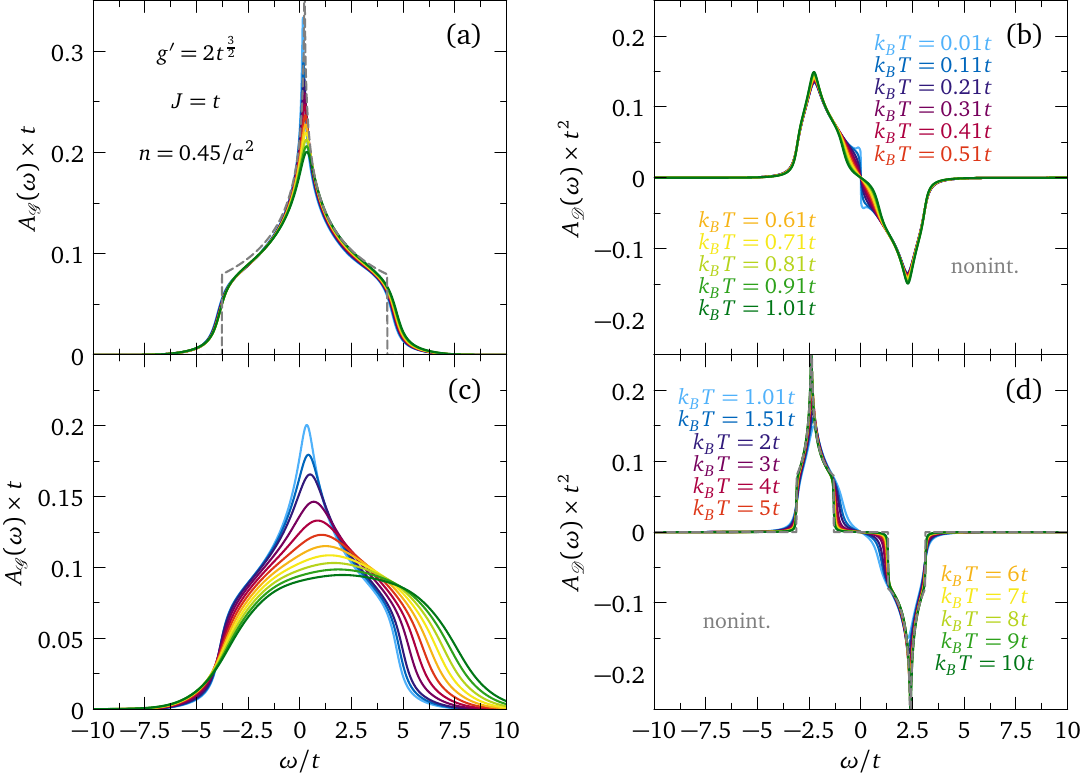}
\caption{\label{fig:DOS_YSYK_square_T} Spectral function of fermions, $A_{\mathscr{G}}=\pi^{-1}\mathrm{Im}\left\{\mathscr{G}^R(\omega)\right\}$, and of bosons $A_{\mathscr{D}}=\pi^{-1}\mathrm{Im}\left\{\mathscr{D}^R(\omega)\right\}$, normalized by fermion hopping $t$ and as a function of frequency $\omega/t$, for interaction $g'=2 t^{3/2}$, boson stiffness $J=t$, and density $n=0.45/a^2$, at different temperatures. Dashed gray curves display the noninteracting limits. 
}
\end{figure*}

Fig\@. \ref{fig:DOS_YSYK_square_T} shows the spectral function of fermions, $A_{\mathscr{G}}=\pi^{-1}\mathrm{Im}\left\{\mathscr{G}^R(\omega)\right\}$, and of bosons $A_{\mathscr{D}}=\pi^{-1}\mathrm{Im}\left\{\mathscr{D}^R(\omega)\right\}$, normalized by fermion hopping $t$ and as a function of frequency $\omega/t$, for different temperatures. Fig\@. \ref{fig:DOS_YSYK_square_T}(a) displays how the sharp peak corresponding to the van Hove singularity on the square lattice, highlighted in the noninteracting case by the dashed gray curve, is broadened and decreased in height by increasing temperature. At the highest $T$, the peak is completely suppressed as shown by the green curve in Fig\@. \ref{fig:DOS_YSYK_square_T}(c). The opposite evolution occurs for bosons: as shown in Fig\@. \ref{fig:DOS_YSYK_square_T}(b,d), the boson spectral function approaches its noninteracting value (dashed gray curve) at high temperatures, consistently with the overall decrease in magnitude of the boson self-energy shown in Fig\@. \ref{fig:Pi_YSYK_square_T}. 

\subsection{Semi-analytical estimations of self-energies at zero temperature}\label{App:semianalyt}

On the square lattice, at $T=0$ and on the imaginary axis, we have the fermionic Green's function 
\begin{equation}\label{eq:G_int_Mats}
\mathscr{G}(i\omega)=\frac{2}{\pi}\frac{1}{i\omega +\mu-\Sigma(i\omega)}K_E\left[\frac{4 t}{i \omega +\mu-\Sigma(i\omega)}\right].  
\end{equation}
Eq\@. (\ref{eq:G_int_Mats}) defines an imaginary-valued and $\omega$-odd function. 
In the following we directly employ Eq\@. (\ref{eq:G_int_Mats}) at $T=0$; the presence of the elliptic integral $K_E(z)$ will not give us full analytical expression for its various forms of convolutions in frequency, but will allow us to extract useful scaling properties with respect to frequency $i\omega$ and coupling $g'$. 
The dynamical part of the boson self-energy (\ref{eq:saddle_point_imag_gen_Pi}) for $\Sigma(i\omega)=0$ is
\begin{equation}\label{eq:deltaPi_T0}
\delta \Pi(i \Omega)=\Pi(i\Omega)-\Pi(0)=2 (g')^2 \int_{-\infty}^{+\infty} \frac{\omega}{2\pi} \mathscr{G}(i \omega)\left[\mathscr{G}(i \omega+i \Omega)-\mathscr{G}(i \omega)\right]. 
\end{equation}
Inserting Eq\@. (\ref{eq:G_int_Mats}) into Eq\@. (\ref{eq:deltaPi_T0}), we have 
\begin{multline}\label{eq:deltaPi_omega}
\delta \Pi(i\Omega)=\Pi(i\Omega)-\Pi(0)=-2 \frac{(g')^2}{t^2} \int_{-\infty}^{+\infty} \frac{d \omega/t}{2\pi} \frac{2 t}{\pi}\frac{1}{i\omega +\mu}K_E\left[\frac{4 t}{i \omega +\mu}\right] \\ \times \left\{\frac{2 t}{\pi}\frac{1}{i\omega+i\Omega +\mu}K_E\left[\frac{4 t}{i \omega+i\Omega +\mu}\right]-\frac{2t}{\pi}\frac{1}{i\omega +\mu}K_E\left[\frac{4 t}{i \omega +\mu}\right]\right\}. 
\end{multline}
Once evaluated numerically, the integral (\ref{eq:deltaPi_omega}) shows a linear dependence on $\left|\Omega\right|$ at small frequency, consistent with the marginal susceptibility of Landau-damped bosons \cite{Varma-1989, Patel-2023}. Fitting the low-energy evolution of Eq\@. (\ref{eq:deltaPi_omega}) with a linear function, inputing the value of the $T=0$ chemical potential for $n=0.45/a^2$, and for $g'=2 t^{3/2}$,  we obtain $\delta \Pi(i\Omega)\approx 1.037/t \left|\frac{\omega}{t}\right|$, which is displayed by the dashed gray line in Fig\@. \ref{fig:Pi_YSYK_square_T}(b).

The static boson self-energy can be calculated exactly with Eq\@. (\ref{eq:G_int_Mats}) at $\mu=0$: 
\begin{align}\label{eq:Pi0}
\Pi(0)&=-2 (g')^2 \int_{-\infty}^{+\infty} \frac{d \omega}{2\pi} \left[\mathscr{G}(i\omega)\right]^2= -2 (g')^2 \int_{-\infty}^{+\infty} \frac{d \omega}{2\pi} \left\{\frac{2}{\pi}\frac{1}{i\omega }K_E\left[\frac{4 t}{i \omega}\right]\right\}^2=\frac{(g')^2}{\pi} \frac{G_{4,4}^{3,3}\left(1\left|
\begin{array}{c}
 \frac{1}{2},\frac{1}{2},\frac{1}{2},\frac{1}{2} \\
 0,0,0,0 \\
\end{array}
\right.\right)}{4 \pi ^2} \nonumber \\ & \approx 1.75715 \frac{(g')^2}{\pi t}.
\end{align}
At finite chemical potential, the associated form of Eq\@. (\ref{eq:Pi0}) requires numerical evaluation: 
\begin{equation}\label{eq:Pi0_mu}
\Pi(0)=-2 (g')^2 \int_{-\infty}^{+\infty} \frac{d \omega}{2\pi} \left[\mathscr{G}(i\omega)\right]^2 = -2 (g')^2 \int_{-\infty}^{+\infty} \frac{d \omega}{2\pi} \left\{\frac{2}{\pi}\frac{1}{i\omega +\mu}K_E\left[\frac{4 t}{i \omega+\mu }\right]\right\}^2.
\end{equation}
For $g'=2 t^{3/2}$, and using the $T=0$ value of $\mu$ for $n=0.45/a^2$, we obtain $\Pi(0)\approx 2 t^2$ from Eq\@. (\ref{eq:Pi0_mu}), which semi-quantitatively agrees with the low-temperature numerics in Fig\@. \ref{fig:Pi_YSYK_square_T}(a), as displayed by the dashed gray line. 
We can qualitatively estimate the boundary of the crossover regime as the frequency/temperature $\left\{\omega_{\mathrm{cr}},T_{\mathrm{cr}}\right\}$ equal to the square root of the static boson self-energy, i.e.\@, the frequency/temperature above which the bosons are not Landau-overdamped: 
\begin{equation}\label{eq:omega_T_cr}
\hbar \omega_{\mathrm{cr}}=k_B T_{\mathrm{cr}}=\sqrt{\Pi(0)}.
\end{equation}
In particular, for $g'=2t^{3/2}$ and $n=0.45/a^2$, we have $\hbar \omega_{\mathrm{cr}}=k_B T_{\mathrm{cr}}\approx t$, which is in qualitative agreement with Fig\@. \ref{fig:Pi_YSYK_square_T}(c).

The boson propagator at $T=0$ and on the imaginary axis reads
\begin{equation}\label{eq:D_Mats_int}
\mathscr{D}(i\Omega)=\frac{2}{\pi}\frac{1}{\Omega+^2-\Pi(i\Omega)+\left(m_b^0\right)^2+4 J}K_E\left[\frac{4 J}{\Omega^2-\Pi(i\Omega)+\left(m_b^0\right)^2+4 J}\right].  
\end{equation}
Let us work at quantum criticality where $\left(m_b^0\right)^2-\Pi(0)=0$. Then Eq\@. (\ref{eq:D_Mats_int}) is
\begin{equation}\label{eq:D_Mats_QC}
\mathscr{D}(i\Omega)=\frac{2}{\pi}\frac{1}{\Omega+^2-\delta\Pi(i\Omega)+4 J}K_E\left[\frac{4 J}{\Omega^2-\delta\Pi(i\Omega)+4 J}\right].  
\end{equation}
The resulting fermion self-energy is 
\begin{multline}\label{eq:Sigma_MFL_square}
\Sigma(i\omega)=\frac{(g')^2}{2\pi} \int_{-\infty}^{+\infty} d\Omega  \mathscr{G}(i\omega+ i\Omega) \mathscr{D}(i\Omega) \\= \frac{(g')^2}{2\pi} \int_{-\infty}^{+\infty} d\Omega  \frac{2}{\pi}\frac{1}{i\omega +i\Omega +\mu}K_E\left[\frac{4 t}{i \omega+i\Omega +\mu }\right] \frac{2}{\pi}\frac{1}{\Omega^2-\delta\Pi(i\Omega)+4 J}K_E\left[\frac{4 J}{\Omega^2-\delta \Pi(i\Omega)+4 J}\right]. 
\end{multline}
for $g'=2t^{3/2}$ and $n=0.45/a^2$. Evaluating the integral in Eq\@. (\ref{eq:Sigma_MFL_square}) numerically and fitting it with the MFL-like expression 
\begin{equation}\label{eq:Sigma_MFL_fit}
\Sigma(i\omega)\approx \lambda \omega \ln\left(\frac{\left|\omega\right|}{\omega_c}\right),
\end{equation}
we obtain $\lambda \approx 0.0996 \ldots$ and $\omega_c \approx 6.825 \ldots$; these values are qualitatively in agreement with the numerics in Fig\@. \ref{fig:Sigma_YSYK_square_T_fixedn}(a,b), as shown by the dashed gray curves. 

\section{Analysis of the linear magnetotransport coefficients and of the chemical potential}\label{App:analysis_magneto}

The numerical data in Figs.\@. \ref{fig:rho_RH_cotThetaH_square_T_vargp_fixedn}, \ref{fig:rho_RH_cotThetaH_square_T_gp2_fixedn}, and \ref{fig:Sigma_YSYK_square_T_fixedn} can be qualitatively understood by referring to the different regimes encountered in our YSYK model (\ref{eq:saddle_point_imag_gen}) for increasing temperature, based on the evolution of the local spectral function (\ref{eq:spectr_local}) with increasing $T$. The two transport functions, $\Phi_{(0)}^{xx}(\omega)$ and $\Phi_{(1)}^{x y}(\omega)$, follow Eqs\@. (\ref{eq:Phi_xx_square_final}) and (\ref{eq:Phi_xy_square_final}), respectively, for the square lattice. In the following, we analyze each regime separately close to the QCP, commenting on what changes if we detune from quantum criticality. For simplicity and clarity, here we perform the analysis at fixed chemical potential, commenting on the ensuing changes when one takes into account the temperature dependence of $\mu(n,T)$.

\subsection{Low-temperature MFL regime}

\begin{figure}[h]
  \includegraphics[width=0.65\linewidth]{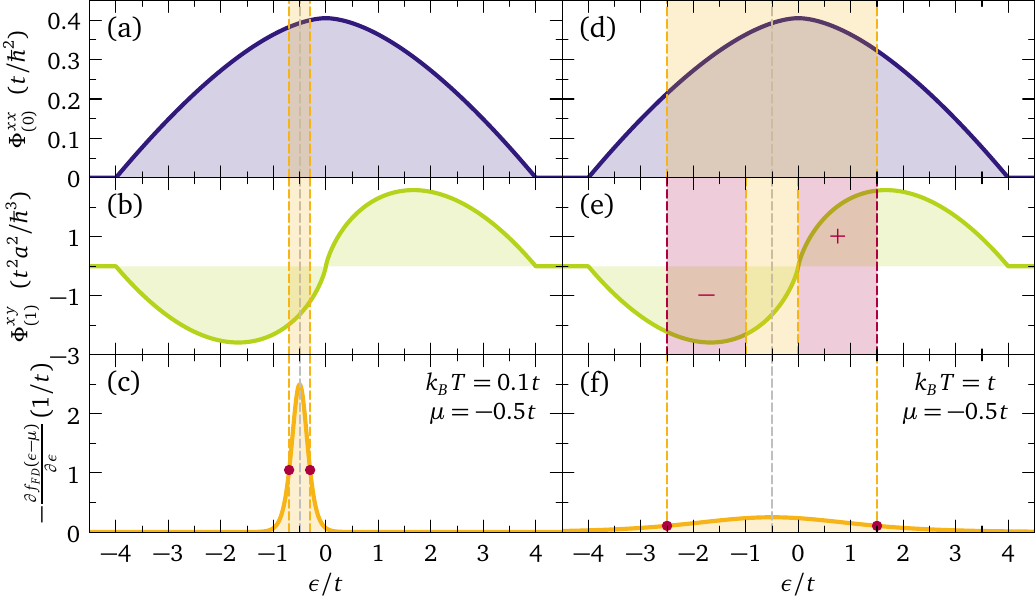}
\caption{\label{fig:scheme_Hall_T}
(a,d) Longitudinal transport function according to Eq\@. (\ref{eq:Phi_xx_square_final}) and (b,e) Hall transport function from Eq\@. (\ref{eq:Phi_xy_square_final}), as a function of normalized frequency $\epsilon/t$; derivative of the Fermi-Dirac distribution $f_{FD}(\epsilon-\mu)$ at chemical potential $\mu=-0.5 t$, for (c) $k_B T=0.1 t$ and (d) $k_B T=t$; red dots mark the value of the derivative at $\epsilon=\mu \pm 2 k_B T$. In the case of $k_B T= t  >\mu$, the corresponding Hall transport function in panel (e) has opposite sign in the regions $\left(-2k_B T+\mu, \mu\right)$ and $\left(0, 2k_B T\right)$, so these contributions approximately cancel each other in the integral over $\epsilon$ for the Hall conductivity; see discussion around Eq\@. (\ref{eq:sigmaxy_crossover_T0}). 
}
\end{figure}

The low-temperature MFL regime is characterized by a fermionic self-energy that respects the MFL scaling form \cite{Varma-1989}, a linear-in-temperature longitudinal resistivity $\rho_{\mathrm{xx}}(T)=1/\sigma_{\mathrm{xx}}(T) \sim T \ln(T)$ up to logarithmic corrections, and a quadratic-in-temperature Hall resistivity $\rho_{\mathrm{xy}}(T)=1/\sigma_{\mathrm{xy}}(T) \sim T^2$. These feature are due to the bosons providing a marginal susceptibility to the fermions through their Landau-damped bosonic propagator
\begin{equation}\label{eq:D_MFL_int}
\mathscr{D}^R(\omega)\approx\frac{2}{\pi}\frac{1}{-\delta \Pi^R(\omega)-4 J}K_E\left[\frac{4 J}{-\delta \Pi^R(\omega)+4 J}\right],
\end{equation}
dominated by the dynamical bosonic self-energy $\Pi^R(\omega)\propto -\left|\omega\right|$ at small energies. Here we have assumed $m_b(T)=\sqrt{m_b^{0}-\Pi^R(0)}\approx 0$ for $T\rightarrow 0^+$. The rest of the self-consistent loop follows as sketched in App\@. \ref{App:Numerics}, and leads to a fermionic retarded self-energy consistent with the MFL phenomenology \cite{Varma-1989}; in particular, at low temperatures and frequencies, $
\mathrm{Im}\left\{\Sigma^R(\omega)\right\}=-\lambda \pi \omega \mathrm{sign}\left\{\omega\right\}/2$.
To further understand this regime, we can employ the following reasoning. For $T \rightarrow 0^+$, the derivative of the Fermi-Dirac distribution $-\partial f_{FD}(\omega)/\partial\omega \approx \delta(\omega)$ samples essentially the $\omega=0$ component of the spectral function (\ref{eq:spectr_local}), which is itself strongly peaked around $\epsilon=\mu$ due to the low-frequency imaginary part of the self-energy $-\mathrm{Im}\left\{\Sigma^R(\omega)\right\}\approx \pi/2 \lambda \mathrm{max}\left\{k_B T, \left|\omega\right|\right\} \propto T$ (see Fig\@. \ref{fig:Sigma_YSYK_square_T_fixedn}). Since $k_B T \ll 4t$, we are also sampling a small quasiparticle energy range $\epsilon \in \left[\mu-k_B T, \mu+k_B T\right]$ around the chemical potential $\mu$. A schematics of the derivative of the Fermi-Dirac distribution and transport functions in this regime is provided in Fig\@. \ref{fig:scheme_Hall_T}. 

An important feature of the self-consistently calculated MFL fermionic self-energy is that it is slightly particle-hole asymmetric, as shown by the finite intercept of the real part at $\omega=0$ in Fig\@. \ref{fig:Sigma_YSYK_square_T_fixedn}(a,c), and by the asymmetry with respect to $\omega=0$ of the imaginary part in Fig\@. \ref{fig:Sigma_YSYK_square_T_fixedn}(b,d). This means that $\mathrm{Re}\left\{\Sigma^R(0)\right\} \neq 0$ contributes to the $\omega=0$ properties, which in turn influence the $T=0$ limit of the Hall coefficient $R_H(0)$, as described below -- see also Eq\@. (\ref{eq:RH_1stB_MFL_T0}). Therefore, at low $T$ we still have to keep the first-order terms in the expansion of the transport functions around $\epsilon=\mu$, as in Eqs\@. (\ref{eq:transport_Phi_linearized}). Using the latter, the longitudinal conductivity \emph{per spin} from Eq\@. (\ref{eq:sigma_xx_def}) can be approximated by
\begin{multline}\label{eq:sigmaxx_MFL_T0}
\sigma_{\alpha \alpha}^{(0)}(T)\approx e^2 \hbar \pi  \int_{-\infty}^{+\infty} d \epsilon \Phi_{(0)}^{\alpha \alpha}(\epsilon)\left[A(0,\epsilon)\right]^2 \approx e^2\hbar \pi \int_{-\infty}^{+\infty} d \epsilon \left[\Phi_{(0)}^{\alpha \alpha}(\mu) +\left. \frac{d \Phi_{(0)}^{\alpha \alpha}(\epsilon)}{d \epsilon}\right|_{\epsilon=\mu}(\epsilon-\mu)\right] \left[A(0,\epsilon)\right]^2 \\ = e^2 \hbar \frac{\Phi_{(0)}^{\alpha \alpha}(\mu)-\left. \frac{d \Phi_{(0)}^{\alpha \alpha}(\epsilon)}{d \epsilon}\right|_{\epsilon=\mu}\mathrm{Re}\left\{\Sigma^R(0)\right\}}{2\left|\mathrm{Im}\left\{\Sigma^R(0)\right\}\right|}\approx e^2 \hbar \frac{\Phi_{(0)}^{\alpha \alpha}(\mu)-\left. \frac{d \Phi_{(0)}^{\alpha \alpha}(\epsilon)}{d \epsilon}\right|_{\epsilon=\mu}\mathrm{Re}\left\{\Sigma^R(0)\right\}}{\pi \lambda k_B T}, \, \alpha=\left\{x,y\right\}.
\end{multline}
At fixed $\mu$, Eq\@. (\ref{eq:sigmaxx_MFL_T0}) gives $\sigma_{\alpha \alpha}^{(0)}(T)\propto 1/T$. The fixed-$\mu$ approximation holds at the lowest temperatures, but it becomes increasingly inaccurate for temperatures above the MFL regime. In the same way, the Hall conductivity per spin from Eq\@. (\ref{eq:sigma_xy_def}) is well approximated by
\begin{multline}\label{eq:sigmaxy_MFL_T0}
\frac{\sigma_{x y}^{(1)}(0)}{B}=\left|e\right|^3\hbar \int_{-\infty}^{+\infty} d \epsilon \Phi_{(1)}^{x y}(\epsilon)\left[A(0,\epsilon)\right]^3 \approx \left|e\right|^3\hbar   \int_{-\infty}^{+\infty} d \epsilon \left[\Phi_{(1)}^{xy}(\mu) +\left. \frac{d \Phi_{(1)}^{xy}(\epsilon)}{d \epsilon}\right|_{\epsilon=\mu}(\epsilon-\mu) \right] \left[A(0,\epsilon)\right]^3 \\ =\left|e\right|^3 \hbar  \frac{3}{8 \pi^2 \left|\mathrm{Im}\left\{\Sigma^R(0)\right\}\right|^2}=\left|e\right|^3 \hbar \left[-\Phi_{(1)}^{xy}(\mu) +\left. \frac{d \Phi_{(1)}^{xy}(\epsilon)}{d \epsilon}\right|_{\epsilon=\mu} \mathrm{Re}\left\{\Sigma^R(0)\right\}\right] \frac{3}{2 \pi^4 \lambda^2 (k_B T)^2},
\end{multline}
which gives $\sigma_{x y}^{(1)}(0)/B\propto 1/(k_B T)^2$ at fixed $\mu$. Notice that at half filling, \textit{i.e}, $n=0.5/a^2$ or $\mu=0$, no Hall conductivity exists because $\Phi_{(1)}^{x y}(\mu)=0$ and the self-energy is symmetric with respect to $\omega=0$ in this situation. 

Using Eqs\@. (\ref{eq:sigmaxx_MFL_T0}), (\ref{eq:sigmaxy_MFL_T0}), and (\ref{eq:transport_Phi_linearized}), we obtain Eq\@. (\ref{eq:RH_1stB_MFL_T0}) for the low-temperature Hall coefficient in MFL regime. In the fixed-$\mu$ approximation, $R_H(T)$ is $T$-independent. However, notice that the limit $\lim_{T\rightarrow 0}R_H(T)$ depends on interactions (in our case $g'$) away from particle-hole symmetry through $\mathrm{Re}\left\{\Sigma^R(0)\right\} \neq 0$, as we find in Fig\@. \ref{fig:rho_RH_cotThetaH_square_T_vargp_fixedn}.

Lastly, the temperature-dependent cotangent of the Hall angle can be estimated from Eqs\@. (\ref{eq:sigmaxx_MFL_T0}), (\ref{eq:sigmaxy_MFL_T0}), and (\ref{eq:cotThetaH_gen}), and it produces Eq\@. (\ref{eq:cotThetaH_gen}). The latter yields a linear-in-$T$ dependence at fixed $\mu$, because of $\left|\mathrm{Im}\left\{\Sigma^R(0)\right\}\right|\propto k_B T$ at small $T$. 

\subsection{Intermediate-temperature crossover regime}

Once $T$ reaches values of the order of the chemical potential $\mu$, the bosonic propagator becomes
\begin{equation}\label{eq:D_crossover_int}
\mathscr{D}^R(\omega)=\frac{2}{\pi}\frac{1}{-(\omega+i 0^+)^2-\delta \Pi^R(\omega)+\left[m_b(T)\right]^2-4 J}K_E\left\{\frac{4 J}{-(\omega+i 0^+)^2-\delta\Pi^R(\omega)+\left[m_b(T)\right]^2+4 J}\right\}.
\end{equation}
where the renormalized boson mass $m_b(T)=\sqrt{m_b^{0}-\Pi^R(0)}< m_b^0$ is finite but still lower than its bare value. It can be shown that the bosons still provide the marginal susceptibility that ultimately yields $T$-linear longitudinal resistivity, analogously to scattering off bosons at energies higher than the typical boson frequency \cite{Mahan-2000}; the same phenomenology is observed in cuprates in Electron Energy-Loss Spectroscopy (EELS) experiments \cite{Chen-2024}.

Let us further analyze this crossover regime.  
At temperatures $2 k_B T \gtrapprox \mu$, the derivative of the Fermi-Dirac distribution is broader than the spectral function $A(\omega,\epsilon)$, so we can approximately substitute $-\partial f_{FD}(\omega)/\partial\omega \approx -\partial f_{FD}(\epsilon)/\partial\epsilon$ \cite{Morpurgo-2024}.
The typical maximum frequency sampled by such derivative is $\epsilon \approx 2 k_B T$. 
At the same time, due to the fermionic self-energy $\mathrm{Im}\left\{\Sigma^R(k_B T \right\}$, the spectral function (\ref{eq:spectr_local}) is considerably broadened, and it has a FWHM comparable to its center value $\epsilon=\mu+\mathrm{Re}\left\{\Sigma^R(\omega)\right\}$ for $k_B T \gtrapprox \mu$. Now, in the transport functions $\Phi_{(0)}^{xx}(\epsilon)$ and $\Phi_{(0)}^{x y}(\epsilon)$ we are sampling a relatively large quasiparticle energy range $\epsilon \in \left[\mu-k_B T, \mu+k_B T\right]$ around the chemical potential $\mu$. This has a qualitatively different effect on the longitudinal and Hall transport functions, as shown in the schematics of Fig\@. \ref{fig:scheme_Hall_T}(d,e).
Since the longitudinal transport function does not change sign, and has a finite support $\epsilon \in \left[-4t,4t\right]$ with $k_B T \gtrapprox \mu/2 \ll 4t$, we can still approximate $\Phi_{(0)}^{\alpha \alpha}(\epsilon)\approx \overline{\Phi_{(0)}^{\alpha \alpha}(\mu)}$ as in MFL regime, in the relevant energy interval $\epsilon \in \left[\mu-k_B T, \mu+k_B T\right]$. Then
\begin{equation}\label{eq:Phi_xx_crossover}
\Phi_{(0)}^{\alpha \alpha}(\epsilon) \approx \overline{\Phi_{(0)}^{\alpha \alpha}(\mu)} \Theta\left(2 k_B T-\left|\epsilon\right|\right).
\end{equation}
Hence, at fixed $\mu$, the longitudinal resistivity from Eqs\@. (\ref{eq:sigma_xx_def}) and (\ref{eq:Phi_xx_crossover}) can be estimated by
\begin{multline}\label{eq:sigmaxx_crossover_T0}
\sigma_{\alpha \alpha}^{(0)}(T)\approx e^2 \hbar \pi \overline{\Phi_{(0)}^{\alpha \alpha}(\mu)} \int_{-\infty}^{+\infty} d\omega \int_{-4t}^{4t} d\epsilon \left[-\frac{\partial f_{FD}(\epsilon)}{\partial \epsilon}\right]\left[\frac{-\mathrm{Im}\left\{\Sigma^R(\omega)\right\}/\pi}{\left[\omega-\epsilon+\mu-\mathrm{Re}\left\{\sigma^R(\omega)\right\}\right]^2+\left[\mathrm{Im}\left\{\Sigma^R(\omega)\right\}\right]^2}\right]^2  \\ \approx\hbar e^2 \frac{\overline{\Phi_{(0)}^{\alpha \alpha}(\mu)}}{2 \left|\mathrm{Im}\left\{\Sigma^R(k_B T)\right\} \right|} \underbrace{\int_{-\infty}^{+\infty} d \epsilon \left[-\frac{\partial f_{FD}(\epsilon)}{\partial \epsilon}\right]}_{1} \propto \frac{1}{k_B T}.
\end{multline}
At the second equality in Eq\@. (\ref{eq:sigmaxx_crossover_T0}) we have performed the integral over $\omega$ of the spectral function in the limit $t \rightarrow +\infty$, approximating the self-energy $\mathrm{Im}\left\{\Sigma^R(\omega)\right\}\approx \Sigma^R(k_B T)\propto k_B T$. Hence, the longitudinal resistivity is still linear in $T$, although with a slightly different slope than in MFL regime.

On the other hand, since the Hall transport function changes sign at $\epsilon=0$, we can approximate it with the double rectangular function
\begin{equation}\label{eq:Phi_xy_crossover}
\Phi_{(1)}^{xy}(\epsilon) \approx \overline{\Phi_{(1)}^{x y}(\mu)} \left\{\Theta\left(\epsilon+2 k_B T \right)-\Theta(\epsilon)-\Theta(\epsilon)+\Theta\left(\epsilon-2 k_B T \right)\right\}. 
\end{equation}
From Eq\@. (\ref{eq:Phi_xy_crossover}), we see that, once $k_B T > \mu/2$, the area integrated in the interval $\epsilon \in \left[-k_B T, -\mu/2\right]$ is approximately compensated by its opposite contribution in $\epsilon \in \left[\mu/2, k_B T\right]$ -- see also Fig\@. \ref{fig:Phixx_xy_example}(b). The Hall conductivity at fixed $\mu$ from Eq\@. (\ref{eq:sigma_xy_def}) is then
\begin{multline}\label{eq:sigmaxy_crossover_T0}
\frac{\sigma_{x y}^{(1)}}{B}\approx \left|e\right|^3 \hbar \int_{-\infty}^{+\infty} d\omega \int_{-\infty}^{+\infty} d\epsilon \Phi_{(1)}^{xy}(\epsilon)  \left[-\frac{\partial f_{FD}(\epsilon)}{\partial \epsilon}\right] \left[\frac{-\mathrm{Im}\left\{\Sigma^R(\omega)\right\}/\pi}{\left[\omega-\epsilon+\mu-\mathrm{Re}\left\{\sigma^R(\omega)\right\}\right]^2+\left[\mathrm{Im}\left\{\Sigma^R(\omega)\right\}\right]^2}\right]^2  \\ \approx \left|e\right|^3 \hbar \frac{3}{8\pi^2} \frac{\overline{\Phi_{(1)}^{x y}(\mu)}}{2 \left|\mathrm{Im}\left\{\Sigma^R(k_B T)\right\} \right|^2} \underbrace{\int_{-\mu}^{\mu} d \epsilon \left[-\frac{\partial f_{FD}(\epsilon)}{\partial \epsilon}\right]}_{\boxed{K_2}}= \left|e\right|^3 \hbar \frac{3}{8\pi^2} \frac{\overline{\Phi_{(1)}^{x y}(\mu)}}{2 \left|\mathrm{Im}\left\{\Sigma^R(k_B T)\right\} \right|^2} \tanh\left(\frac{\mu}{2 k_B T}\right) \propto \frac{1}{(k_B T)^3}. 
\end{multline}
At the second equality in Eq\@. (\ref{eq:sigmaxy_crossover_T0}) we have performed the integral over $\omega$ of the cube of the spectral function in the limit $t \rightarrow +\infty$, approximating the self-energy $\mathrm{Im}\left\{\Sigma^R(\omega)\right\}\approx \Sigma^R(k_B T)\propto k_B T$, and taking into account that the integral over $\epsilon$ is restricted to the interval $\epsilon \in \left[-\mu,\mu \right]$ due to the sign change of the transport function (\ref{eq:Phi_xy_crossover}); the resulting integral $\boxed{K_2}$ yields $\tanh\left[\mu/(2 k_B T)\right]=\mu/(2k_B T)+\mathscr{o}\left[(k_B T)^{-3}\right]$, with the last step valid for $k_B T \gg \mu$.

Using Eqs\@. (\ref{eq:sigmaxx_crossover_T0}) and (\ref{eq:sigmaxy_crossover_T0}), the Hall coefficient is then 
\begin{equation}\label{eq:RH_1stB_crossover_T}
R_H(T)\approx \frac{1}{\left|e\right|} \frac{3 \overline{\Phi_{(1)}^{x y}(\mu)}}{2 \pi ^2 \hbar \left[\overline{\Phi_{(0)}^{x x}(\mu)} \right]^2} \tanh\left(\frac{\mu}{2k_B T}\right). 
\end{equation}
Eq\@. (\ref{eq:RH_1stB_crossover_T}) predicts a constant Hall coefficient at low temperature, consistently with Eq\@. (\ref{eq:RH_1stB_MFL_T0}) in MFL regime, but at higher temperatures $k_B T \gtrapprox \mu/2$ we have $R_H(T)\propto 1/(k_B T)$. Let us emphasize that this scaling results from the fact that we are on a square lattice, with Hall transport function $\Phi_{(1)}^{x y}(\epsilon)$ that changes its sign at $\epsilon=0$; therefore, we expect a similar scaling for other lattice configurations with similar sign-changing Hall transport functions. 
We also stress that Eq\@. (\ref{eq:RH_1stB_crossover_T}) is independent from the self-energy $\Sigma^R(\omega)$, which means that a qualitatively similar crossover is expected for any type of local interaction.

Lastly, the temperature-dependent cotangent of the Hall angle is 
\begin{equation}\label{eq:cotthetaH_1stB_crossover_T}
\cot\left[\theta_H(T)\right]=\frac{1}{\hbar B}\frac{3 \hbar}{4 \pi^2}\frac{\overline{\Phi_{(0)}^{x x}(\mu)}}{\overline{\Phi_{(1)}^{x y}(\mu)}} \frac{ \left|\mathrm{Im}\left\{\Sigma^R(0)\right\}\right|}{\tanh\left[\mu/(2k_B T)\right]}. 
\end{equation}
Eq\@. (\ref{eq:cotthetaH_1stB_crossover_T}) qualitatively predicts a linear-in-$T$ Hall angle for $k_B T \lessapprox \mu/2$, and a quadratic-in-$T$ Hall angle for $k_B T \gtrapprox \mu/2$. 

All the results in this section are further modified by the temperature dependence of the chemical potential $\mu(T)$, which occurs in fixed-density calculations. In the scanned parameter space, $\mu(T)$ decreases with $T$ for $n<0.5/a^2$, which lowers the value of the exponent $\alpha$ with respect to $\alpha=2$ predicted for fixed-$\mu$ calculations. 
The latter observation also implies that the exponent $\alpha$ is maximized by minimizing the $T$ dependence of $\mu(T)$: this feature occurs at lower doping $\Delta n$, that is, close to particle-hole symmetry, as shown by Fig\@. \ref{fig:mb_mu_square_T_varg_fixedn}(d).

\subsection{High-temperature regime at weak coupling}\label{eq:classical_metal}

At a temperature $k_B T \gtrapprox \sqrt{\Pi^R(0)}$ and coupling $(g')^2 \lessapprox 1$, the boson self-energy becomes negligible and we end up with essentially free bosons: 
\begin{equation}\label{eq:D_highT_int}
\mathscr{D}^R(\omega)\approx \frac{2}{\pi}\frac{1}{-(\omega+i 0^+)^2+\left(m_b^0\right)^2-4 J}K_E\left\{\frac{4 J}{-(\omega+i 0^+)^2+\left(m_b^0\right)^2+4 J}\right\}.
\end{equation}
The saddle-point equations for fermions and bosons then decouple, and the problem is equivalent to free bosons scattering off the fermions. Then, the fermions have a self-energy due to inelastic scattering off free bosons; in particular $\left|\mathrm{Im}\left\{\Sigma^R(0)\right\}\right|< 8t$ is less than the fermion bandwidth $W=8t$. Ultimately, by virtue of the same mechanism of acoustic-phonon scattering above the Debye temperature, the system behaves effectively as a ``classical metal'' with $T$-linear longitudinal resistivity \cite{Mahan-2000}. 

\subsection{High-temperature regime at strong coupling}\label{App:bad}

At a temperature $k_B T \gtrapprox \sqrt{\Pi^R(0)}$ and coupling $(g')^2 \gg 1$, the boson self-energy is negligible and we have 
\begin{equation}\label{eq:D_highT_bad}
\mathscr{D}^R(\omega)\approx \frac{2}{\pi}\frac{1}{-(\omega+i 0^+)^2+\left[m_b(T)\right]^2-4 J}K_E\left\{\frac{4 J}{-(\omega+i 0^+)^2+\left[m_b(T)\right]^2+4 J}\right\},
\end{equation}
with $\left[m_b(T)\right]^2 \propto T^2$. The fermions have a self-energy $\left|\mathrm{Im}\left\{\Sigma^R(0)\right\}\right|> 8t$ greater than their bandwidth, which makes for a bad metal \cite{Patel-2018b}. The Green's function (\ref{eq:G_int_def}) for fermions becomes essentially local: 
\begin{equation}\label{eq:Dyson_fermions_bad}
\mathscr{G}(i \omega_n)=\int \frac{d\vec{k}}{(2 \pi)^2} \frac{1}{i\omega_n-\epsilon_{\vec{k}}-\Sigma(i\omega_n)}\approx \int \frac{d\vec{k}}{(2 \pi)^2} \frac{1}{i\omega_n-\Sigma(i\omega_n)}=\frac{1}{i\omega_n-\Sigma(i\omega_n)}.
\end{equation}
We see that the fermion dispersion $\epsilon_{\vec{k}}$ does not matter in this strong-coupling bad-metal regime, which implies that the analysis here is universal for any lattice bandstructure. When the self-energy of fermions is so large, we can approximate $\mathscr{G}(i \omega_n) \approx 1/\Sigma(i\omega_n)$, and look for power-law solutions of the self-energy with the result $\Sigma(i\omega_n) \propto -i \left|\omega_n\right|^{1/2}$ similarly to completely local SYK-like and Yukawa-SYK models \cite{Sachdev-1993,Schmalian-2000,Georges-2000,Westfahl-2003,Song-2017,Gu-2020,Kim-2020,Patel-2018,Esterlis-2019,Wang-2020a,Chowdhury-2020a,long-paper,short-paper,Chowdhury-2022}. This scaling leads to $T$-linear longitudinal resistivity in bad-metal regime (Parcollet-Georges mechanism) \cite{Parcollet-1999,Georges-2000,Georges-2001,Cha-2020}. We defer a detailed analysis of the bad-metal regime to future works.

\twocolumngrid

\end{document}